\documentclass[iop]{emulateapj}
\usepackage{graphicx}
\usepackage{amssymb}
\usepackage{threeparttable}

\mathchardef\mhyphen="2D

\slugcomment{Submitted to ApJ 2013 April 9; Accepted 2013 May 21}

\shorttitle{Pa$\alpha$ LF of H {\sc ii} Regions}
\shortauthors{Liu et al.}

\begin{document}


\title{The P\lowercase{a}$\alpha$ Luminosity Function of H \lowercase{{\sc ii}} Regions in Nearby Galaxies from HST/NICMOS$^{\ast}$}


\author{
Guilin Liu\altaffilmark{1,2},
Daniela Calzetti\altaffilmark{1},
Robert C. Kennicutt Jr.\altaffilmark{3},
Eva Schinnerer\altaffilmark{4},
Yoshiaki Sofue\altaffilmark{5},
Shinya Komugi\altaffilmark{6},
Fumi Egusa\altaffilmark{7}, 
and
Nicholas Z. Scoville\altaffilmark{8}
}

\affil{$^1$Astronomy Department, University of Massachusetts, Amherst, MA 01003, USA}
\affil{$^2${\em Current address:} Center for Astrophysical Sciences, Department of Physics \& Astronomy, Johns Hopkins University, Baltimore, MD 21218, USA; Email: liu@pha.jhu.edu}
\affil{$^3$Institute of Astronomy, University of Cambridge, Madingley Road, Cambridge, CB3 0HA, UK}
\affil{$^4$MPI for Astronomy, K\"onigstuhl 17, 69117 Heidelberg, Germany}
\affil{$^5$Department of Physics, Meisei University, 2-1-1 Hodokubo, Hino, Tokyo 191-8506, Japan}
\affil{$^6$Joint ALMA Office, Alonso de C\'{o}rdova 3107, Vitacura, Santiago 763-0355, Chile}
\affil{$^7$Institute of Space and Aeronautical Science, Japan Aerospace Exploration Agency, 3-1-1 Yoshino-dai, Chuo-ku, Sagamihara, Kanagawa 252-5210, Japan}
\affil{$^8$California Institute of Technology, MC 249-17 Pasadena, CA 91125, USA}

\altaffiltext{$\ast$}{Based on observations taken with the NASA/ESA
Hubble Space Telescope obtained at the Space Telescope Science Institute,
which is operated by AURA, Inc., under NASA contract NAS5-26555. These observations
are associated with program GO-11080.}


\begin{abstract}

The H {\sc ii} region luminosity function (LF) is an important tool for deriving the 
birthrates and mass distribution of OB associations, and is an excellent tracer of 
the newly formed massive stars and associations.
To date, extensive work (predominantly in H$\alpha$) 
has been done from the ground, which is hindered by dust extinction and the severe 
blending of adjacent (spatially or in projection) H {\sc ii} regions. Reliably 
measuring the properties of H {\sc ii} regions requires a linear resolution $<$40 pc, 
but analyses satisfying this requirement have been done only in a handful of
galaxies, so far. As the first space-based work using a galaxy sample, we have
selected 12 galaxies from our HST NICMOS Pa$\alpha$ survey and studied the luminosity 
function and size distribution of H {\sc ii} regions both in individual galaxies and 
cumulatively, using a virtually extinction-free tracer of the ionizing photon rate.
The high angular resolution and low sensitivity to diffuse emission of 
NICMOS also offer an advantage over ground-based imaging by enabling a 
higher degree of de-blending of the H {\sc ii} regions. 
We do not confirm the broken power-law LFs found in ground-based studies. 
Instead, we find that the LFs, both individual and co-added, follow a 
single power law $dN(L)/d\ln L \propto L^{-1}$, consistent with the mass function 
of star clusters in nearby galaxies, and in agreement with the results of the 
existing analyses with HST data. The individual and co-added size distribution of
H {\sc ii} regions are both roughly consistent with $dN(D)/d\ln D \propto D^{-3}$,
but the power-law scaling is probably contaminated by blended regions or complexes.

\end{abstract}





\section{INTRODUCTION}

As a well-established tracer of the newly formed massive stars and star 
associations and clusters, the distribution of H {\sc ii} regions has sparked a 
long interest since the pioneering investigations in nearby spiral galaxies, as 
reviewed by \citet{Kennicutt92}. Extensive statistical analyses of galactic H {\sc ii} 
regions and complexes have been undertaken during the recent couple of decades, 
predominantly focusing on the luminosity functions (LFs), an important tool for 
deriving the birthrates and mass distribution of OB associations. These studies, 
mostly based on ground-based imaging in the hydrogen H$\alpha$ emission line, 
have yielded a collection of remarkable results. Notably, \citet{Kennicutt89} 
analysed a sample of 30 galaxies with Hubble type ranging from Sb to Irr and 
found the locus of differential LFs to follow a power law expressed as 
$dN/dL\propto~L^{-2\pm0.5}$ for luminosities $L_{\rm H\alpha}\gtrsim 10^{37}$ erg s$^{-1}$. 
They also found that a subsample ($\sim$20\% of the total) of the galaxies is better 
described by a double power-law LF, which they termed ``type {\sc ii}'', with a break 
in the slope for luminosities $L_{\rm H\alpha} < 10^{38.7}$ erg s$^{-1}$.' 
The H$\alpha$ analyses of H {\sc ii} region LFs were later extended to early-type 
(Sa) spirals \citep{Caldwell91}, H {\sc i}-selected galaxies \citep{Helmboldt05} 
and low surface brightness galaxies \citep{Helmboldt09}.

Two conventions for the representation of the LFs exist in the literature, 
one for number counts binned linearly in luminosity, and the other binned 
logarithmically. We adopt the logarithmic formalism, $dN/d\ln L\propto L^{\alpha}$, 
equivalent to $dN/dL\propto L^{\alpha-1}$ (note $\alpha$ is generally negative).
The existence of the slope break at an H$\alpha$ luminosity of $L_{\rm br}\sim10^{38.5\mhyphen38.7}$ erg s$^{-1}$
is confirmed in the LF of M101 by \citet{Scowen92}, M51 by \citet{Rand92}, eight 
other galaxies by \citet{Rozas96, Rozas99, Rozas00}, and more recently, in the co-added
LF derived from the 18,000 H {\sc ii} regions in 53 galaxies by \citet{Bradley06}. 
This high luminosity break is hence considered robust, and termed the Str\"{o}mgren 
luminosity by \citet{Beckman00}, in accordance to their hypothesis that the break 
results from a transition from ionization-bounded to density-bounded H {\sc ii} regions. 
Conversely, \citet{Oey98} suggest that the luminosity limit $L_{\rm H\alpha}\sim10^{38.5}$ erg s$^{-1}$ 
is linked to the regions transitioning from stochastic sampling of the stellar initial 
mass function (IMF) to full sampling. The ground-based H$\alpha$ LFs generally show a 
``type {\sc ii}'' profile when a steep slope at the bright end ($\alpha_{\rm br+}\sim-1$) 
is connected to a flatter one in the less luminous regime ($\alpha_{\rm br-}\sim-$(0.3--0.5), 
see \S ~\ref{sec:h2lf_ind} for Pa$\alpha$ LFs of individual galaxies). Here and throughout
this section, H {\sc ii} regions brighter (fainter) than $L_{\rm br}$ are
called {\em super-break} ({\em sub-break}), and the quantities that describe their 
behavior are marked with a subscript ``br$+$'' (``br$-$'').'

The hypothesis that the break in the slope has a physical nature 
rather than being an observational artifact has been under suspicion for long, especially 
when considering that the resolution of ground-based images, dominated by seeing, 
ranges from 0\farcs8 to 3\arcsec--4\arcsec, corresponding to up to $\sim$400 pc at 
a distance of 20 Mpc. The ground-based work is potentially hindered by blending of adjacent (spatially 
or in projection) H {\sc ii} regions. This results in an artificially pumped-up bright end 
of the number distribution together with a depleted faint end. As shown by \citet{Pleuss00}, 
measuring the properties of H {\sc ii} regions reliably requires a linear resolution of 
$\sim$40 pc or better, translating to an angular resolution of $\lesssim$0.4\arcsec\ for a 
galaxy located 20 Mpc away. The Hubble Space Telescope (HST), which has a $<$0.3\arcsec~
resolution (actual value depending on the instrument), is thus an ideal facility to tackle 
the H {\sc ii} region statistics. 

The high angular resolution of the HST data motivated \citet{Scoville01} and \citet{Pleuss00} 
to scrutinize the previous conclusions drawn for M51 and M101, respectively, using HST H$\alpha$ 
images. Interestingly, neither of the two galaxies shows evidence for the broken power-law H$\alpha$ 
LF obtained in ground-based studies. Both galaxies exhibit a single power-law LF characterized by 
$\alpha\simeq-1$, or equivalently, $\alpha_{\rm br-}\simeq-1$, which is much steeper than previously 
determined $\alpha_{\rm br-}$ values. These values of $\alpha_{\rm br-}$ are consistent with  
results on the luminosity and mass functions of star clusters \citep[e.g., a recent analysis of HST WFC3 
data of M83 and M51,][]{Chandar10,Chandar11}. In addition, both galaxies contain only a small amount 
($\lesssim 1\%$ of the total) of super-break H {\sc ii} regions. This result indicates 
that super-break H {\sc ii} regions may not exist after all, but are simply blends of 
multiple regions and associated diffuse gas. If this is correct, the break at $L_{\rm br}$ is actually 
where the power law scaling truncates. The discrepancy between these HST analyses of two galaxies 
and the more extensive previous ground-based work brings urgency to a more systematic investigation 
using the high angular resolution data of the HST. Although the more recent HST study by
\citet{Lee11} finds evidence for the existence of a break luminosity in the H {\sc ii} LF of M51, 
the break occurs at a luminosity fainter than the type {\sc ii} LF break by a factor of $\sim30$
($L_{\rm H\alpha}=10^{37.1}$ erg s$^{-1}$) and is thus likely to have a different origin.
Meanwhile, the upper cutoff of the H {\sc ii} LF of M31 is below the type {\sc ii} LF break
luminosity, disqualifying this galaxy for testing the above discrepancy \citep{Azimlu11}.

Far less investigated than the impact of blending is the effect of dust attenuation on the 
derivation of the LFs. Variable dust attenuation across galactic disks induces scatter on the 
measurements. Furthermore, since brighter H {\sc ii} regions are generally associated with larger 
extinctions \citep[e.g.,][]{Calzetti07}, the presence of dust can hinder accurate derivations of 
H {\sc ii} LFs and also the verification of the presence of a luminosity truncation, as discussed 
in \citet{Scoville01}. Less susceptible to dust attenuation, near-IR tracers of current star 
formation activity, such as the Paschen or Brackett emission lines, can be used as good alternatives 
to H$\alpha$ and other optical tracers. Such measurements are only available for very limited samples, 
but an HST/NICMOS Pa$\alpha$ survey of the central $\sim$1\arcmin~field of 84 nearby galaxies has 
been recently completed (PI: Daniela Calzetti, GO: 11080). The Pa$\alpha$ hydrogen line emission, 
with a central rest-frame wavelength of 1.8756 $\mu$m, suffers much less from dust attenuation than 
H$\alpha$. The visual extinctions of individual H {\sc ii} regions in the almost face-on spirals 
like M51 can reach values as high as $A_V\simeq6$ mag \citep{Scoville01}, implying that the H$\alpha$ 
line is dimmed by two orders of magnitude, but Pa$\alpha$ only a factor of $\sim$2. In general, we 
expect a 20\% dimming factor for Pa$\alpha$ to be more typical in nearby disk galaxies \citep{Calzetti07}.

In parallel to the LF investigations, the size distribution of the H {\sc ii} 
regions, physically related to the LF, has also been studied at length
\citep[e.g.,][]{Hodge89,Knapen98,Petit98,Youngblood99,Hodge99}. More recently, \citet{Oey03} 
compiled and refitted the data in the literature and suggest that the differential 
nebular size distribution follows a power law of the form $N(D)dD \propto D^{-4}dD$,
with flattening at diameters below $\sim$260 pc. Although the regions that enter in
the power-law distribution are clearly blends of multiple smaller objects, 
their investigation provides important clues to the clustering properties of 
H {\sc ii} regions.

HST is relatively insensitive to the low surface brightness emission from the diffuse ionized gas, 
due to its intrinsic high angular resolution \citep{Kennicutt07}. Thus, our HST Pa$\alpha$ images are 
only sensitive to the compact emission from associations of relatively massive ($\gtrsim$30~$M_{\odot}$), 
young ($\lesssim$8--10 Myr) stars. This feature, in addition to its high angular resolution, further 
makes characterizing H {\sc ii} regions more reliable, as the associated diffuse gas also plays an 
important role in preventing H {\sc ii} regions from being de-blended \citep{Pleuss00}. 

We select 12 spirals out of the 84 targets in our HST/NICMOS program to construct our sample 
so that every member contains at least 50 well-detected H {\sc ii} regions (see \S~\ref{sec:sample} 
for further details). Our sample ranges from Sb to Sd in morphological type (see Table~\ref{tbl:sample}); 
these are later types than the sample as a whole, and is a result of the bias introduced by the 
requirement that each galaxy contains an appreciable number of H {\sc ii} regions. This is our 
best sample to infer the intrinsic (i.e., minimally affected by blending and dust attenuation) 
luminosity function of H {\sc ii} regions with high accuracy. This work is the first systematic 
investigation of the H {\sc ii} region statistics utilizing the HST Pa$\alpha$ images of a 
significant galaxy sample.

\begin{figure*}
\centering
\includegraphics[scale=0.8]{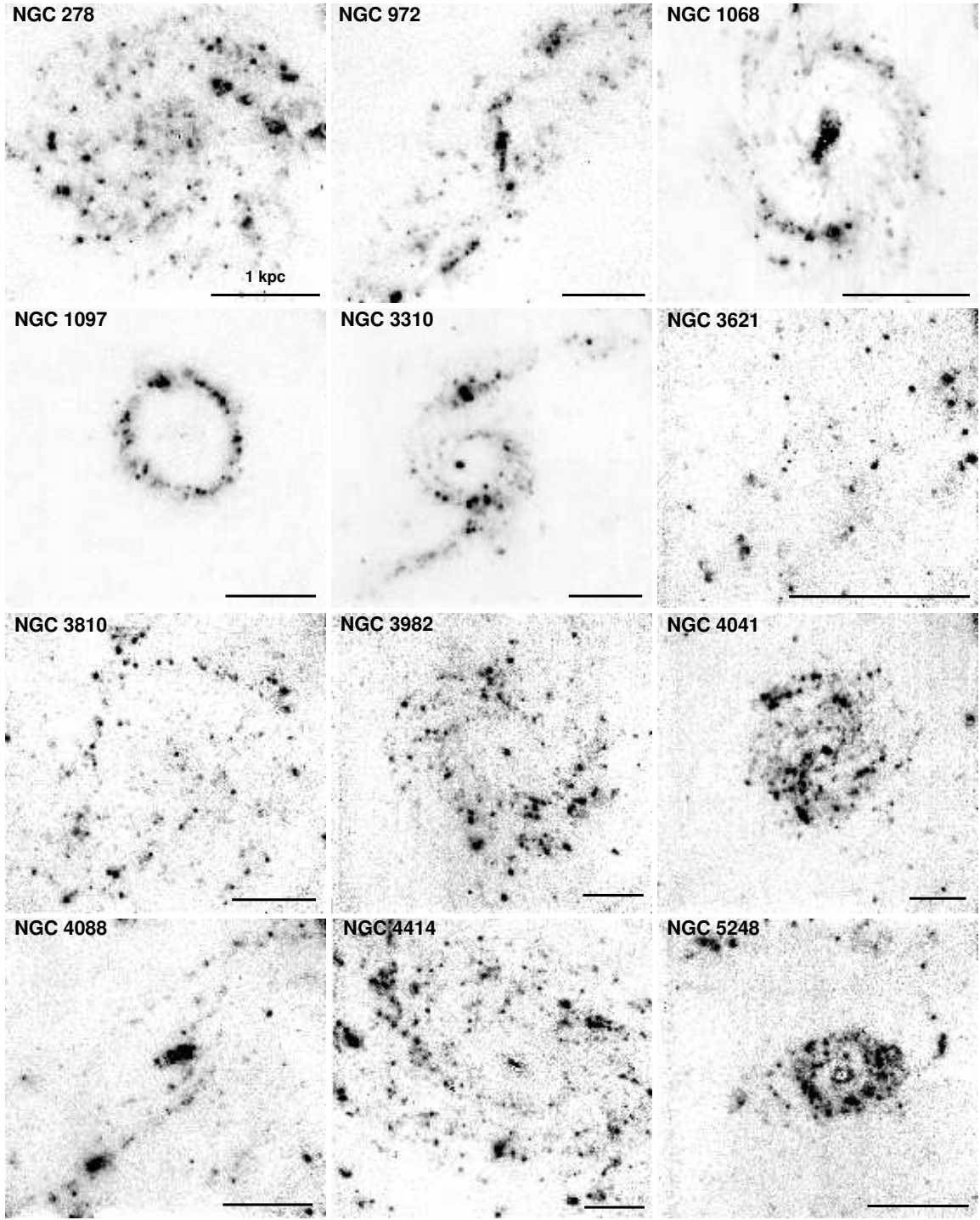}
\caption{The HST/NICMOS Pa$\alpha$ images of the 12 star-forming spiral galaxies that form 
our sample. 
The underlying stellar continuum has been subtracted from the line emission images. The data 
have an angular resolution of 0.26\arcsec\ and a field of view of 51\arcsec\ for each target. 
The bar at the bottom right corner of each panel shows the linear scale of 1 kpc.}
\label{fig:h2lf_images}
\end{figure*}

The paper is organized as follows: \S 2 describes the basic properties of our galaxy 
sample, provides some detail on the data reduction and the identification and photometry
of H {\sc ii} regions; \S 3 presents our analysis of H {\sc ii} luminosity function
and size distribution in individual galaxies; \S 4 is devoted to the analysis of co-added 
H {\sc ii} region distributions by combining the whole sample, and its comparison to 
previous ground-based studies in H$\alpha$; the results are discussed in \S 5
and summarized in \S 6.

\section{Data Sample}

\subsection{Sample Description}
\label{sec:sample}

The galaxies analyzed in the present paper are selected from a recent HST/NICMOS survey of the 
central $\sim$1\arcmin~field of 84 nearby galaxies evenly distributed across all Hubble and bar 
types (HST/GO--11080). Targeting the redshifted hydrogen Pa$\alpha$ emission line using the NIC3 
Camera, these observations were completed in February 2008. 

In order to optimize our analysis, we down-select a sub-sample from the original list of 84 galaxies 
according to the following criteria:

\begin{enumerate}

\item Highly inclined spirals ($i>70^{\circ}$) are excluded to avoid severe 
blending of overlapping (in projection) H {\sc ii} regions. 

\item Galaxies with heliocentric recession velocities higher than 1600 km s$^{-1}$ are 
rejected. For larger redshifts, the throughput of F187N filter is low, leading to a large 
transmittance correction which yields uncertain Pa$\alpha$ fluxes.

\item To ensure the fidelity of the statistical analysis, a galaxy is rejected if it contains fewer than 
50 H {\sc ii} regions detected with signal-to-noise ratio $S/N\geqslant3$ within the field of view (FoV) 
of the NIC3 camera. If the galaxy hosts a central active galactic nucleus (AGN), these H {\sc ii} regions need to be $>$200 pc 
from the galaxy center to avoid contamination from it. As a reference, our observations cover the central 
5~kpc or less of each galaxy. 

\item Only spirals located within 23 Mpc are considered, implying a worst spatial resolution of 29 pc 
(for the 0.26\arcsec\ resolution of our NIC3 images, see next section), well below the 40 pc maximum 
size suggested by \citet{Pleuss00} for reliably measuring H {\sc ii} region properties.

\end{enumerate}

The above selection procedure filters out 6/7 of the original sample, leaving 12 spiral
galaxies. The Hubble type of these selected galaxies ranges from Sab to Sd, which, as mentioned 
in the introduction, is due to the third selection criteria that biases the sample towards 
later type spirals. The basic properties of our sample galaxies are 
summarized in Table~\ref{tbl:sample}.

\begin{center}
\begin{deluxetable*}{l l c c c c c r r}
\setlength{\tabcolsep}{0.02in} 
\tablecaption{Characteristics of the sample galaxies and their HST Pa$\alpha$ data. \label{tbl:sample}}
\tablehead{
\colhead{Galaxy} & \colhead{Morph} & \colhead{$T$} & \colhead{$i$} & \colhead{Dist} 
& \colhead{Ref} & \colhead{FoV} & \colhead{Res} & \colhead{AGN type}\\ 
\colhead{(1)} & \colhead{(2)} & \colhead{(3)} & \colhead{(4)} & \colhead{(5)} 
& \colhead{(6)} & \colhead{(7)} & \colhead{(8)} & \colhead{(9)} 
}

\startdata

NGC 278  & SAB(rs)b  				& 2.9 & 12.8 & 11.8 & 1 & 2.9 & 15 & 	    \\
NGC 972  & Sab   					& 2.0 & 65.9 & 15.5 & 2 & 3.8 & 20 &	    \\
NGC 1068  & (R)SA(rs)b    			& 3.0 & 21.1 & 10.1 & 2 & 2.5 & 13 & Seyfert 1.8\tablenotemark{a}\\
NGC 1097  & SB(s)b-LINER b  			& 3.3 & 37.0 & 14.2 & 2 & 3.5 & 18 & Seyfert 1\tablenotemark{b}\\
NGC 3310  & SAB(r)bc pec-H {\sc ii}	& 4.0 & 31.2 & 17.5 & 3 & 4.3 & 22 &      \\
NGC 3621  & SA(s)d  					& 6.9 & 65.6 & 7.11 & 4 & 1.8 &  9 &      \\
NGC 3810  & SA(rs)c    				& 5.2 & 48.2 & 15.3 & 2 & 3.8 & 19 &      \\
NGC 3982  & SAB(r)b? 				& 3.2 & 29.9 & 21.2 & 5 & 5.2 & 27 & Seyfert 1.9\tablenotemark{a}\\
NGC 4041  & SA(rs)bc?   				& 4.0 & 22.0 & 22.7 & 1 & 5.6 & 29 &      \\
NGC 4088  & SAB(rs)bc  				& 4.7 & 69.4 & 14.3 & 2 & 3.5 & 18 &      \\
NGC 4414  & SA(rs)c?    				& 5.2 & 54.0 & 21.4 & 6 & 5.3 & 27 & trans. obj. 2\tablenotemark{a}\\
NGC 5248  & SAB(rs)bc  				& 4.0 & 56.4 & 12.7 & 2 & 3.1 & 16 &      

\enddata

\tablecomments{
(1) Galaxy name. 
(2) Morphology, from the NASA/IPAC Extragalactic Database (NED, http://ned.ipac.caltech.edu).
(3) Numerical Hubble type \citep[as defined in][]{RC3}, from the HyperLeda database 
\citep[][http://leda.univ-lyon1.fr]{Paturel03}.
(4) Inclination ($^\circ$), from the HyperLeda database.
(5) Distance (Mpc), from the literature listed in column 6.
(6) References for distances: 1. \citet{Tully88}; 2. \citet{Tully09}; 3. \citet{Terry02}; 
4. \citet{Rizzi07}; 5. \citet{Ngeow06}; 6. \citet{Saha06}.
(7) Linear scale of field of view (kpc).
(8) Linear resolution (parsec).
(9) AGN type, if it exists: a. from \citet{Ho97}; b. from NED.
}
\end{deluxetable*}
\end{center}

\begin{center}
\begin{deluxetable*}{c r r r r r r}
\setlength{\tabcolsep}{0.02in} 
\tablecaption{Catalog of the 1907 H {\sc ii} Regions detected with $S/N\geqslant3$ in our sample galaxies. \label{tbl:catalog}}
\tablehead{
\colhead{Galaxy} & \colhead{ID} & \colhead{RA} & \colhead{Dec} 
& \colhead{$\log L_{\rm Pa\alpha}$} & \colhead{$D$} & \colhead{$S/N$}\\ 
\colhead{(1)} & \colhead{(2)} & \colhead{(3)} & \colhead{(4)} & 
\colhead{(5)} & \colhead{(6)} & \colhead{(7)}
}

\startdata

NGC278  &   1  &  00:52:06.7  &  +47:33:00.2  &  36.79  &   33.4  &   3.4 \\
NGC278  &   2  &  00:52:06.7  &  +47:33:01.2  &  37.13  &   39.8  &   5.3 \\
NGC278  &   3  &  00:52:06.7  &  +47:33:02.2  &  37.07  &   30.7  &   5.3 \\
NGC278  &   4  &  00:52:06.7  &  +47:33:03.6  &  37.16  &   46.3  &   5.7 \\
NGC278  &   5  &  00:52:06.7  &  +47:32:57.0  &  36.85  &   43.4  &   3.5 \\
NGC278  &   6  &  00:52:06.7  &  +47:33:01.3  &  38.13  &  112.0  &  17.1 \\
NGC278  &   7  &  00:52:06.6  &  +47:33:05.4  &  37.28  &   62.1  &   6.9 \\
NGC278  &   8  &  00:52:06.6  &  +47:33:03.8  &  36.78  &   33.4  &   3.5 \\
NGC278  &   9  &  00:52:06.6  &  +47:33:03.1  &  37.43  &   53.2  &   7.8 \\
NGC278  &  10  &  00:52:06.6  &  +47:33:04.6  &  36.92  &   41.4  &   4.3 \\
NGC278  &  11  &  00:52:06.6  &  +47:33:02.7  &  37.84  &   67.7  &  13.0 \\
NGC278  &  12  &  00:52:06.6  &  +47:33:00.7  &  37.02  &   37.0  &   4.5 \\
NGC278  &  13  &  00:52:06.6  &  +47:33:03.6  &  37.13  &   43.4  &   5.4 \\
NGC278  &  14  &  00:52:06.6  &  +47:33:01.1  &  37.35  &   43.4  &   7.0 \\
NGC278  &  15  &  00:52:06.5  &  +47:33:02.2  &  37.84  &   68.0  &  12.8 

\enddata

\tablecomments{
(1) Galaxy name. 
(2) Identification number of the H {\sc ii} region, starting
from one for each galaxy. 
(3) Right ascension of the identified region (hh:mm:ss.s).
(4) Declination of the region (dd:mm:ss.s). 
(5) Pa$\alpha$ luminosity of the region (erg s$^{-1}$), in logarithmic form.
Uncertainties on $L_{\rm Pa\alpha}$ can be estimated as explained in 
the text (Section \ref{sec:h2_iden}).
(6) Equivalent diameter which matches the area of a circle $\pi D^2/4$ to that 
of the region (parsec).
(7) Signal-to-noise ratio of the region.
Only a portion of this table is shown here to demonstrate 
its form and content. The full data set is available via this link: 
http://www.pha.jhu.edu/{\raise.17ex\hbox{$\scriptstyle\mathtt{\sim}$}}liu/papers/h2lf.tab2.dat}

\end{deluxetable*}
\end{center}	
		
\subsection{Observations and Data Reduction}
\label{sec:reduc}

The 12 sample galaxies were observed between February 13, 2007 and February 3, 2008 with
the NICMOS/NIC3 camera on board the HST. NIC3 images have a 51\arcsec$\times$51\arcsec\ 
FoV and a pixel scale of 0.2\arcsec~with an undersampled point spread function (PSF). In 
this survey, each observation consists of images taken in two narrowband filters: one 
centered on the Pa$\alpha$ hydrogen recombination line (1.87~$\mu$m, F187N), and the other 
on the adjacent narrow-band continuum (F190N). 
Each set of exposures were made with a 6-position small ($<$1\arcsec~step) dither to maximize 
recovery of spatial information and rejection of cosmic rays and their persistence. Exposures 
of 160 and 192 seconds per dither position in the narrow-band filters targeting the line emission 
and the adjacent continuum, respectively, reach a 1$\sigma$ detection limit of
4.5$\times$10$^{-17}$~erg~s$^{-1}$~cm$^{-2}$~arcsec$^{-2}$ in the continuum-subtracted 
Pa$\alpha$ images. Assuming the IMF given by \citet{Kroupa01} and applying the relation between
star formation rate and Pa$\alpha$ luminosity \citep{Calzetti07a}, we find the limiting SFR per 
unit area to be $\sim$0.0015 M$_{\odot}$~yr$^{-1}$~kpc$^{-2}$.


The data reduction steps consist of removing the NICMOS {\it Pedestal} Effect with the
{\it pedsub} task in the STSDAS package of IRAF. The offsets between the dither images
are then found using the {\it crossdriz} and {\it shiftfind} tasks by computing the 
cross-correlation functions of the individual maps. After that, we apply the data 
quality files to create a static pixel mask for each dither image before these
256$\times$256 pixel maps are drizzled onto a larger (1024$\times$1024) and finer
(0.1\arcsec~per pixel) grid frame using the {\it drizzle} package. A single median 
image is then made by combining all the drizzled images using the offsets determined 
in the previous step. In order to reject cosmic rays, the median image is ``blotted'' 
or reverse-drizzled back to the original dimensions of each original image, and also 
shifted. By creating the derivative images of the outputs with the {\it deriv} task, the
value of each pixel essentially represents the largest gradient across that pixel. 
A comparison between the median and blotted images enables us to produce a cosmic 
ray mask using {\it driz\_cr} task. The static mask file is then multiplied with each 
individual bad pixel mask file. Equipped with these ``master'' mask files, we drizzled 
each input image onto a single output image, applying the shifts and the mask files 
previously created. More details are available in the {\sl HST Dither Handbook}\footnote{http://www.stsci.edu/hst/HST\_overview/documents}, which we 
followed to derive our final images.

After applying all the steps above on both the F187N and F190N images, we aligned each 
filter pair  to each other, scaled the F190N image to the F187N using the filter transmission
curves, and subtracted the former from the latter to produce a stellar-continuum-free line 
emission image. We note that the line emission maps suffer slightly from residual shading, by 
exhibiting a pair of vertical banding features. We remove the artifact column by column, by 
calculating the median value of each column of the image and subtracting it. This strategy works 
remarkably well in the case of sparse emission blobs (the typical case in our images). The global 
image background is also removed by this subtraction. The transmittance correction for the
line emission images is finally performed, adopting the recession velocities from the 
literature and the F187N throughput curve from the HST instrument handbook. The final
continuum-subtracted Pa$\alpha$ images have PSFs of 0.26\arcsec~(FWHM). 
The Pa$\alpha$ data of two galaxies from this HST campaign analyzed by \citet[][NGC 1097]{Hsieh11} 
and \citet[][NGC 6951]{vanderLaan13} were reduced following an identical procedure.

\subsection{H {\sc ii} Region Identification and Photometry}	
\label{sec:h2_iden}

The identification, photometry and cataloging of H {\sc ii} regions in the HST Pa$\alpha$ images
are carried out using the IDL program {\sl HIIphot}. This software, developed by \citet{Thilker00}, is 
specifically designed for automated photometry and statistical analysis of H {\sc ii} regions on 
H$\alpha$ images of nearby galaxies, and is directly applicable to our Pa$\alpha$ data. 
Its capability to deal with crowded fields and adapt to irregular source morphologies 
results in accurate photometric characterization of H {\sc ii} regions.
 
The {\sl HIIphot} algorithm can be summarized as follows. On a pure hydrogen emission line 
image of a galaxy, it identifies H {\sc ii} regions by smoothing the image with kernels of different 
sizes, utilizes object recognition techniques to fit models (``seeds'') to peaks that are identified 
via a minimum assigned $S/N$ ratio, and then assigns pixels to identified H {\sc ii} regions. After 
that, an iterative growing procedure allows for departures from these ``seeds''. The boundaries are 
built up at successively fainter isophotal levels that keep growing until either boundaries of other 
regions are encountered, or a pre-set limiting gradient of surface brightness is reached implying its 
arrival at the background level. Two-dimensional interpolations are then performed by fitting the 
background pixels around the designated H {\sc ii} regions to estimate the local background behind 
the source. The final information on the detected sources is recorded in a file for further analysis.

The astrometric and photometric information of all the 1907 H {\sc ii} regions detected with 
$S/N\geqslant3$ are tabulated in Table \ref{tbl:catalog}. Note that the calibration of our HST data
is accurate within $\sim$5\%, the total uncertainty introduced by instrumental artifacts and continuum 
subtraction is $\sim$10\%, thus the relative uncertainty of the derived Pa$\alpha$ luminosity can
be estimated as $(0.05^2+0.10^2+(S/N)^{-2})^{1/2}$, which is about 35\% for an H {\sc ii} region 
detected with $S/N=3$.

In agreement with the range of electron temperatures generally derived for H {\sc ii} regions in 
metal-rich spiral galaxies \citep{Bresolin04, Li13}, throughout the rest of this section we adopt 
$T_e\sim5,000$--10,000 K. We also adopt $n_e=100$ cm$^{-3}$, implying an intrinsic range 
H$\alpha$/Pa$\alpha$=7.4--8.5, i.e. a mean value of 8 and a dispersion $\lesssim$10\% \citep{Osterbrock06}. 
Although our measured $n_e$ are significantly smaller than what adopted here (see \S~\ref{sec:n_e}), 
we recall that the hydrogen line ratio has a weak dependence on the adopted $n_e$.

Since we do not have dust extinction maps for 
our sample, we use H {\sc ii} regions extinction distributions from the literature as our reference, 
and assume our galaxies behave in a similar way as other galaxies. The dust extinction distribution 
for H {\sc ii} regions in M51, measured by \citet{Scoville01} using HST data both in H$\alpha$ and 
Pa$\alpha$, covers the range $A_V$=0--6 mag, with $\sim$80\% of the regions having $A_V$ values between 
2 and 4 and a mean value of 3.1. Adopting the Galactic extinction curve of \citet{Cardelli89}, this 
extinction translates into an observed H$\alpha$/Pa$\alpha$ line ratio between 0.72--2.47, with a mean 
value of 1.25. \citet{Calzetti07} measure the dust extinction distribution in 33 nearby galaxies, using 
HST Pa$\alpha$ and ground-based H$\alpha$. Albeit the low resolution of the ground H$\alpha$ images may 
reduce the measured extinction, they still obtain a median $A_V$=2.2 mag, which translates into an 
observed H$\alpha$/Pa$\alpha$ line ratio of 2.2. For practical purposes, these measurements mean that 
for an {\it observed} break luminosity in H$\alpha$, $L_{\rm H\alpha, br}\sim10^{38.6}$~erg~s$^{-1}$ the 
corresponding $L_{\rm Pa\alpha, br}\sim$10$^{38.3\mhyphen38.5}$~erg~s$^{-1}$, only 0.1--0.3 dex lower 
than the unextincted break luminosity in Pa$\alpha$.

\section{P\lowercase{a}$\alpha$ Luminosity Function and Size Distribution}

\subsection{Individual Galaxies: Pa$\alpha$ Luminosity Functions}
\label{sec:h2lf_ind}

Similar to the H$\alpha$ H {\sc ii} region LF, the bright end of which is often represented by 
a truncated power law, we express the Pa$\alpha$ LF as, following \citet{McKee97} 
and \citet{Scoville01},
\begin{equation}
dN(L_{\rm Pa\alpha})/d\ln L_{\rm Pa\alpha}=N_{\rm up}(L_{\rm Pa\alpha}/L_{\rm up})^{\alpha},
\end{equation}
where the power index $\alpha<0$, $L_{\rm up}$ is the brightest H {\sc ii} region, and $N_{\rm up}$ 
is roughly the number of regions with luminosities $0.5<L_{\rm Pa\alpha}/L_{\rm up}<1$ in case 
$\alpha\approx-1$. Further discussion on the physical interpretation of the parameters is deferred 
to the original papers. 

\begin{figure*}
\centering
    \includegraphics[origin=c,scale=0.26,trim=0cm 10mm 1cm 5mm]{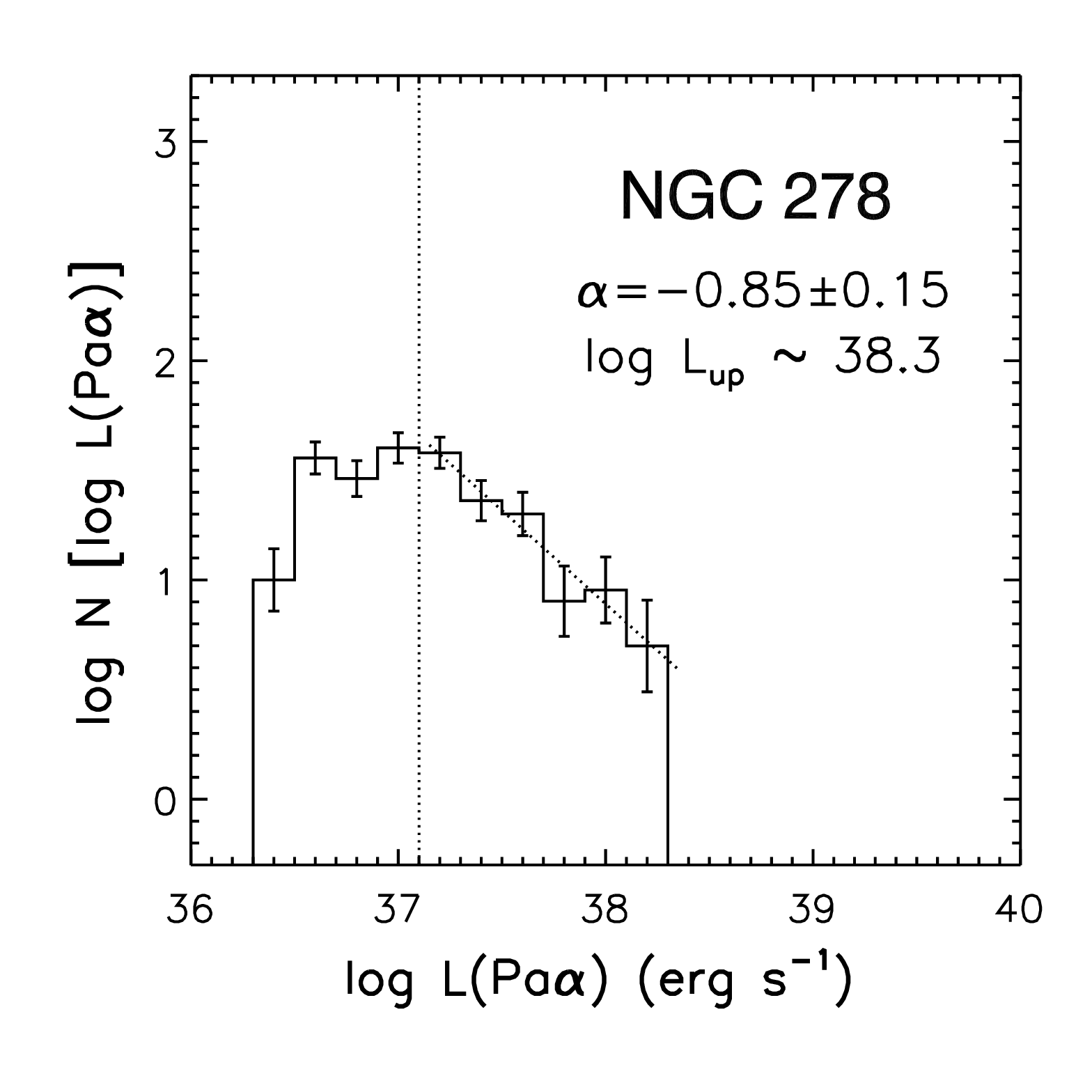}%
    \includegraphics[origin=c,scale=0.26,trim=0cm 10mm 1cm 5mm]{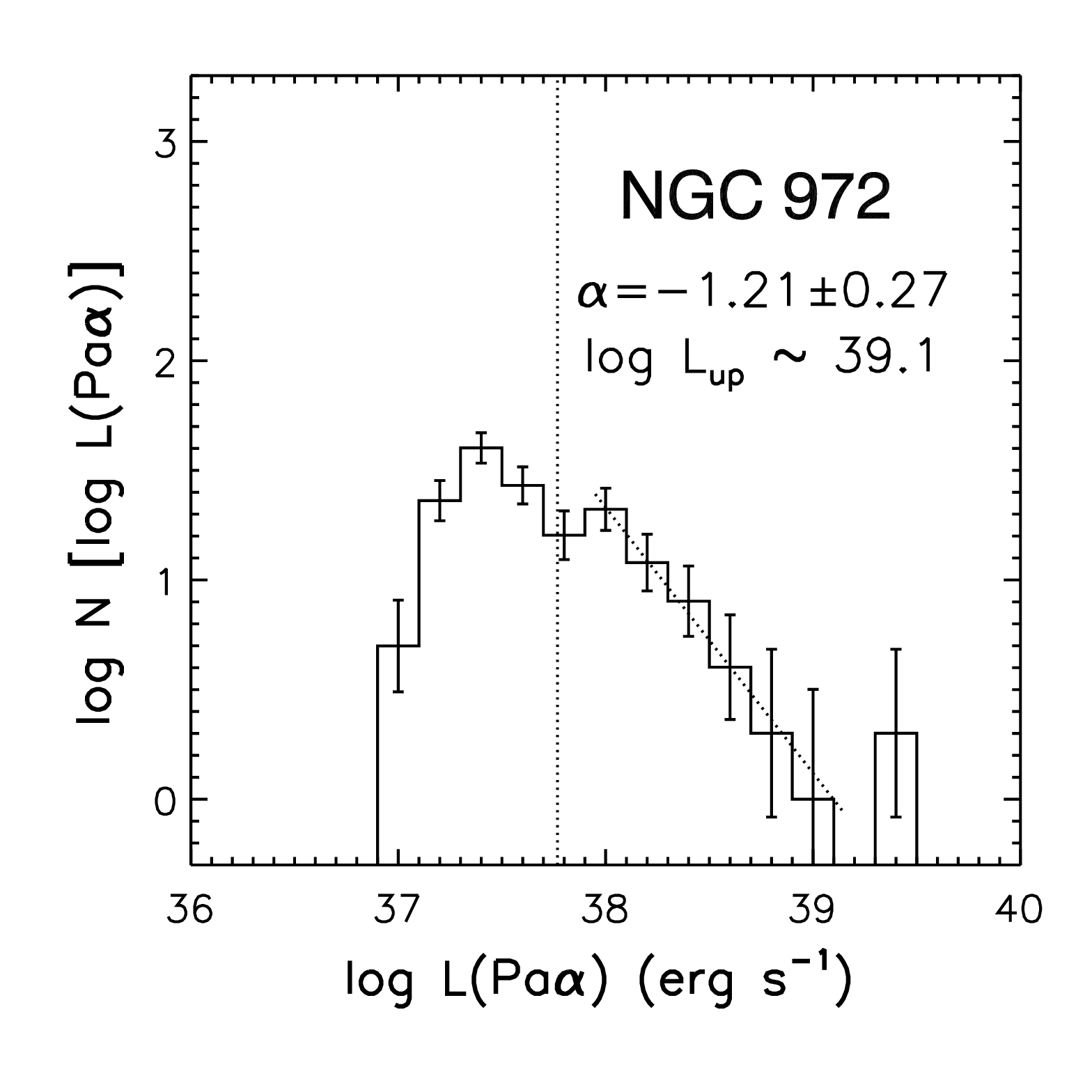}%
    \includegraphics[origin=c,scale=0.26,trim=0cm 10mm 1cm 5mm]{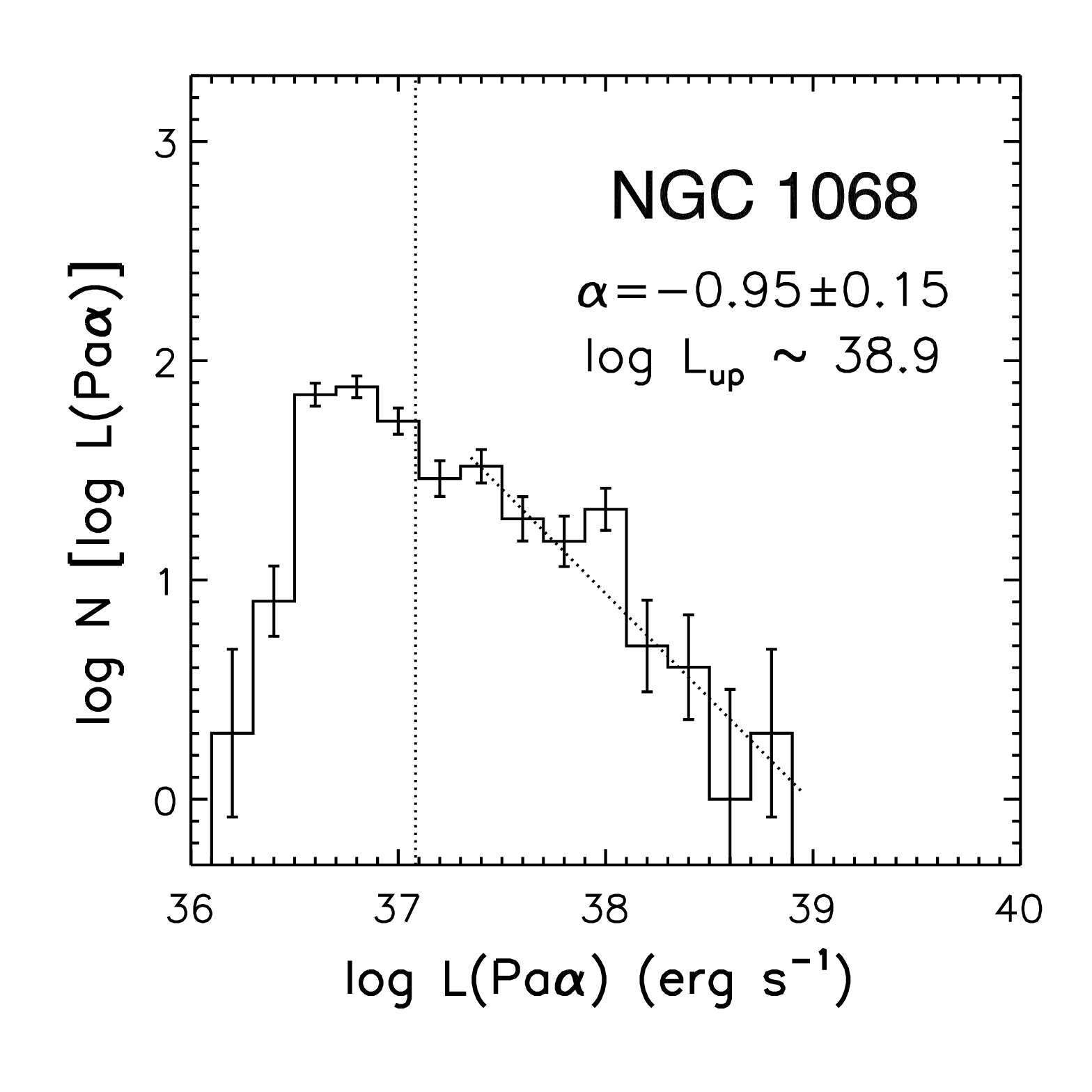}
    \includegraphics[origin=c,scale=0.26,trim=0cm 10mm 1cm 5mm]{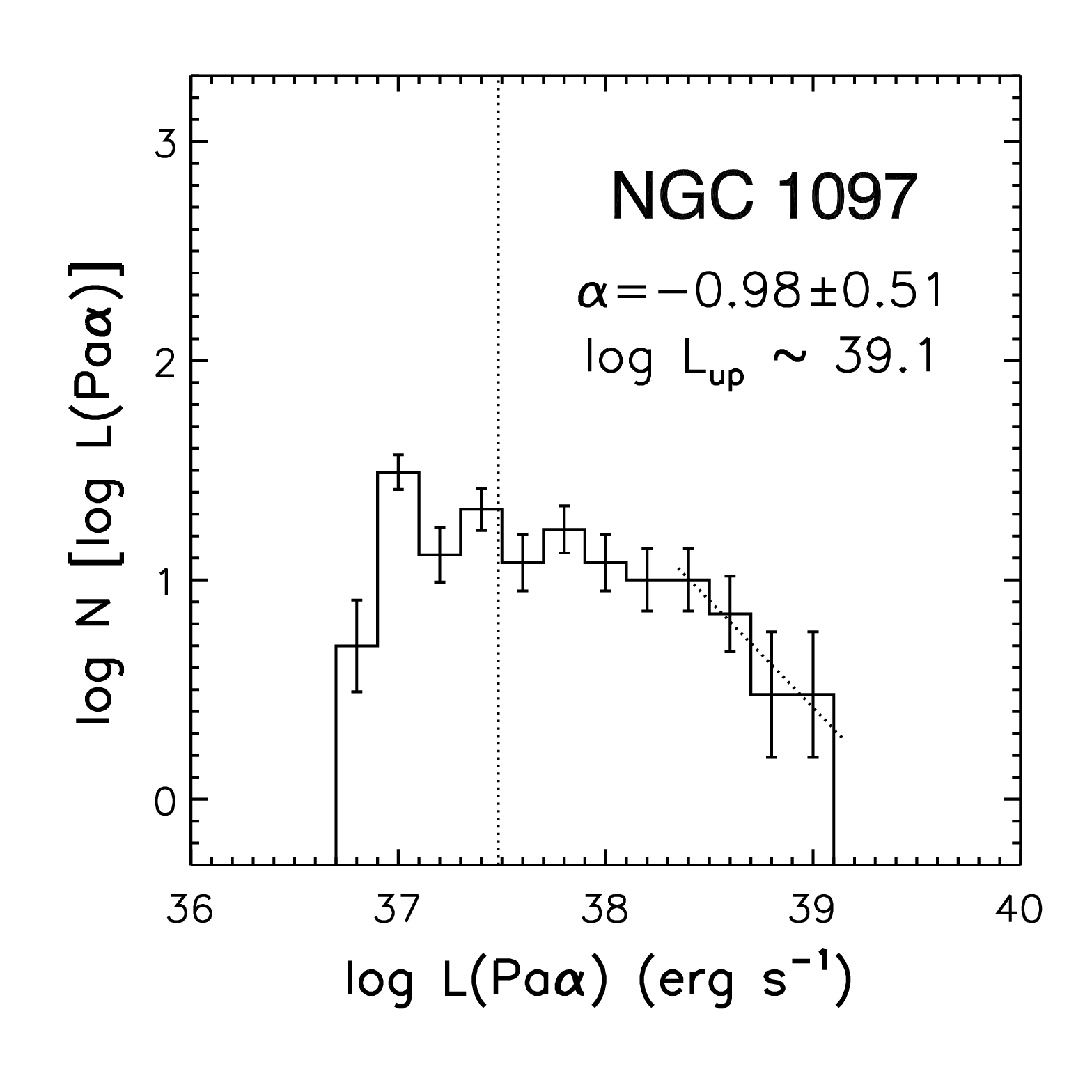}%
    \includegraphics[origin=c,scale=0.26,trim=0cm 10mm 1cm 5mm]{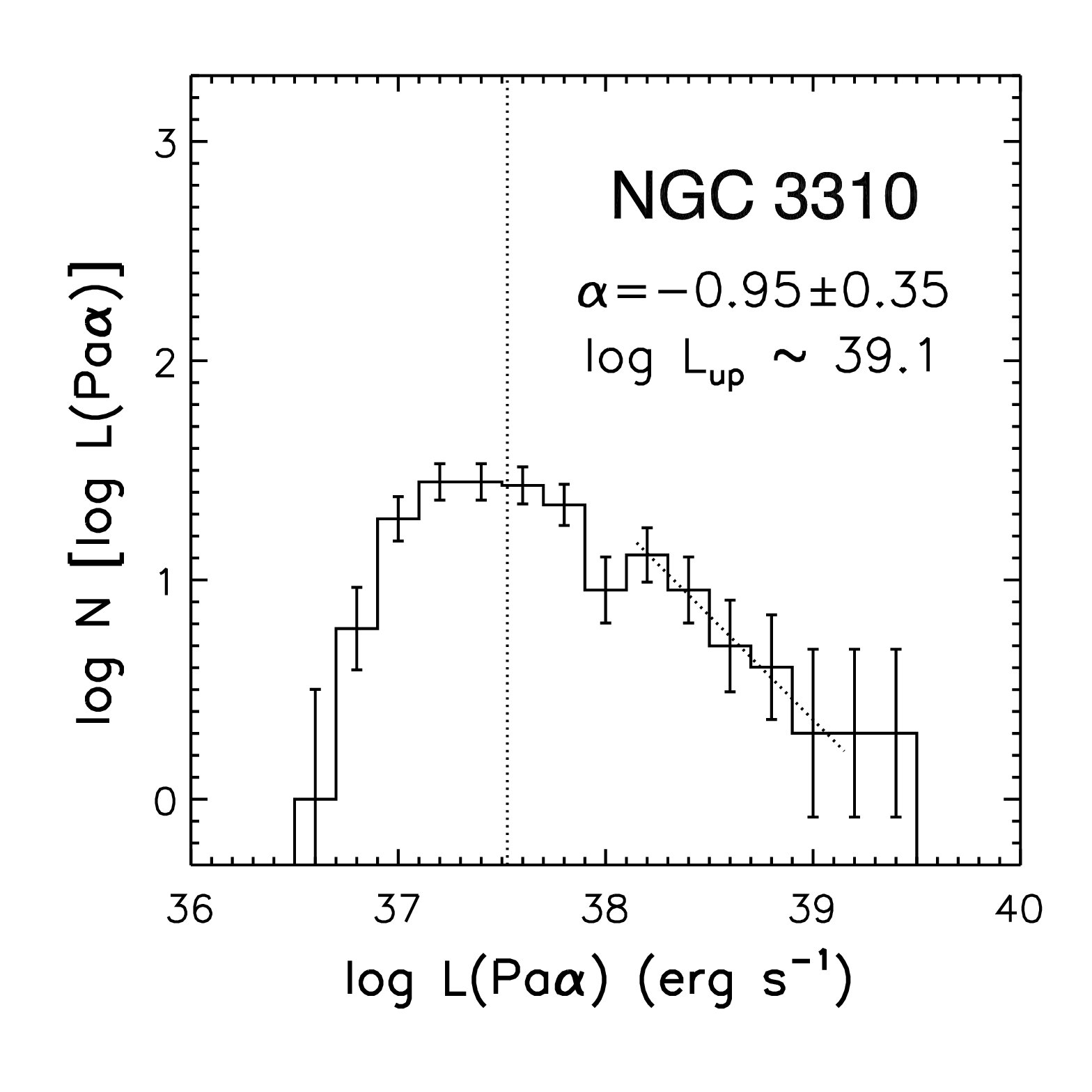}%
    \includegraphics[origin=c,scale=0.26,trim=0cm 10mm 1cm 5mm]{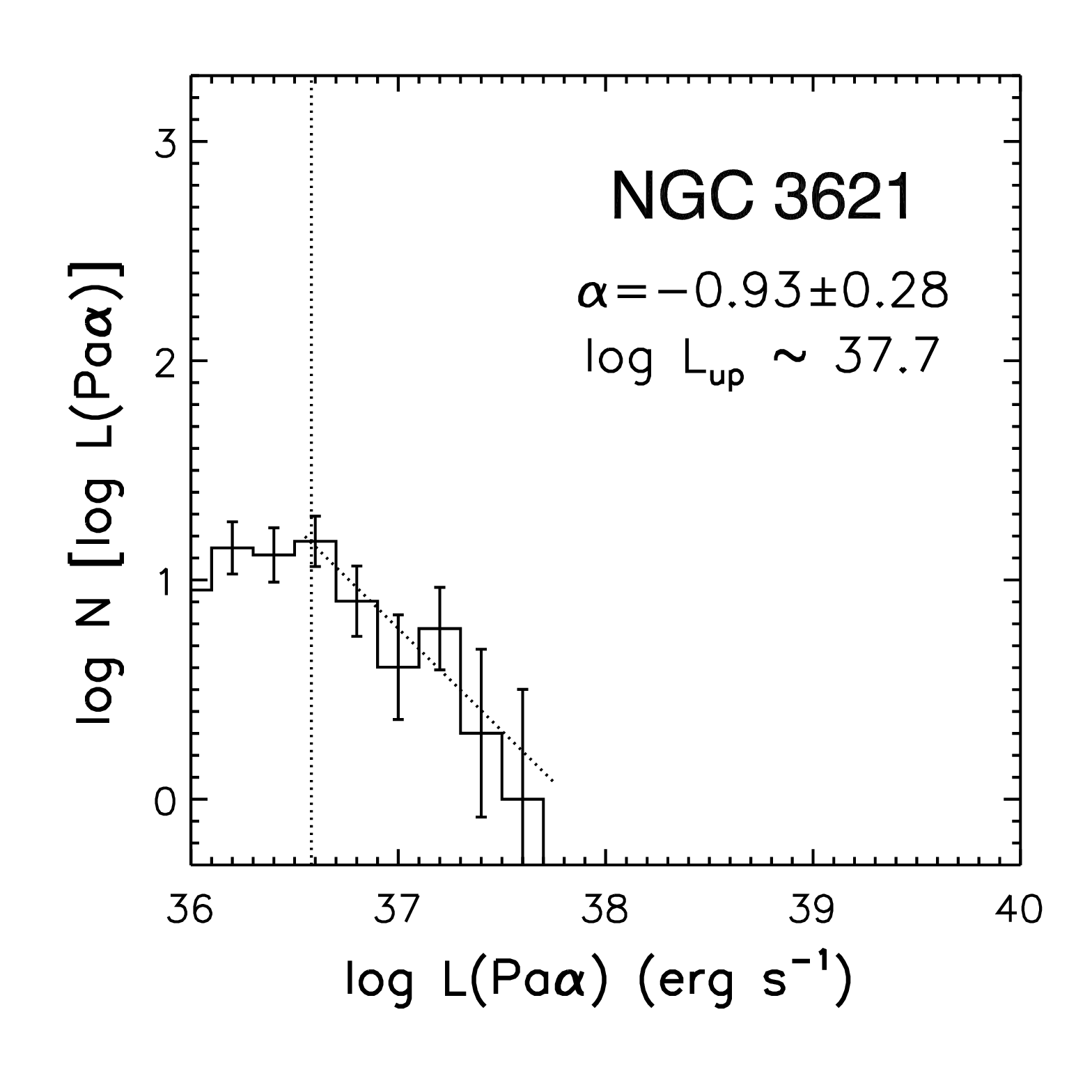}
    \includegraphics[origin=c,scale=0.26,trim=0cm 10mm 1cm 5mm]{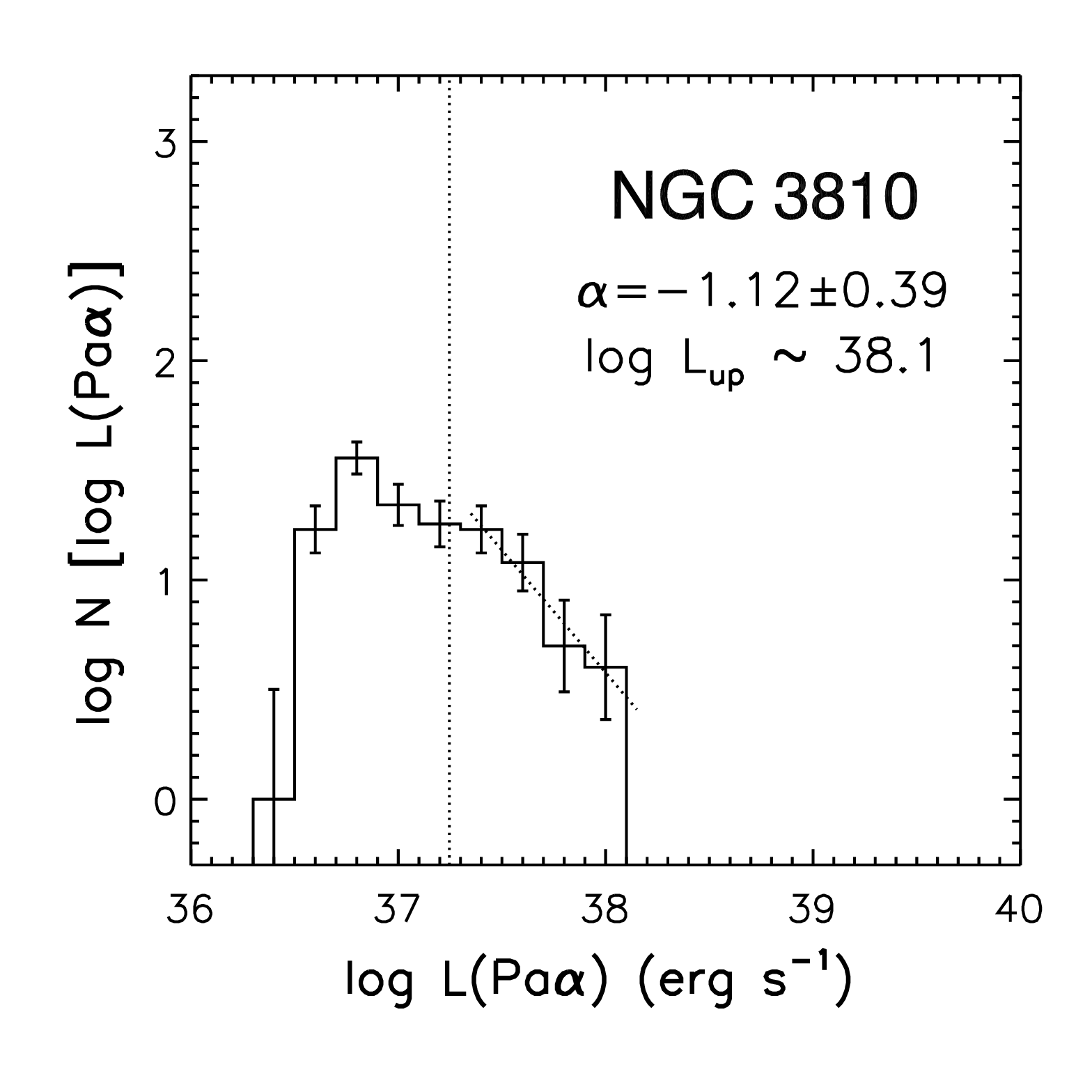}%
    \includegraphics[origin=c,scale=0.26,trim=0cm 10mm 1cm 5mm]{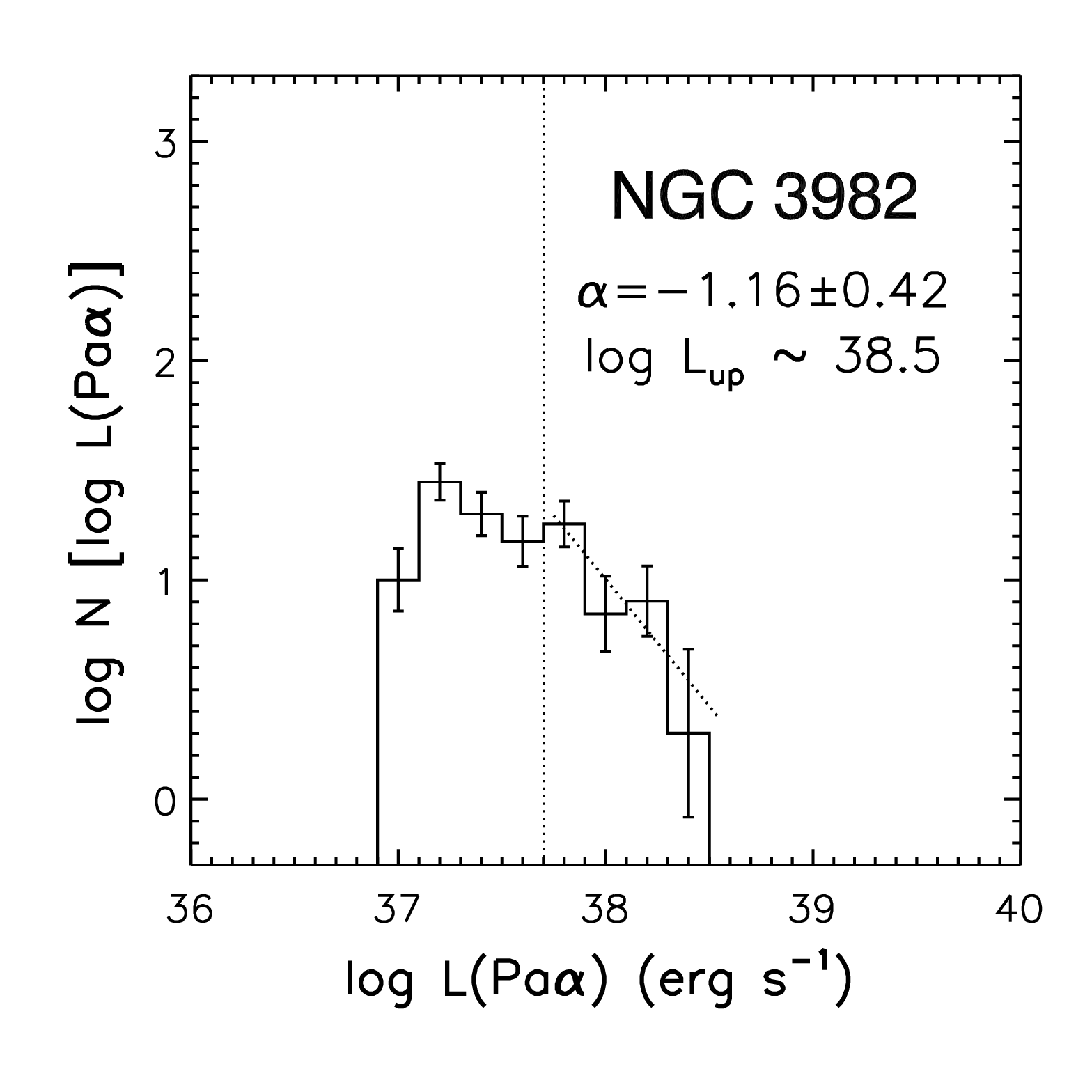}%
    \includegraphics[origin=c,scale=0.26,trim=0cm 10mm 1cm 5mm]{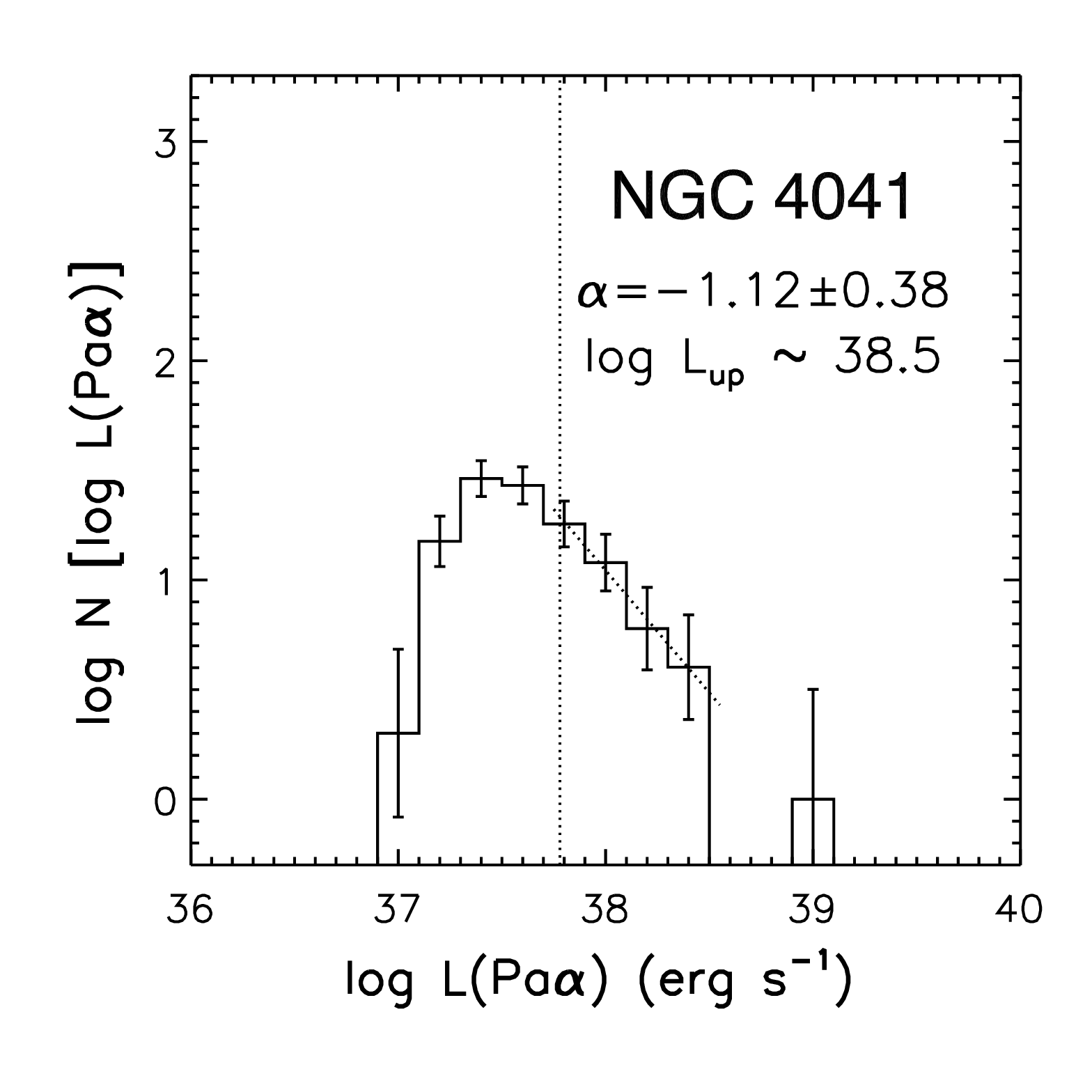}
    \includegraphics[origin=c,scale=0.26,trim=0cm 10mm 1cm 5mm]{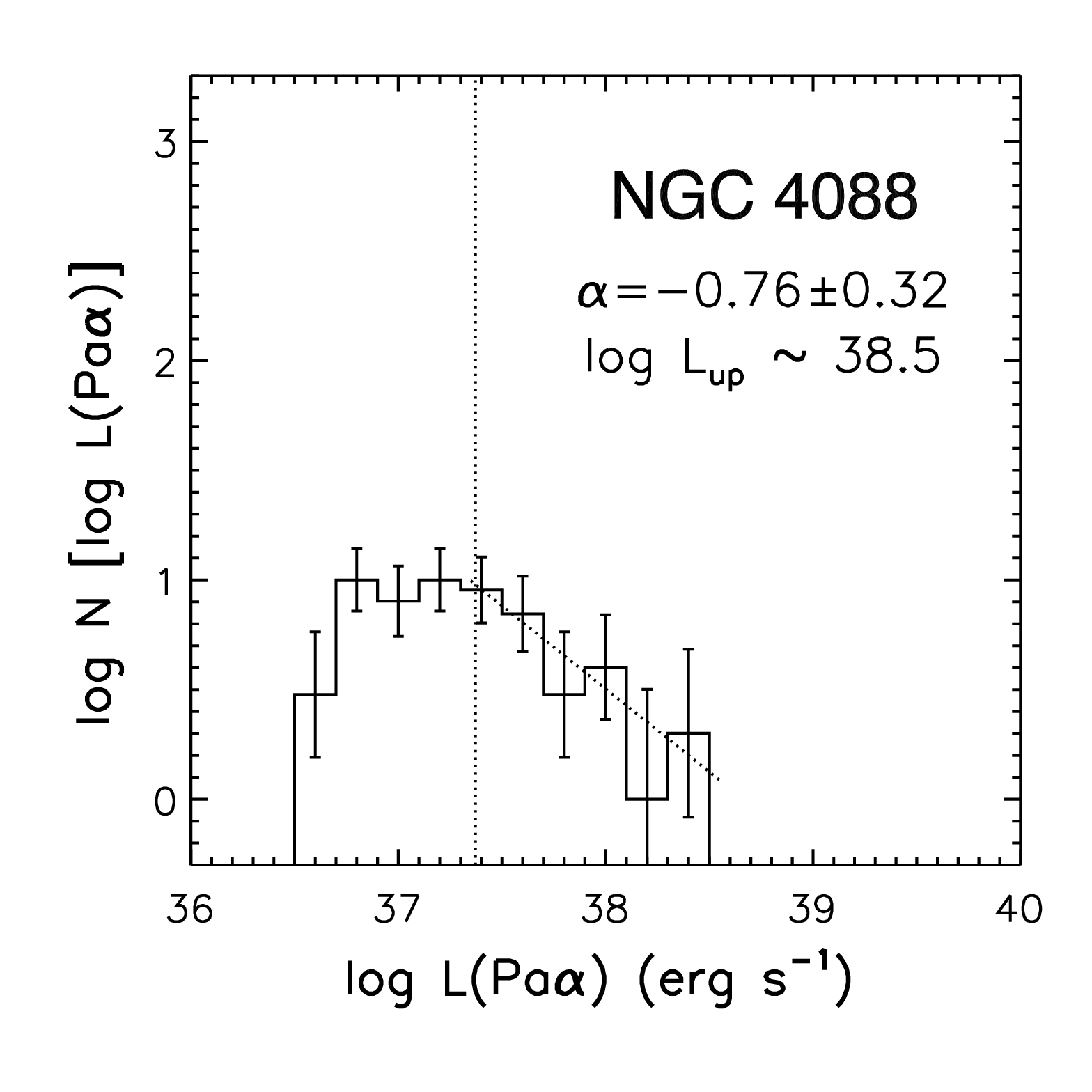}%
    \includegraphics[origin=c,scale=0.26,trim=0cm 10mm 1cm 5mm]{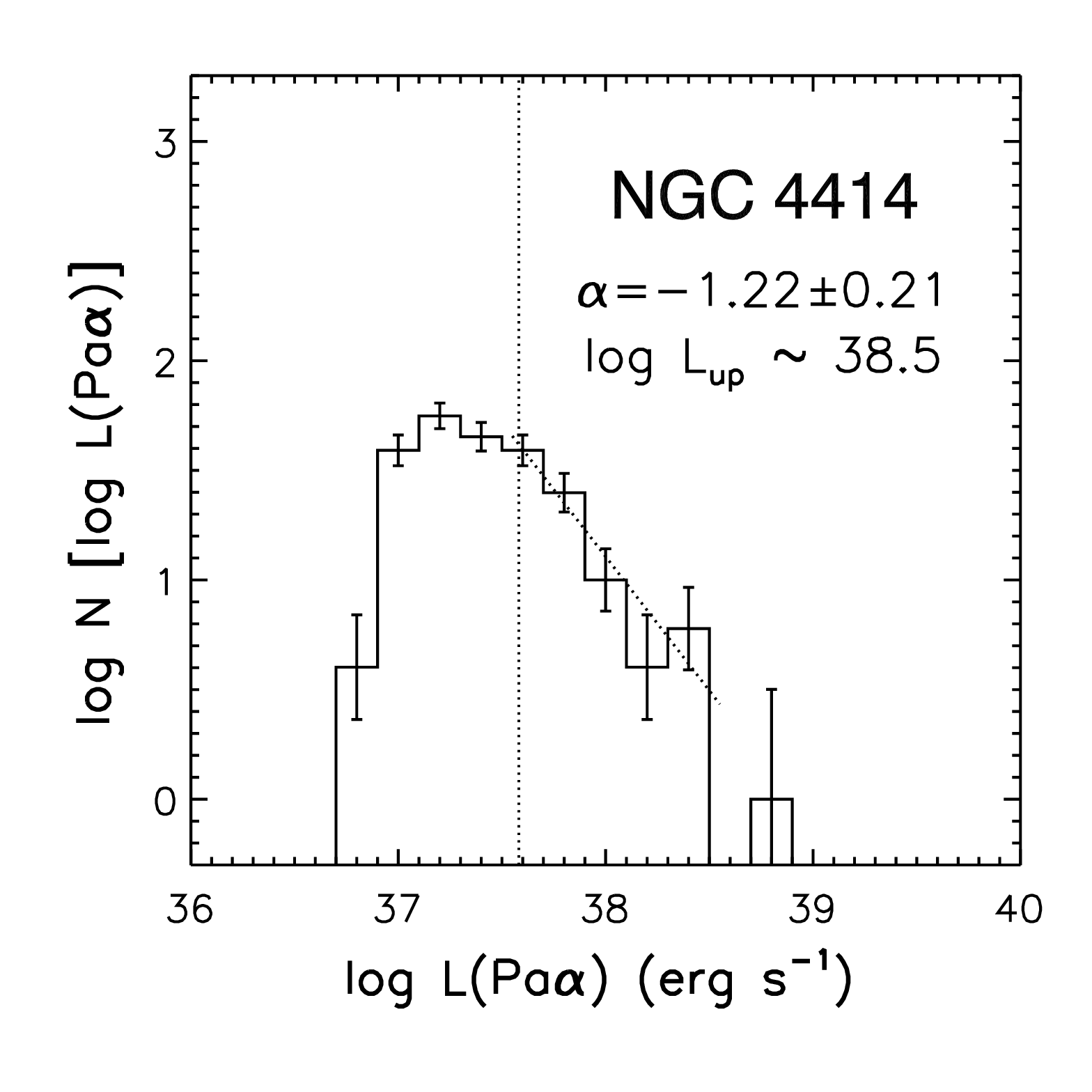}%
    \includegraphics[origin=c,scale=0.26,trim=0cm 10mm 1cm 5mm]{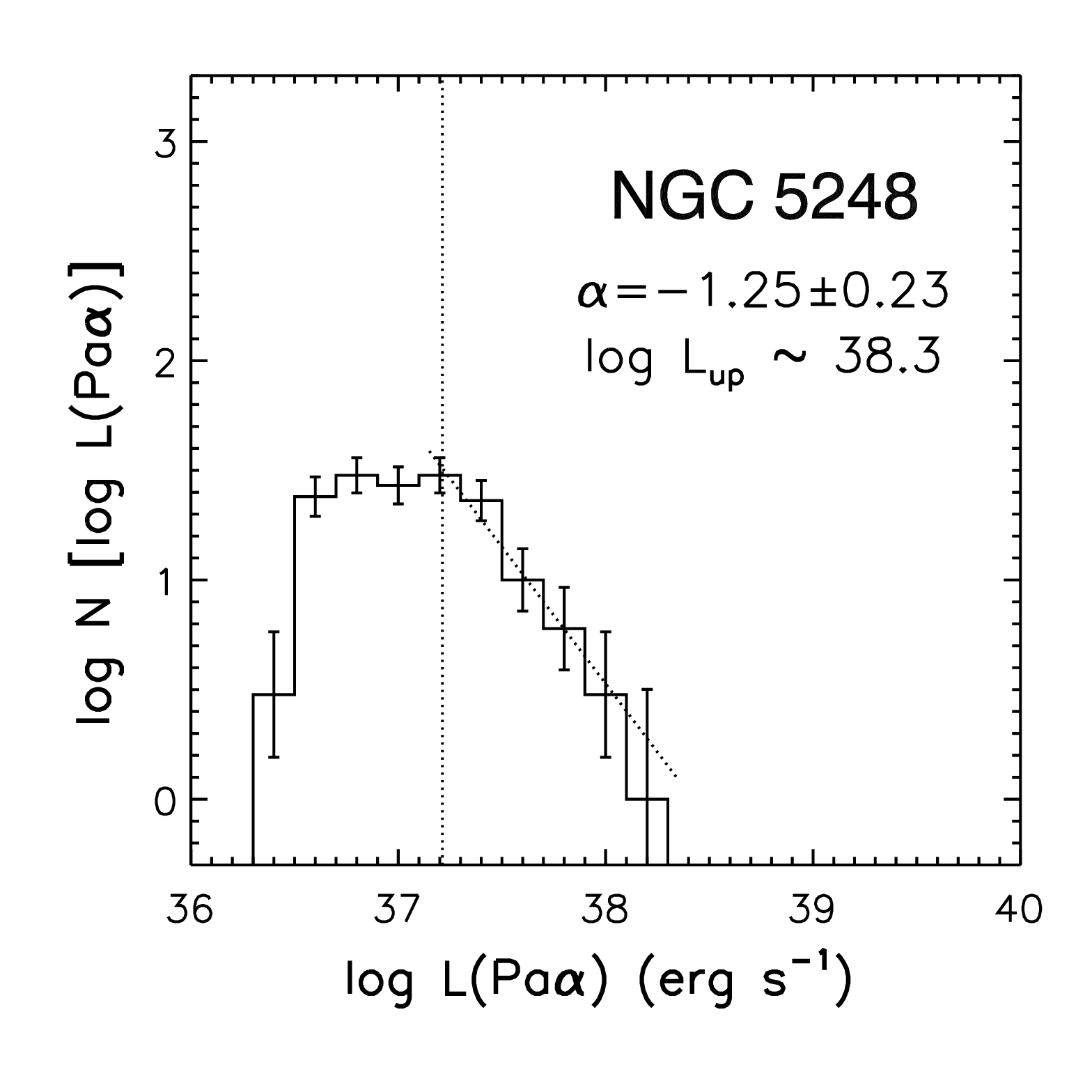}\\
\caption{The Pa$\alpha$ H {\sc ii} region luminosity functions of the 12 sample galaxies. 
Each bin containing $N$ regions is weighted by $\sqrt N$ when the power-law
fits are performed. Plotted are all regions with $S/N\geqslant3$, but power-law fitting
is performed on the bright end of regions above the $S/N=5$ completeness limit depicted 
by the vertical dotted lines.}
\label{fig:lf_gold}
\end{figure*}

The results of fitting Equation 1 to the observed Pa$\alpha$ luminosity function in our
sample galaxies are presented in Figure~\ref{fig:lf_gold}. In this figure, all the H {\sc ii} 
regions with $S/N\geqslant3$ are plotted, but only the regions 
above the completeness limit for luminosity are used for the power-law fits. $S/N=5$ represents 
the completeness limit recommended by \citet{Thilker00} and is converted to a luminosity limit 
in Figure~\ref{fig:lf_gold} (vertical dotted lines) by performing power-law fits to the 
well-defined luminosity vs. $S/N$ correlation. One should note that this completeness limit 
of H {\sc ii} region identification and characterization is, unsurprisingly, much higher than 
the detection limit of the HST imaging quoted in the observations and data reduction section (\S \ref{sec:reduc}).
For the whole sample, we find a mean value for the LF power index 
$\langle\alpha\rangle=-1.05$, and a median value $\tilde{\alpha}=-0.98\pm0.19$.

As mentioned in the Introduction, some galaxies are characterized by LFs described by 
a broken, but connected, power law, with a luminosity at the break 
$L_{\rm H\alpha, br}\sim10^{38.6}$~erg~s$^{-1}$. To provide some perspective on this value, 
we recall that a single O-type star with mass 120~$M_{\odot}$ produces an ionizing photon 
rate of $\sim10^{50.1}$~s$^{-1}$ \citep{Schaerer97,Oey98}, which corresponds to an 
intrinsic $L_{\rm H\alpha}\sim10^{38.2}$~erg~s$^{-1}$ \citep{Leitherer95}, and a dust-attenuated 
H$\alpha$ luminosity of $\sim10^{37.2\mhyphen37.5}$~erg~s$^{-1}$, for the dust extinction 
values discussed in \S \ref{sec:h2_iden}. The observed H$\alpha$ luminosity for a
single O-star is thus well below the value of the break luminosity mentioned above, and 
close to the completeness limits of H {\sc ii} region identification shown in Figure~\ref{fig:lf_gold}.

The broken power-law LF does not seem to be present for the H {\sc ii} LF of 
our sample galaxies. All the 12 galaxies exhibit Pa$\alpha$ LFs that follow a single power law with 
$\alpha \simeq -1$ above the completeness limit, with no apparent break around $10^{38.2\mhyphen38.4}$ 
erg s$^{-1}$ (Figure~\ref{fig:lf_gold}), consistent with the results for Galactic radio H {\sc ii} 
regions, which show a single power-law index of $-1.3$ \citep{Smith89} or $-1.0$ \citep{McKee97}. 
This conclusion also coincides the result found from the H$\alpha$ HST/WFPC2 imaging of M51 
\citep[$\alpha=-1.01\pm0.04$, no break,][]{Scoville01}, although it is steeper than what is found 
for M101 \citep[$\alpha=-0.74\pm0.08$, no break,][]{Pleuss00}. 

\begin{figure}
\centering
\includegraphics[scale=.35,clip,trim=0cm 0cm 0cm 0cm]{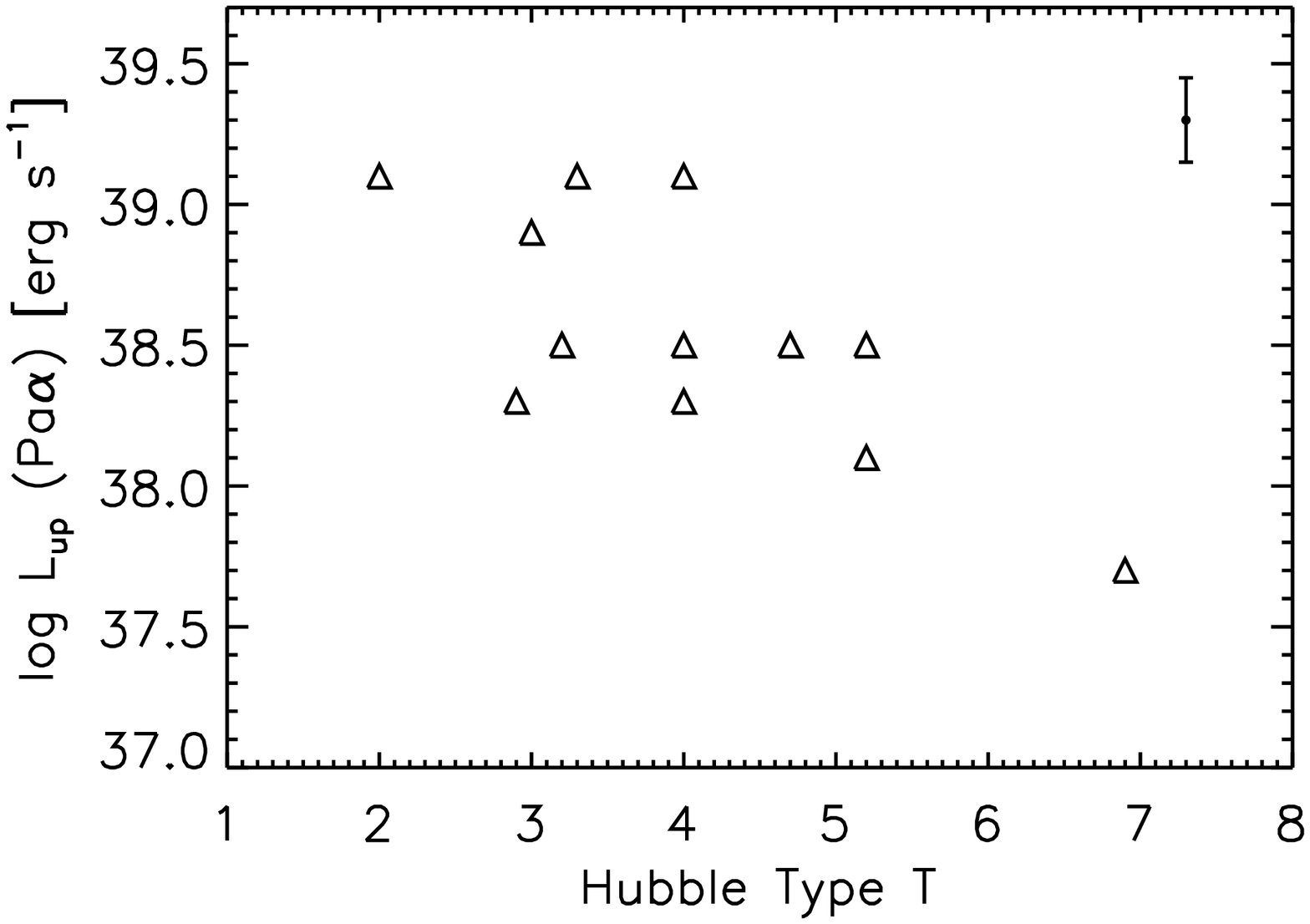}
\includegraphics[scale=.35,clip,trim=0cm 0cm 0cm 0cm]{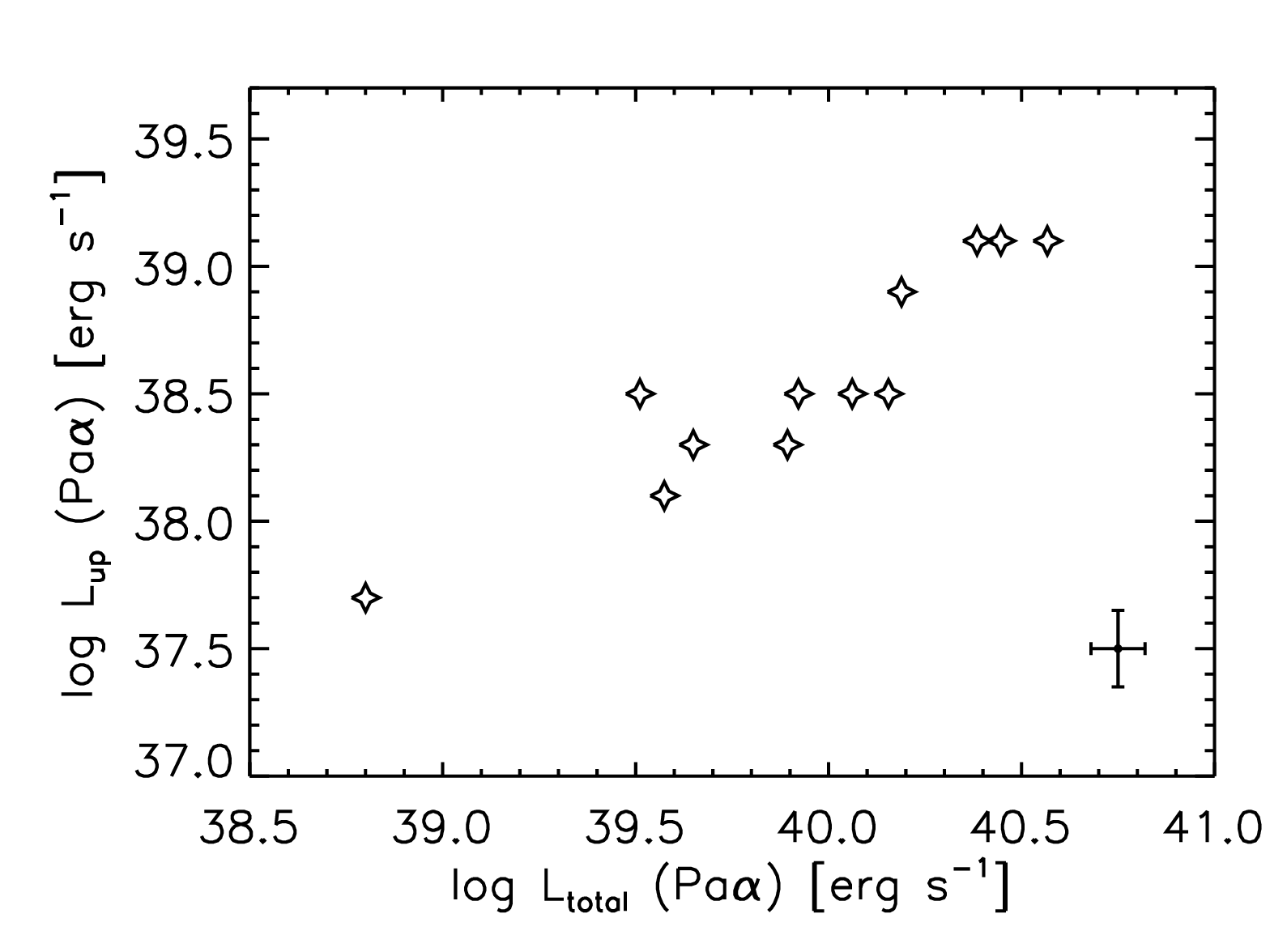}
\caption{The dependence of the truncation luminosity of Pa$\alpha$ H {\sc ii} region luminosity function 
($L_{\rm up}$) on the morphological type and the total Pa$\alpha$ luminosity ($L_{\rm total}$) of the host galaxy. 
The numerical Hubble type $T$ of each galaxy is listed in Table~\ref{tbl:sample}, and $L_{\rm total}$
is the total luminosity of all the H {\sc ii} regions identified in each galaxy with $S/N\geqslant3$
and listed in Table \ref{tbl:catalog}.}
\label{fig:t_lup}
\end{figure}

Figure \ref{fig:lf_gold} also shows the location of a high luminosity truncation, $L_{\rm up}$, where  
the power-law bright end is reached because of small number statistics. We find 6 galaxies with 
$L_{\rm up}=10^{38.3\mhyphen38.5}$ erg s$^{-1}$ and 4 with $L_{\rm up}=10^{38.9\mhyphen39.1}$ erg s$^{-1}$, 
which appear to form a double-peak distribution. The truncation luminosity in individual galaxies is 
possibly correlated to the Hubble type $T$, as plotted in Figure~\ref{fig:t_lup}. A tentative overall
trend is observed, for which $L_{\rm up}$ is higher for galaxies with earlier morphological types, and the 
faintest data point, $L_{\rm up}=10^{37.7}$ erg s$^{-1}$, is contributed by the one with the latest type, 
the only Sd spiral NGC 3621. Our calculation shows that the Kendall rank correlation coefficient 
$\tau$=$-$0.44 with 5\% probability that no correlation is present, but the correlation is tentative 
because it actually hinges on the single Sd galaxy.
Meanwhile, we observe a tight correlation between $L_{\rm up}$ and the total 
luminosity of H {\sc ii} regions ($L_{\rm total}$) in each galaxy (Figure~\ref{fig:t_lup}), indicating a 
$L_{\rm total}$ vs. $T$ correlation which probably gives rise to the above relationship. However, the 
relatively small sample size prevents us from drawing a definitive conclusion from this data set.
No dependence of $L_{\rm up}$ on the distance, inclination or rotation velocity of the sample galaxies 
is observed. 

In Figure \ref{fig:lf_gold} we also find very few H {\sc ii} region candidates brighter than $L_{\rm Pa\alpha,br}$, 
in agreement with previous HST results. On a galaxy-by-galaxy basis, their number is too small to infer any 
property (especially, NGC 278, NGC 3621, NGC 3810 and NGC 5248 do not have H {\sc ii} regions more luminous 
than $\sim$10$^{38.5}$ erg s$^{-1}$ at all), so we will revisit this issue later, once we co-add all the LFs 
together. 

\subsection{Individual Galaxies: Luminosity-size relation}

\begin{figure*}
\centering
    \includegraphics[origin=c,scale=0.26,trim=0cm 10mm 1cm 5mm]{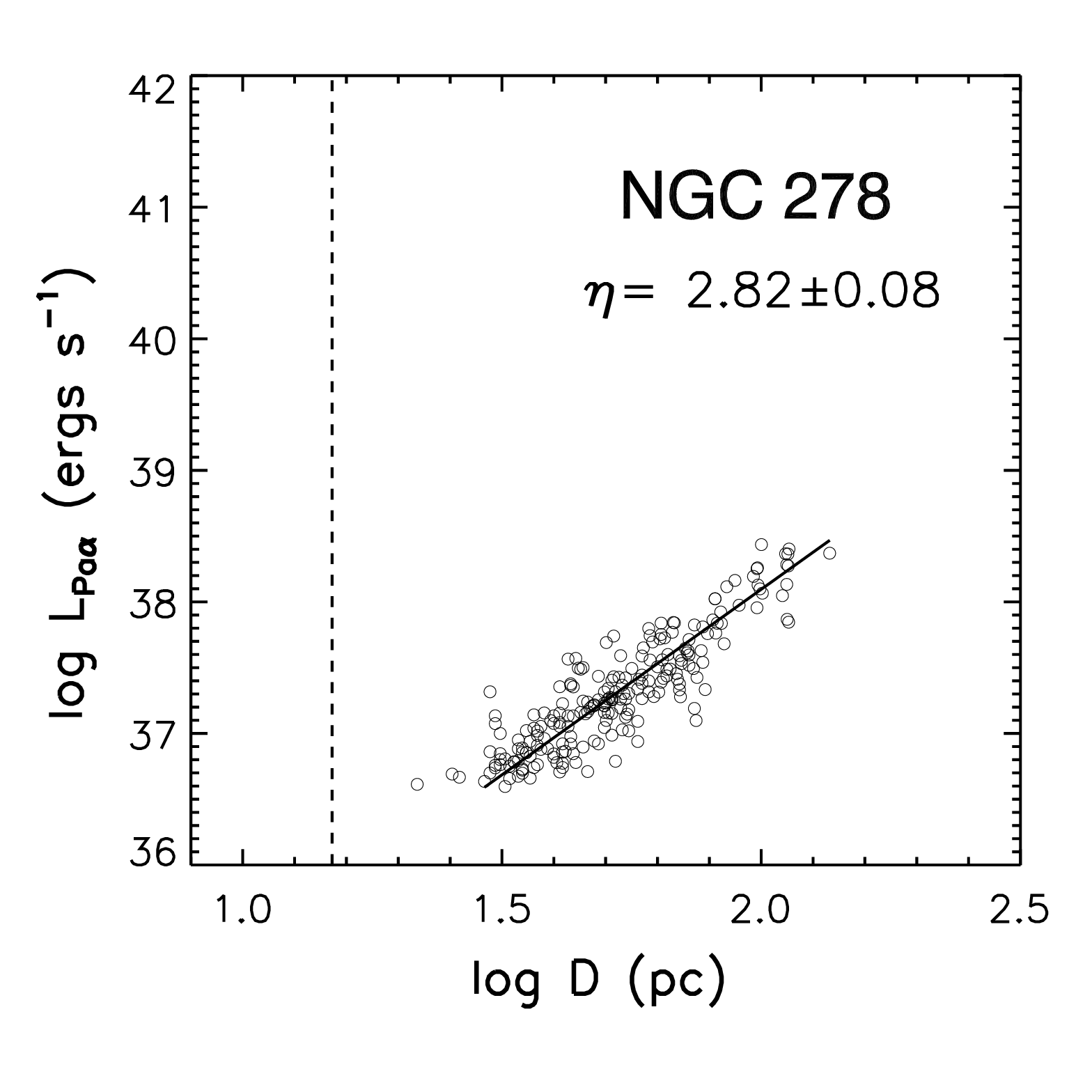}%
    \includegraphics[origin=c,scale=0.26,trim=0cm 10mm 1cm 5mm]{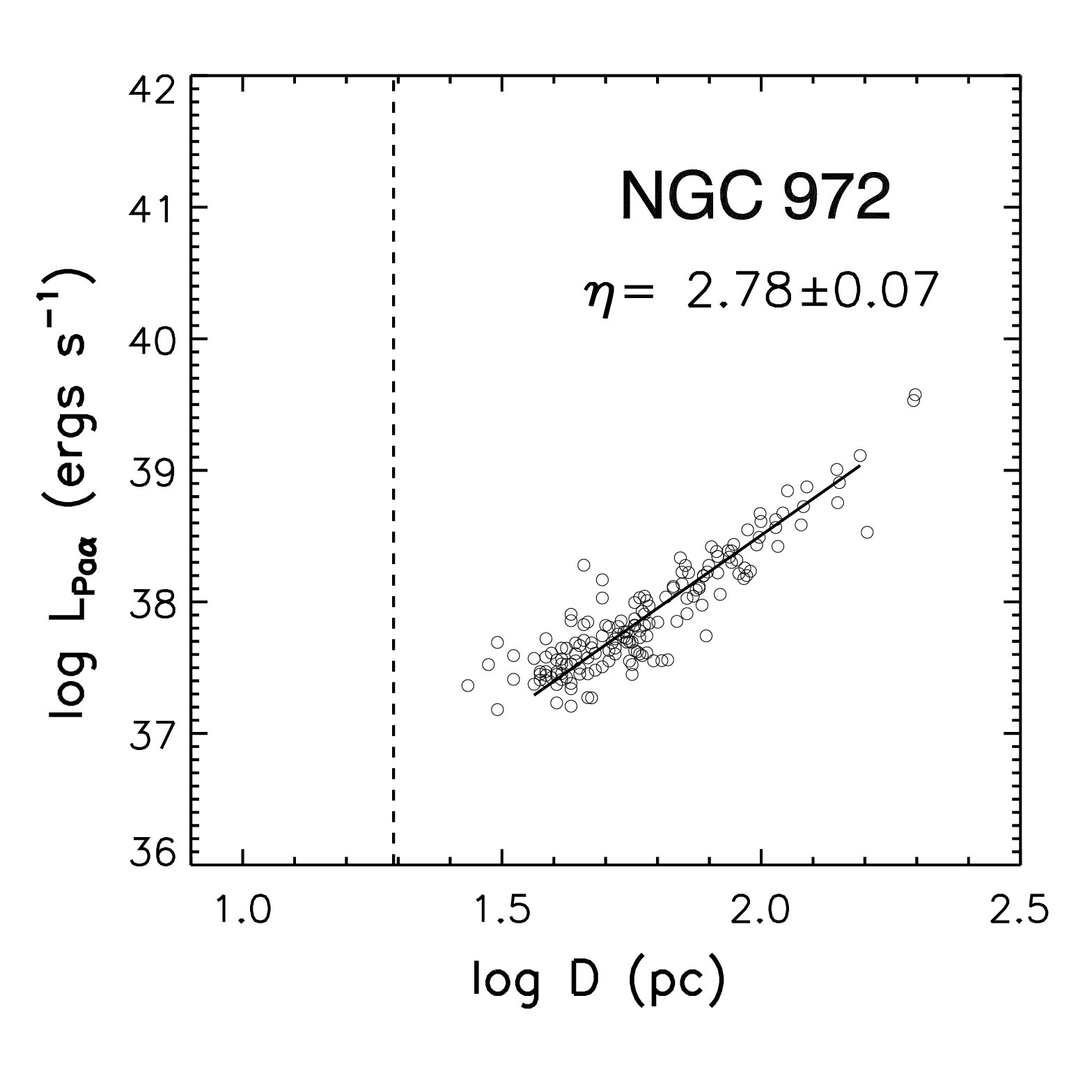}%
    \includegraphics[origin=c,scale=0.26,trim=0cm 10mm 1cm 5mm]{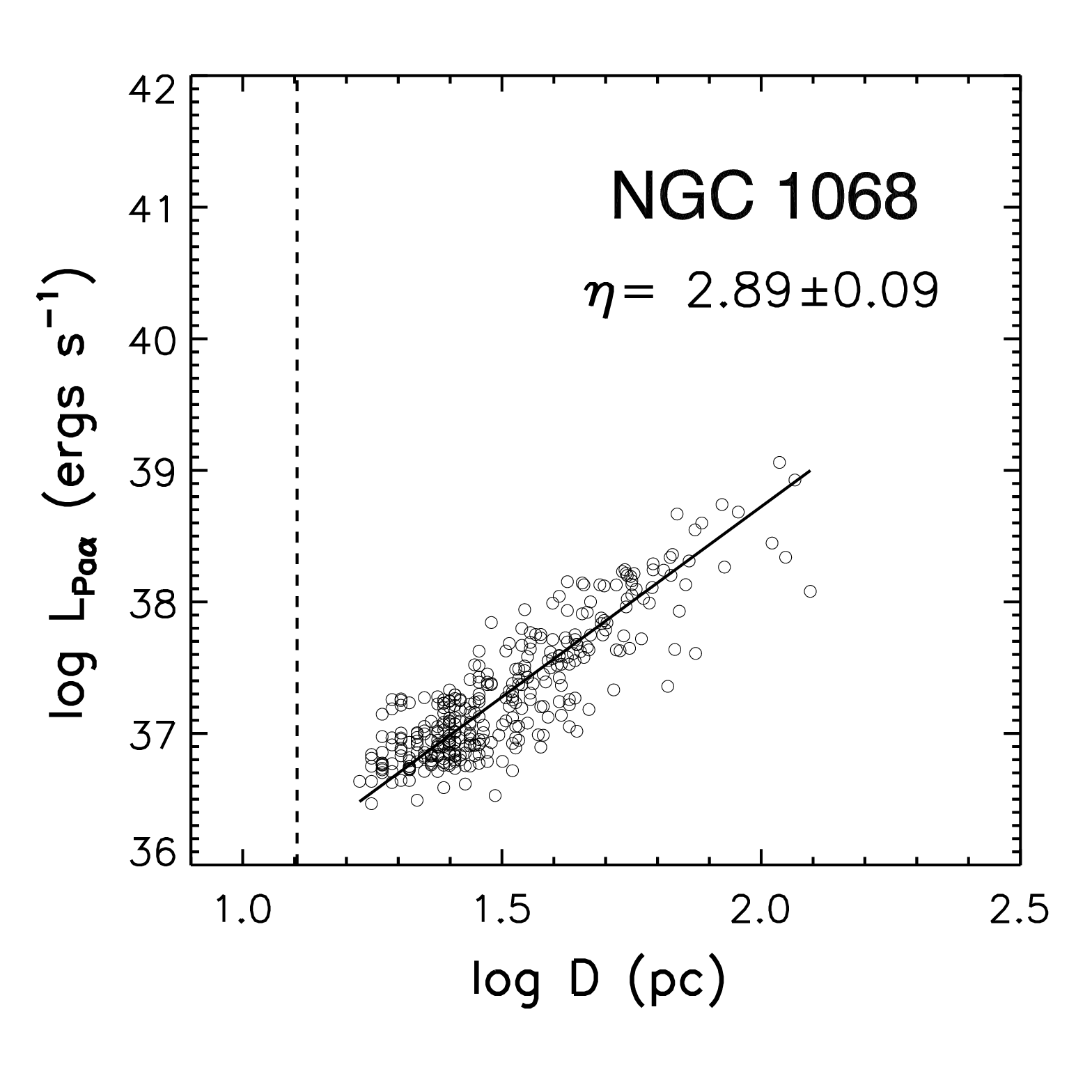}
    \includegraphics[origin=c,scale=0.26,trim=0cm 10mm 1cm 5mm]{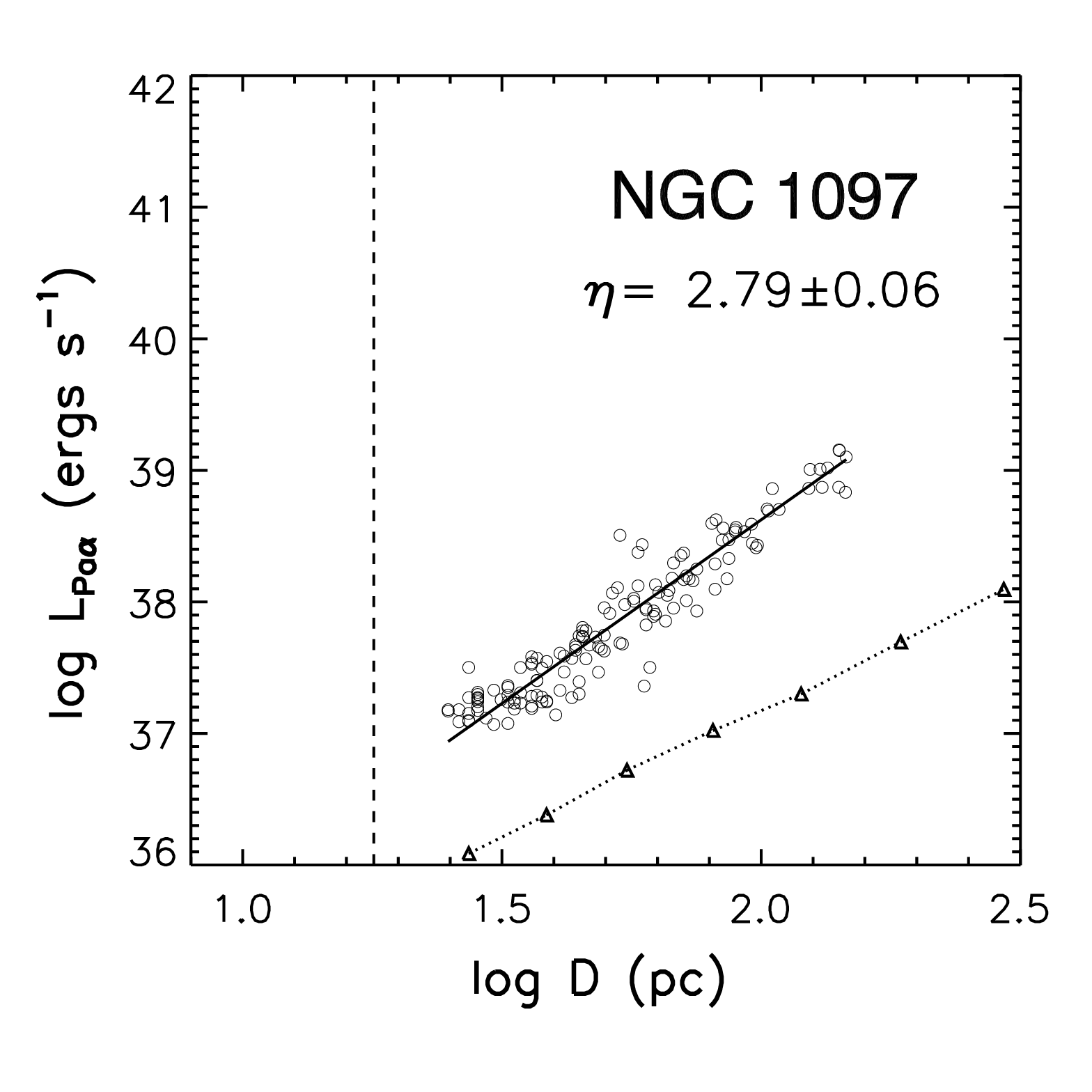}%
    \includegraphics[origin=c,scale=0.26,trim=0cm 10mm 1cm 5mm]{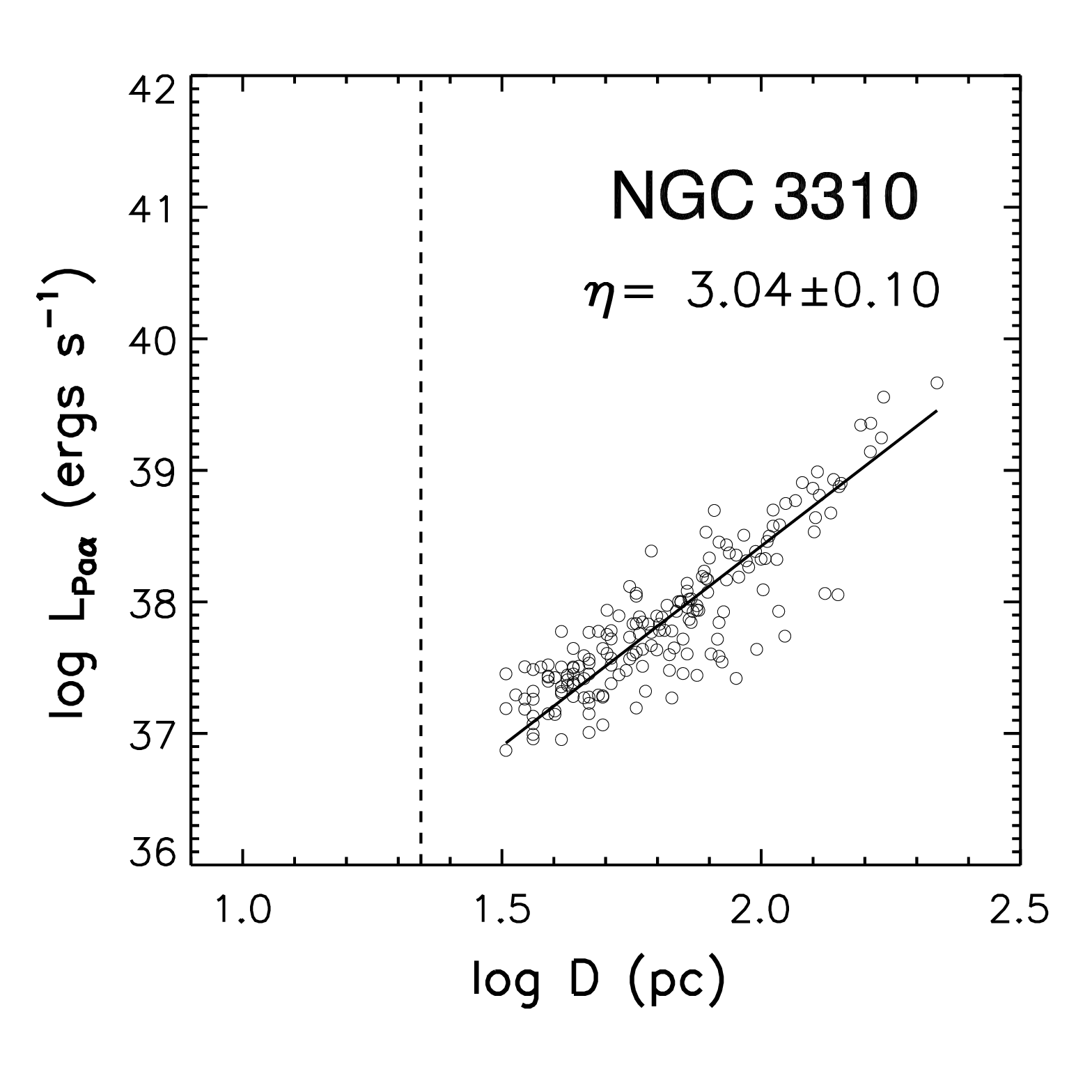}%
    \includegraphics[origin=c,scale=0.26,trim=0cm 10mm 1cm 5mm]{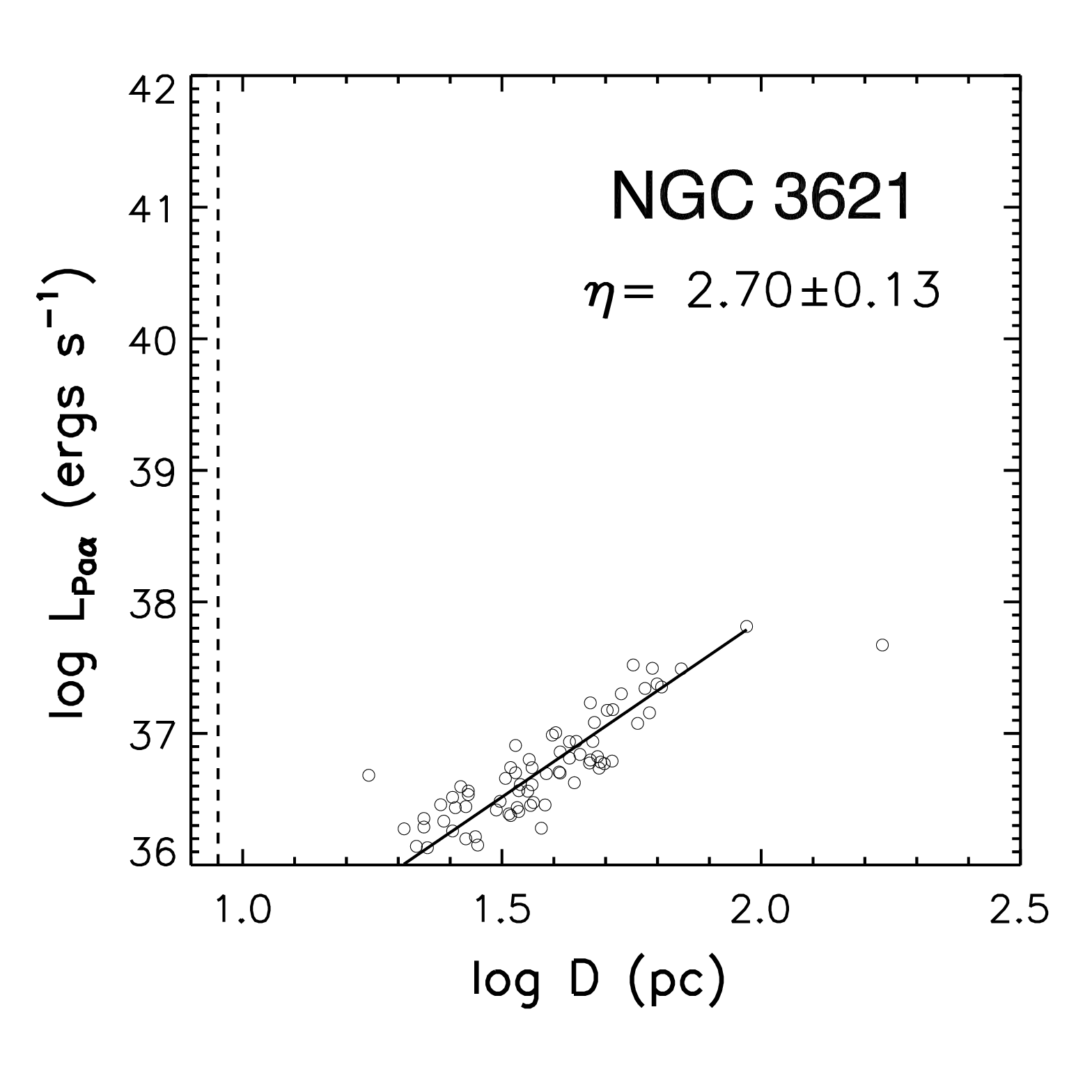}
    \includegraphics[origin=c,scale=0.26,trim=0cm 10mm 1cm 5mm]{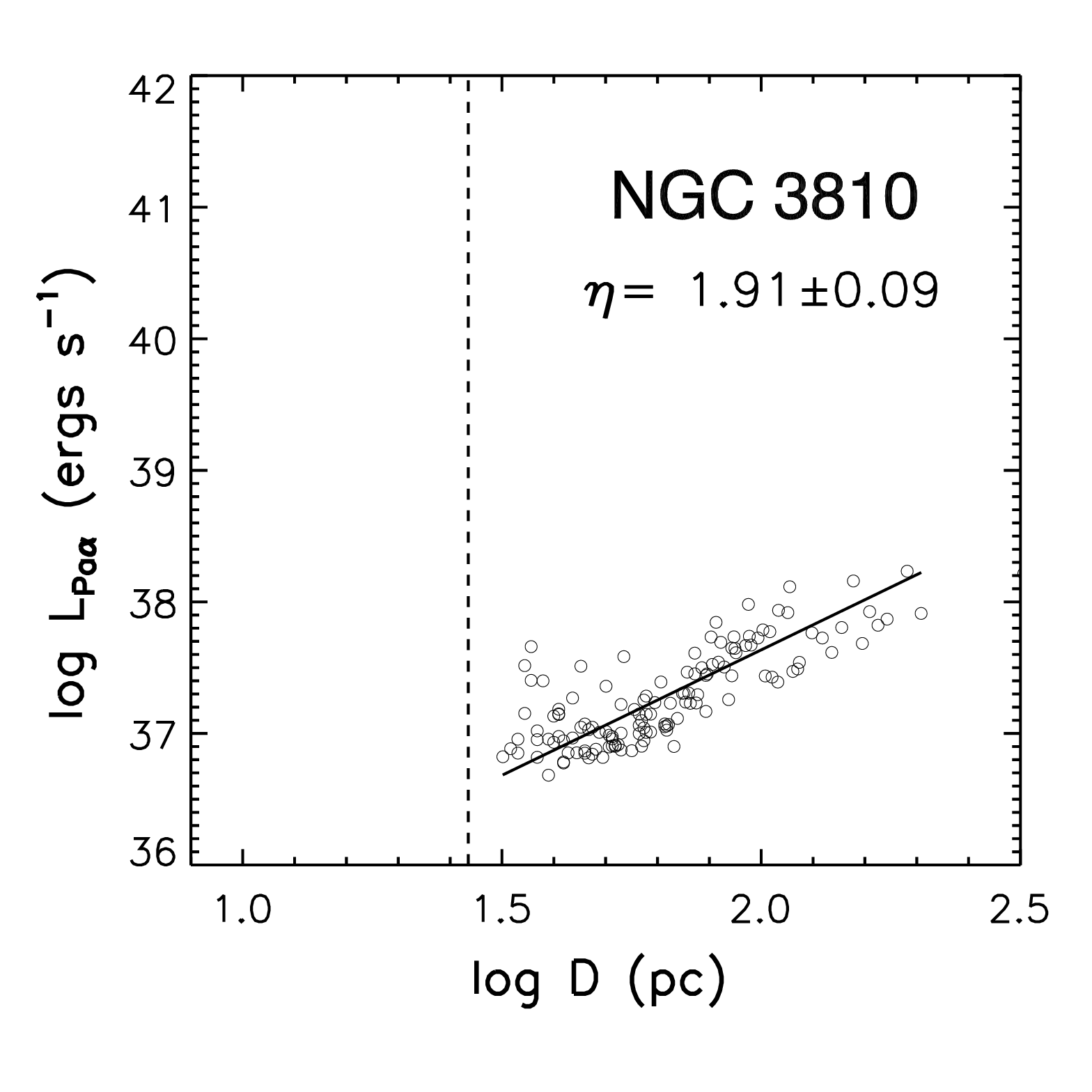}%
    \includegraphics[origin=c,scale=0.26,trim=0cm 10mm 1cm 5mm]{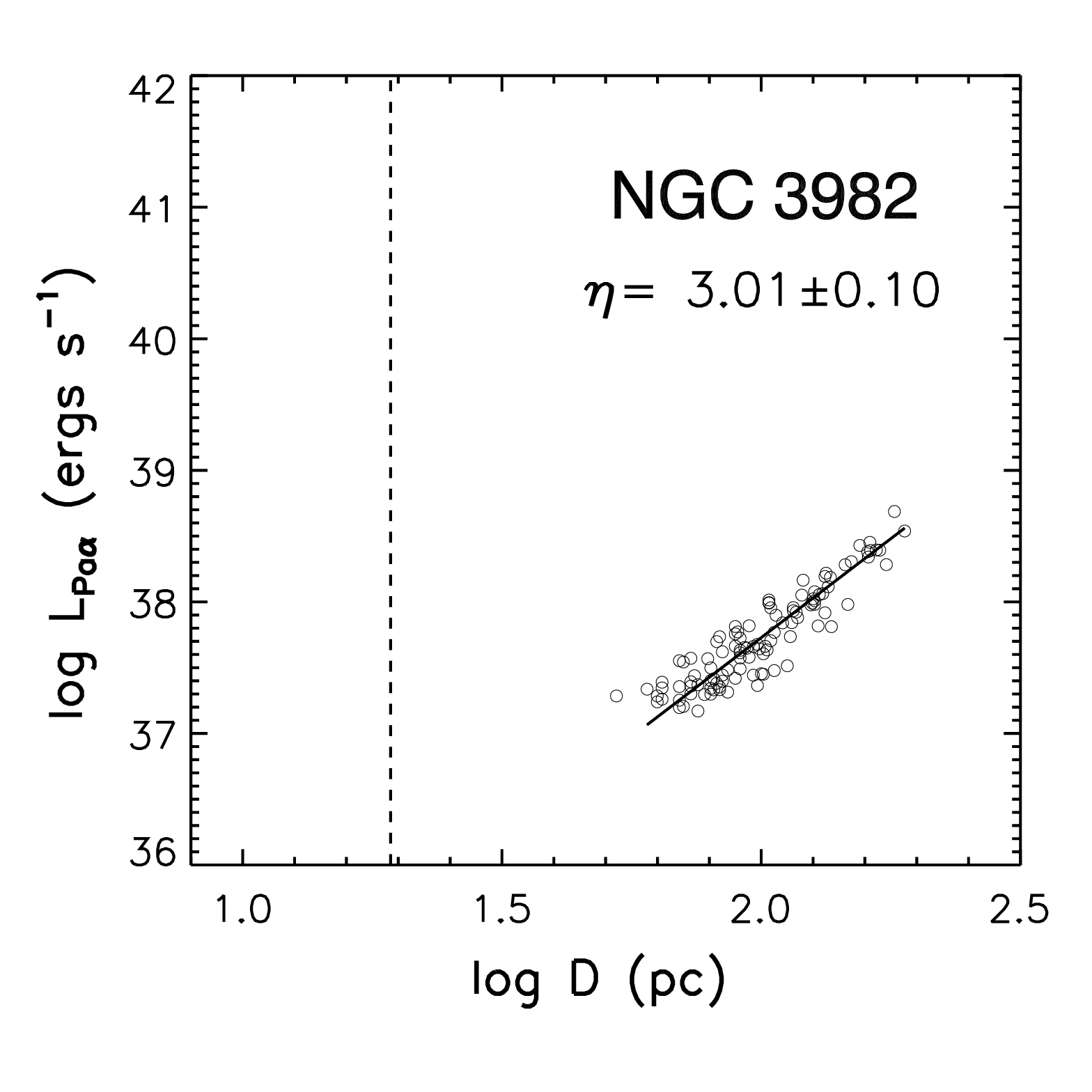}%
    \includegraphics[origin=c,scale=0.26,trim=0cm 10mm 1cm 5mm]{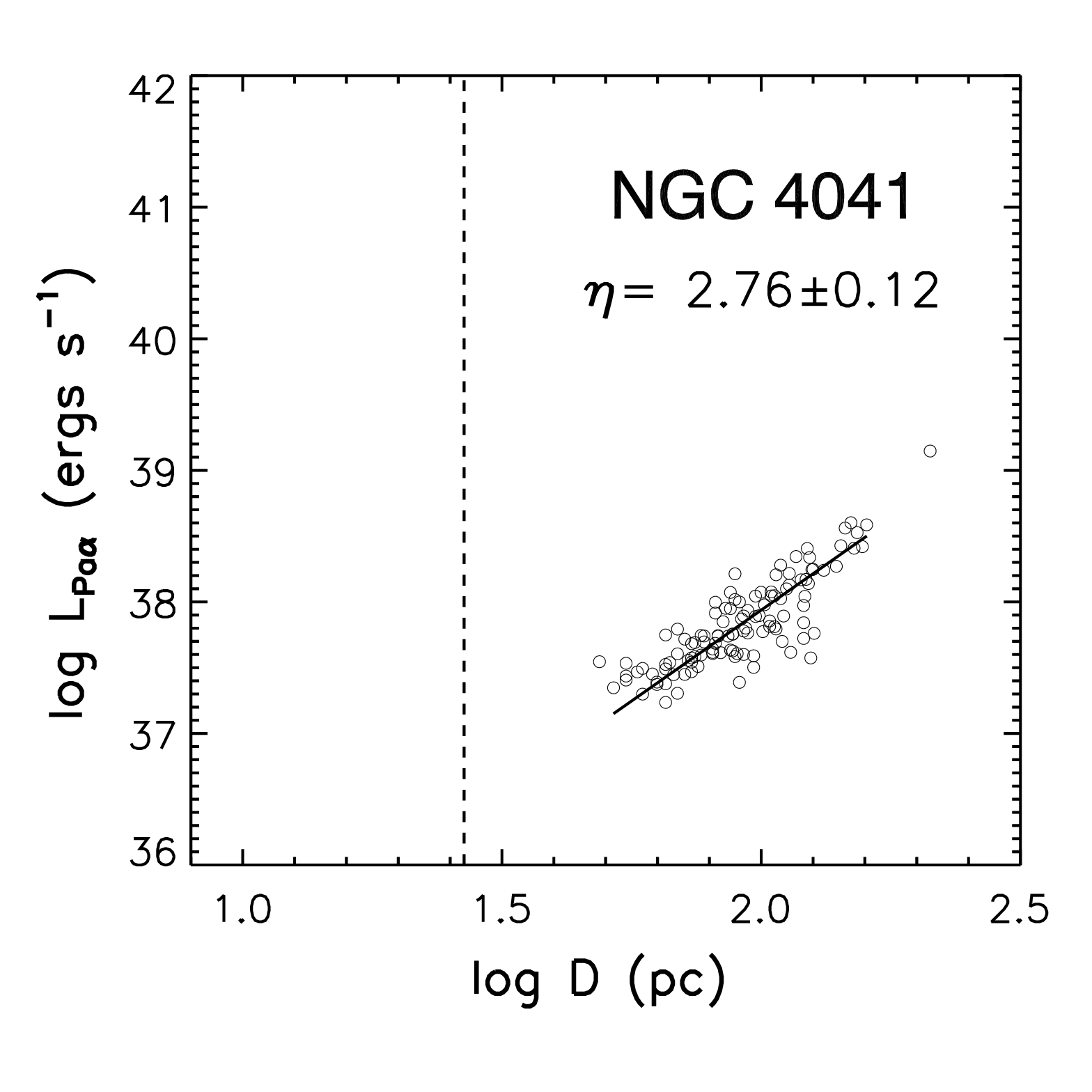}
    \includegraphics[origin=c,scale=0.26,trim=0cm 10mm 1cm 5mm]{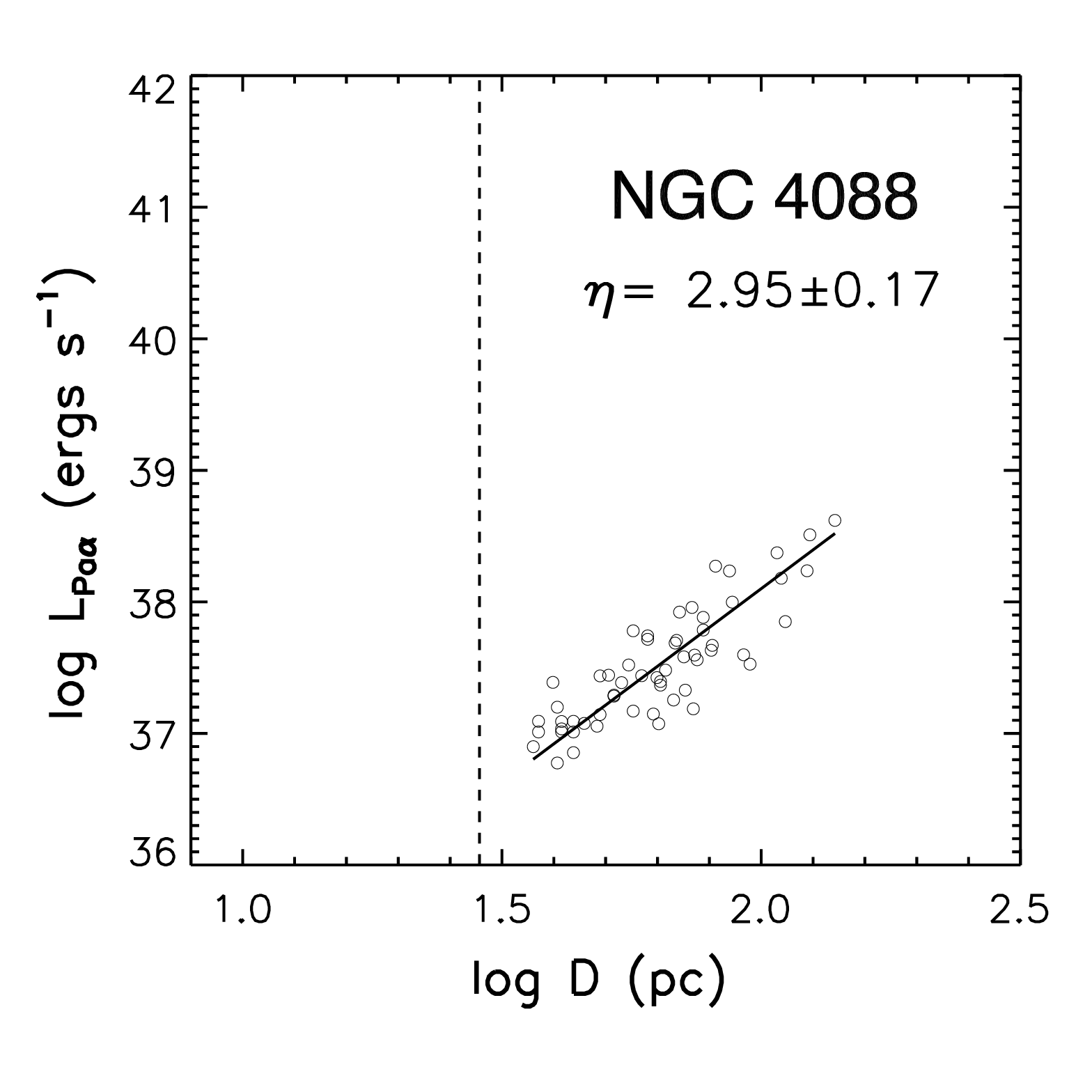}%
    \includegraphics[origin=c,scale=0.26,trim=0cm 10mm 1cm 5mm]{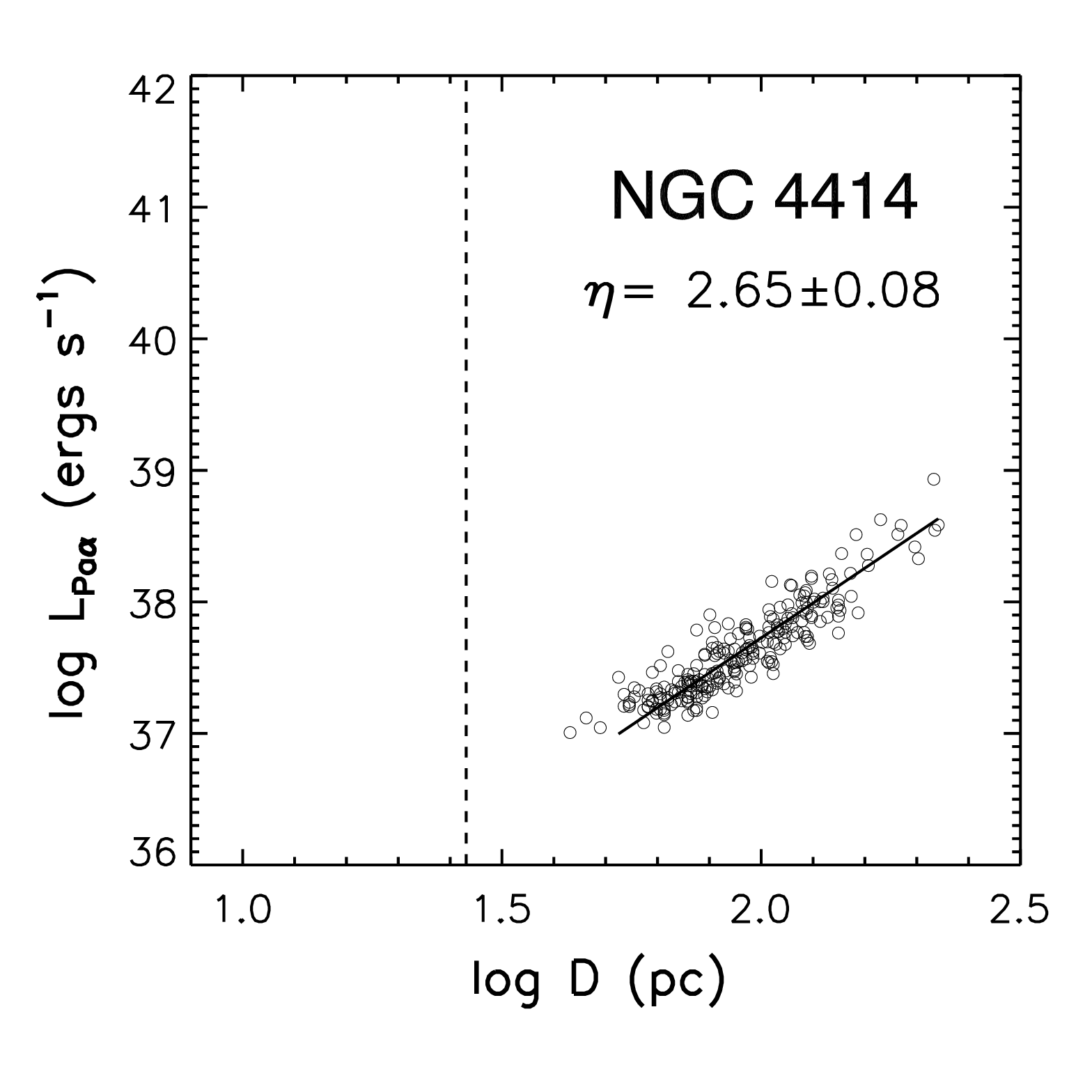}%
    \includegraphics[origin=c,scale=0.26,trim=0cm 10mm 1cm 5mm]{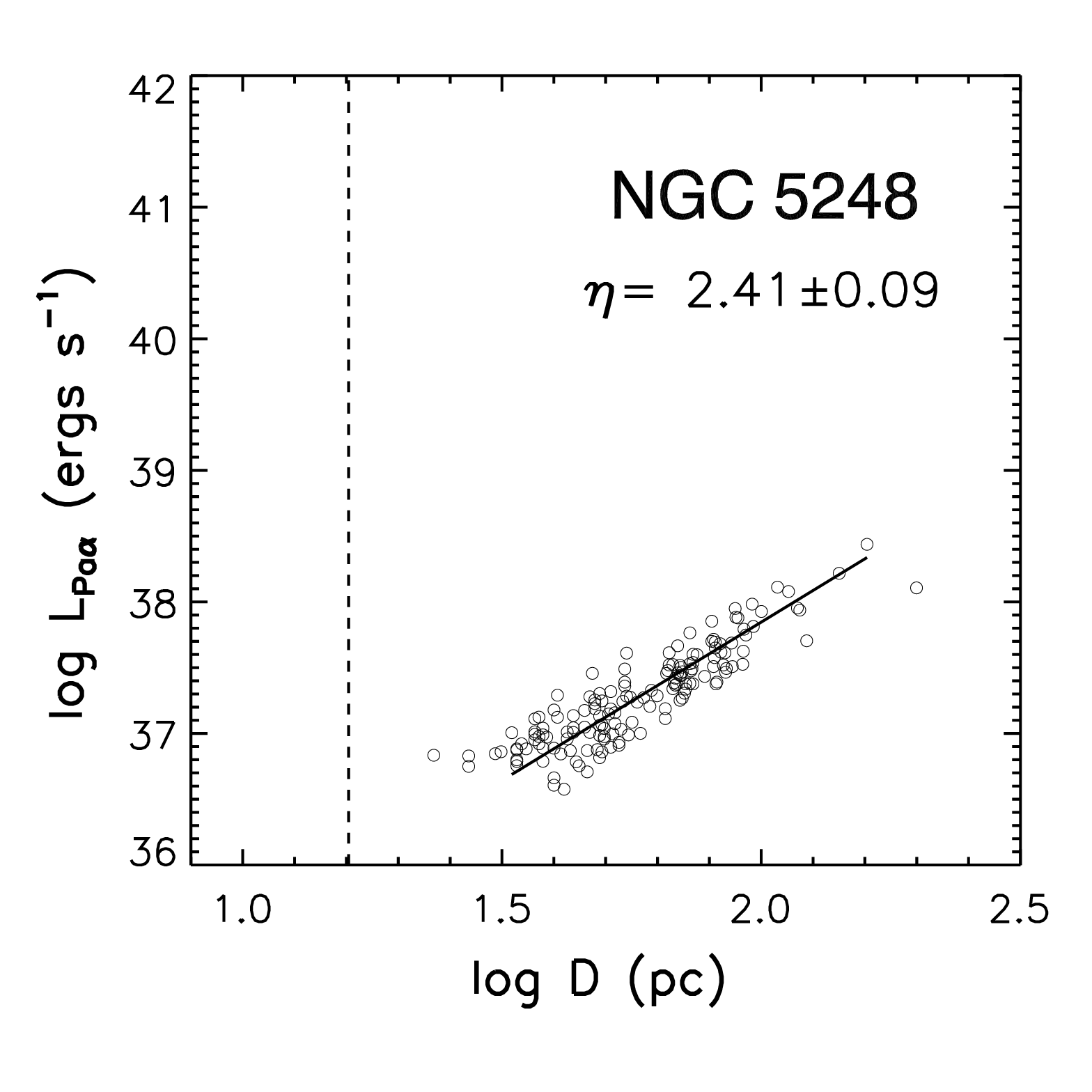}\\
\caption{The observed Pa$\alpha$ luminosities of the H {\sc ii} regions in the individual
galaxies as a function of equivalent diameter ($D$). Power-law fits are performed on 
regions with $S/N\geqslant3$. The linear resolution listed in Table~\ref{tbl:sample}
is shown by a vertical dashed line for each galaxy. In the panel for NGC 1097, the 
limiting envelope for source identification is shown by triangles connected with dotted 
lines (see text).}
\label{fig:ld_gold}
\end{figure*}

In order to characterize the spatial scale of an H {\sc ii} region with irregular morphology, we define
an equivalent diameter ($D$) which matches the area of a circle $\pi D^2/4$ to that of the region.
In general, larger H {\sc ii} regions tend to be brighter, leading to an empirical power-law scaling
relationship between their luminosities and sizes, expressed as $L_{\rm Pa\alpha} \propto D^{\eta}$.
We show the luminosity-size correlation derived from our Pa$\alpha$ images in 
Figure~\ref{fig:ld_gold}, where all $S/N\geqslant3$ regions are fit to the power law $L_{\rm Pa\alpha} \propto D^{\eta}$. 
We notice that all of the 12 galaxies show a value of $\eta$ between 2 and 3, and most (10 out of 12) galaxies 
have $\eta\gtrsim2.5$. The best-fit exponent $\eta$ is independent of the distance of the galaxy. 
However, there exists the possibility that the tight luminosity-diameter scaling relations are not physical,
but regulated by the surface brightness limits in the way we have defined objects. To test this, we 
run simulations by placing artificial sources onto the Pa$\alpha$ maps with Gaussian 
profiles, varying the luminosities and sizes of the sources, and running {\sl HIIphot} using the same 
parameter setting as for the respective galaxies to see if they can be identified. Fainter and more extended
H {\sc ii} regions are harder to detect, which forms an envelope below which the regions are not identified 
by the algorithm. We find that our observed luminosity-diameter relations are well above this limiting envelope 
for source identification, and are therefore physical rather than artificial. This envelope is shown for 
NGC 1097 as an example in Figure~\ref{fig:ld_gold}. 

\subsubsection{Source blending effects}

The existence of tight luminosity-size scaling relations like those of Figure~\ref{fig:ld_gold} 
begs for an explanation of the variations in the observed power-law index $\eta$, which 
generally has values between 2 and 3.  

\citet{Scoville01} argue that the luminosity is linked to $D$ through $L_{\rm H\alpha}\propto D^3$ 
if all the H {\sc ii} regions in a galaxy are resolved and no blending is present, while 
$L_{\rm H\alpha}\propto D^2$ in case of strong blending. Traditional ground-based H$\alpha$ 
imaging is limited to $\sim$1\arcsec--2\arcsec\ by atmospheric seeing. At a distance of 20 Mpc, 
this angular scale translates to 100--200 pc, larger than the diameter of a typical H {\sc ii} region, 
although comparable to the size of the much rarer giant H {\sc ii} regions, like 30 Doradus in the LMC. 
One might expect that higher resolution data like those provided by the HST would break the blends 
with ease. However, \citet{Scoville01} find a $L_{\rm H\alpha}\propto D^2$ relation for M51, even 
when using the 0.1\arcsec\ resolution HST images (corresponding to a linear scale of 4.5 pc at the 
distance of M51). Those authors thus argue that source blending is still strong in M51, despite the 
high angular resolution of the data.

\begin{figure}
\centering
\includegraphics[scale=0.4,clip,trim=0cm 0cm 0cm 0cm]{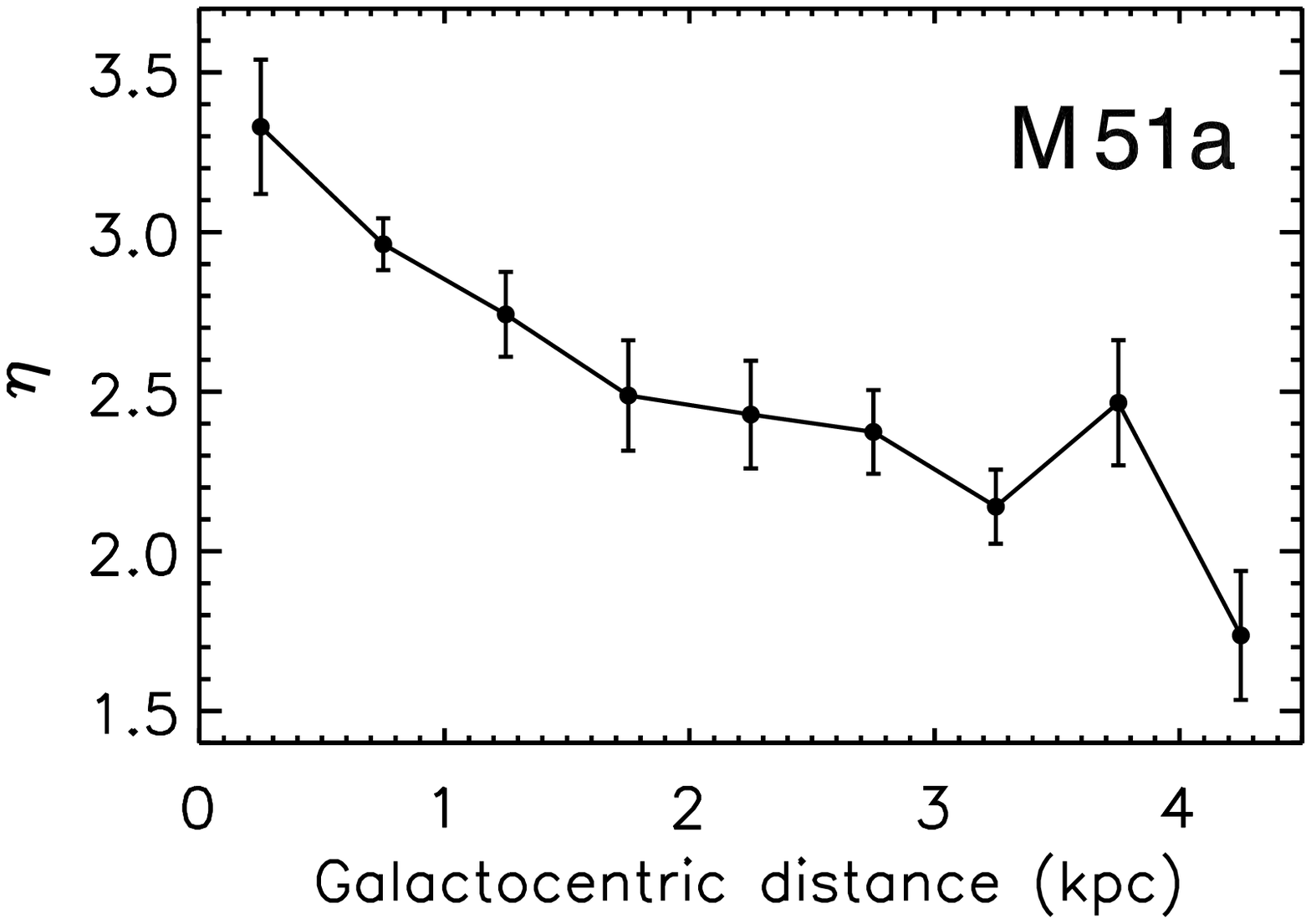}
\caption{The best-fit exponent of the luminosity-size relation ($L\propto D^{\eta}$) of the M51 
H {\sc ii} regions, plotted as a function of the galactocentric distance on the star-forming disk. 
This analysis is performed on the same HST H$\alpha$ image as that used by \citet{Scoville01}.}
\label{fig:m51_eta}
\end{figure}

If source blending were the reason for flattening $\eta$ towards a value of 2, it would be difficult 
to understand two observational facts. (1) Our Pa$\alpha$ observations yield larger $\eta$ values than 
\citet{Scoville01}, suggesting that our data break the H {\sc ii} blends more effectively, despite 
their lower angular resolution (0.26\arcsec) relative to that of \citet{Scoville01} in H$\alpha$ 
(0.1\arcsec). (2) When we divide the H$\alpha$ image obtained by \citet{Scoville01} into nine annuli 
and run {\sl HIIphot} on each of them, we find that the $L$--$D$ relation becomes flatter as a function 
of the galactocentric distance (Figure \ref{fig:m51_eta}). The best-fit value of $\eta$ is 3.3 in the 
central regions, and decreases monotonically to $\sim$2 at $\gtrsim$4 kpc. This result is 
counter-intuitive because the central kpc region of M51 is significantly more crowded than its 
outer regions, as promptly seen in the H$\alpha$ image.

In fact, the interpretation of the $L\propto D^2$ relation as an effect of blending is questionable. 
If individual regions (denoted $i$) have luminosity-diameter relations $L_i \propto D_i^3$, the 
luminosity-diameter relation of the blended regions will also show an overall $L\propto D^3$ trend. 
In reality, as both $L_i$ and $D_i$ will cover a range of values, the blended luminosity-size 
relation is likely to have a more complicated functional form. In order to derive the dependence of 
$\eta$ on region blending, we carry out a simple simulation by generating $10^6$ H {\sc ii} regions 
whose luminosities $L_i$ are randomly distributed between $10^{35}$ and $10^{38}$ erg s$^{-1}$ and 
following $d N(L)\;\propto L^{-2}dL$. We divide the whole sample into subsamples containing regions 
each resulting from $n$ blends. After calculating $L$ and $D$ using the same formalism as \citet{Scoville01}: 
\begin{equation}
L=\sum_{i=1}^{n} L_i \propto \sum_{i=1}^{n} D_i^3,~~{\rm and}~~ D=\left(\sum_{i=1}^{n} D_i^2\right)^{1/2},
\end{equation}
we fit a power law to the simulated data, and find that the exponent $\eta$ actually increases with 
the number $n$ of blended regions. Specifically, $\eta$ is found to range from 3.00 to 3.81 for $n$ 
that progressively increases from 1 to 10. Therefore, source blending steepens the $L$--$D$ relation 
and leads to $\eta>3$, instead of reducing $\eta$ to a value $\sim$2 as suggested by \citet{Scoville01}.

This simulation also provides an explanation for the larger values of $\eta$ we measure in our 
individual galaxies relative to that measured by \citet{Scoville01} in M51: blending actually steepens 
the luminosity-size relation. Furthermore, the radial profile of $\eta$ in M51, with decreasing values 
for increasing galactocentric distance, can be understood as a natural result of decreasing source 
blending effects. However, we still have not explained why $\eta\approx 2$ in many observational studies. 

\subsubsection{Other effects}

In this subsection, we explore other effects that may drive the observed values of $\eta$: (1) 
differences between the H$\alpha$ and Pa$\alpha$ analyses; (2) the H {\sc ii} region identification 
and measurement algorithm; (3) the presence of noise in the data; or (4) a faint-end cut-off in the 
H {\sc ii} LF. These may all play a role in shaping the $L$--$D$ relation, and could counteract the 
steepening effect of source blending. To disentangle the interplay among these factors, we have carried 
out a series of experiments.

In order to test the consistency of the H$\alpha$ and Pa$\alpha$ analyses, we compare the luminosity 
functions and the luminosity-size relations derived from M51a, for which images in both lines are available.
The central $R \leqslant 2.75$ kpc region of M51a is mapped both in H$\alpha$ and Pa$\alpha$ by HST 
\citep{Scoville01}. Running {\sl HIIphot} over this region results in an H$\alpha$ LF with 
$\alpha_{\rm H\alpha}=-1.12\pm0.05$ and a Pa$\alpha$ LF with $\alpha_{\rm Pa\alpha}=-1.09\pm0.05$, which 
are fully consistent with each other. Using {\sl HIIphot} again, this time on the entire FoVs of the HST 
H$\alpha$ and Pa$\alpha$ images of M51a, to reconstruct the $L$--$D$ relation, we find best-fit power 
indices $\eta_{\rm H\alpha}=2.47\pm0.04$ for the entire covered disk with a diameter of $\sim8$ kpc, and 
$\eta_{\rm Pa\alpha}=2.37\pm0.07$, consistent with each other within $1\sigma$. We thus recover a steeper 
slope than either \citet{Scoville01}, who derive $\eta_{\rm H\alpha}=2.1$, and \citet{Gutierrez11}, who 
find $\eta_{\rm H\alpha}=1.924$. Our relation also shows significantly more scatter than that found by 
both \citet{Scoville01} and \citet{Gutierrez11} (both use similar algorithms to identify and recover 
H {\sc ii} regions), but similar to the scatter recovered by \citet{Lee11}, who use {\sl HIIphot} (but 
derives $\eta_{\rm H\alpha}=2.16\pm0.02$). Clearly, the employed algorithm for identifying and measuring 
the H {\sc ii} regions has an effect on the level of scatter recovered in the $L$--$D$ relation. 

When limiting our analysis to the central 5 kpc diameter of M51 (the largest area available for our  
12 program galaxies), we find an H$\alpha$ $L$--$D$ relation with $\eta_{\rm H\alpha}=2.62\pm0.07$. 
This value is consistent with our Pa$\alpha$ results, that yield an average exponent $\langle\eta\rangle=2.70\pm0.27$,
implying consistency between our Pa$\alpha$ analysis and the H$\alpha$ analysis of \citet{Scoville01}. 
We conclude that although our Pa$\alpha$ data are shallower than the M51 H$\alpha$ by a factor of about 
10 \citep{Kennicutt07}, the derived statistical properties of H {\sc ii} regions are consistent with each 
other. The use of different emission lines is ruled out as a driver for $\eta\sim2$.

\begin{figure}
\centering
\includegraphics[scale=0.4,clip,trim=0cm 0cm 0cm 0cm]{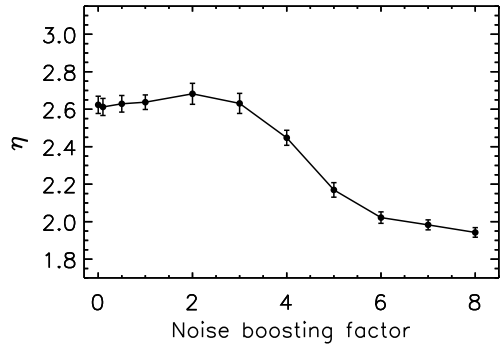}
\caption{The power-law index of the H {\sc ii} region luminosity-size relation as a function of
the noise level, tested using simulated images. The noise is extracted from the Pa$\alpha$ image of
NGC 1097 and multiplied by the noise boosting factor before being added to the simulated H {\sc ii}
regions.}
\label{fig:ld_noise}
\end{figure}

We now proceed to quantify the effect of using a specific algorithm for identifying and measuring H {\sc ii} 
regions (in our case: {\sl HIIphot}) and how this plays out for varying noise levels in the data. We carry 
out the experiment detailed below by running {\sl HIIphot} on a set of simulated Pa$\alpha$ images. We first 
create a noise map with the same size and pixel scale as our Pa$\alpha$ images by mosaicking the blank sky 
regions of the NGC 1097 Pa$\alpha$ map repeatedly. After that, we generate a sample of 800 H {\sc ii} regions 
whose luminosities are randomly drawn from the distribution $dN(L)/d\ln L\propto L^{-1}$ in the range  
$10^{35\mhyphen39}$ erg s$^{-1}$. The upper limit is anchored to the maximum H {\sc ii} region luminosity in 
NGC 1097 but is roughly applicable to other galaxies as well. The luminosity range of the observed H {\sc ii} 
regions covers about two orders of magnitude (Figure \ref{fig:ld_gold}), but we adopt a much broader range so 
that the faint end extends well below the detection limit of the H {\sc ii} regions. This is to avoid that the 
low luminosity end may play a role in the results. Each region is represented by a two-dimensional Gaussian 
profile whose integrated flux is the assigned luminosity and whose FWHM satisfies $L\propto\;{\rm FWHM}^3$ 
(the coefficient is adjusted to roughly match the observed size range). When we assign a random position to 
each region, we create a noiseless simulated image. In order to examine the variation of the $L$--$D$ relation 
at different noise levels, we multiply the noise map by a ``noise boosting factor'' which varies from 0, 0.1, 
0.5 and 1, 2, 3, ... to 8 progressively before adding it to the noiseless image to create a set of final 
simulated images. Using a parameter setting identical to what was applied to NGC 1097, we apply {\sl HIIphot} 
on the simulated images, and the measured values of $\eta$ are plotted in Figure \ref{fig:ld_noise}.

From this figure, we conclude that $\eta<3$  results from a combination of the {\sl HIIphot} algorithm
and the presence of noise in the data. Even when the noise boosting factor is 0 (noiseless data), 
we do not retrieve the input $L\propto D^3$, but find a significantly flatter slope ($\eta=2.62\pm0.05$), 
close to the average value in our sample, $\langle\eta\rangle=2.70\pm0.27$. Thus, the H {\sc ii} region 
retrieval algorithm has a measurable effect on the recovered $\eta$. As the noise increases, $\eta$ starts 
decreasing reaching a value of $\sim$2, when the noise is increased by a factor of 6 or more. From these 
simulations, we conclude that the value of $\eta$ decreases and reaches an asymptote of $\sim$2 as the 
noise in the image increases, or, analogously, as the signal-to-noise ratio of the H {\sc ii} regions 
decreases.  

If the minimum H {\sc ii} region that can be detected in a galaxy is ionized by a cluster containing a 
single O7.5 star, the corresponding Pa$\alpha$ luminosity is $10^{36.2\mhyphen36.3}$ erg s$^{-1}$. A faint-end 
luminosity cutoff has also the property of flattening the $L$--$D$ relation, almost irrespective of the 
level of blending. Running Monte Carlo simulations, we recover smaller values of $\eta$ corresponding to 
$\Delta\eta\sim0.4$--0.5, for the cut-off above, even when the data are noiseless. Higher cut-off 
luminosities yield stronger flattening for $\eta$. 

In summary, the exponent of the $L$--$D$ relation, $\eta$, is subject to variations that are unrelated to 
intrinsic physical changes in the properties of the H {\sc ii} regions, but are most likely induced by 
observational biases, as already noted by \citet{Scoville01}. However, not all observational biases work 
in the direction of yielding smaller-than-true values of $\eta$. For instance, our simulations indicate 
that source blending steepens the $L$--$D$ relation to values $\eta>3$. Conversely, $\eta$ flattens to 
values $<3$, and approaches $\eta\sim2$ in the presence of both decreasing signal-to-noise data and a 
faint-end cut-off to the luminosity function of H {\sc ii} regions. The use of the algorithm {\sl HIIphot} 
to identify and measure the H {\sc ii} regions also has some role in recovering values of $\eta<3$. 
Instead, the use of either H$\alpha$ or Pa$\alpha$ images to perform the analysis has negligible 
impact on the recovered values of $\eta$. 

In light of the above, we can interpret both $\eta\sim2$ as measured by \citet{Scoville01} in M51, and 
the radial profile of $\eta$ in the same galaxy (Figure \ref{fig:m51_eta}). In the center of the galaxy, 
significant blending yields $\eta>3$, but, as the galactocentric distance increases, both source blending 
and the overall signal-to-noise ratio in the image decrease, thus reducing $\eta$. Presence of a intrinsic 
faint-end cut-off in the H {\sc ii} regions of M51 would further help reducing $\eta$. In spite of these
artifacts, the intrinsic radial dependence of the H {\sc ii} region properties and their environment, if 
it exists, could also play a role in shaping the radial profile of $\eta$.

\begin{figure*}
\centering
    \includegraphics[origin=c,scale=0.26,trim=0cm 10mm 1cm 5mm]{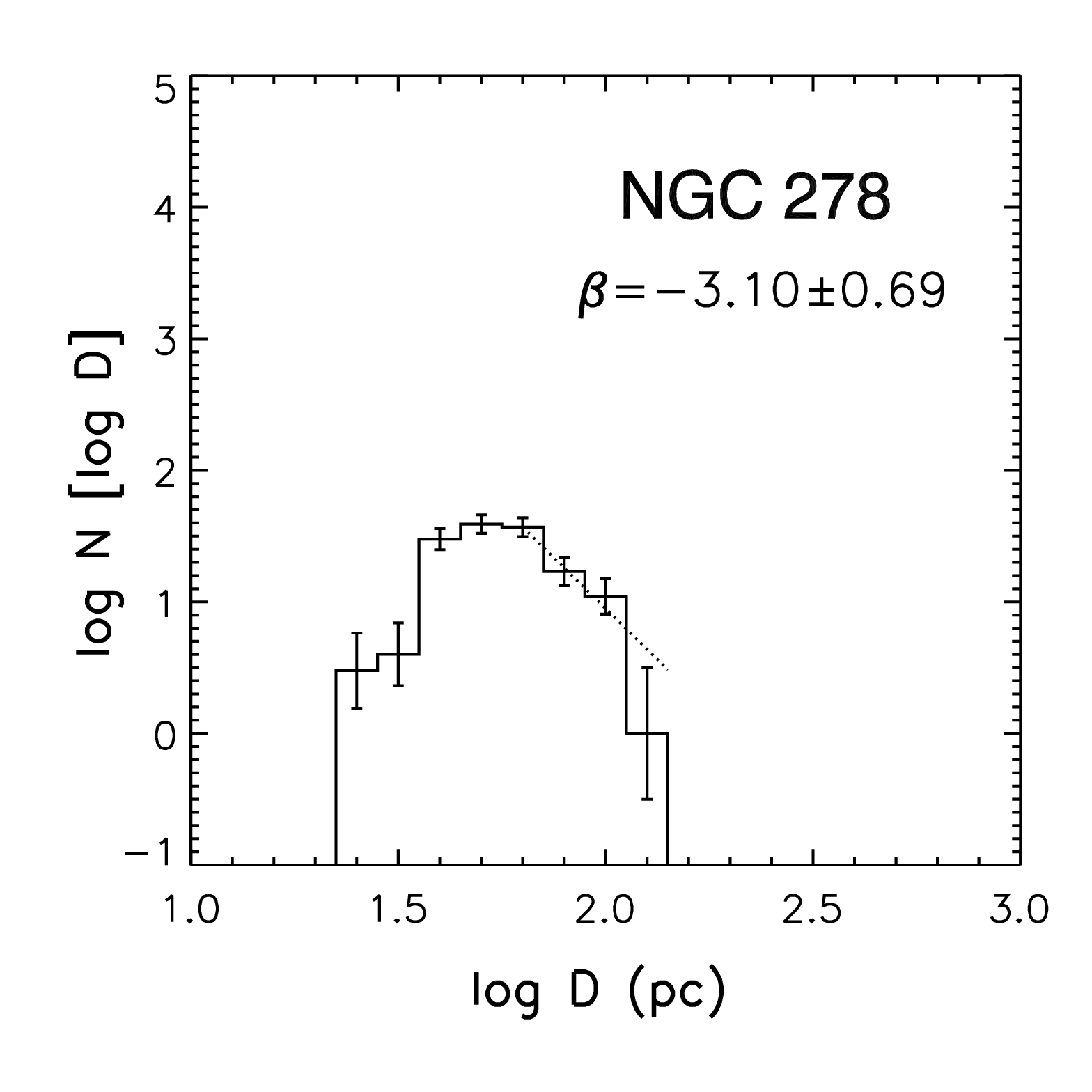}%
    \includegraphics[origin=c,scale=0.26,trim=0cm 10mm 1cm 5mm]{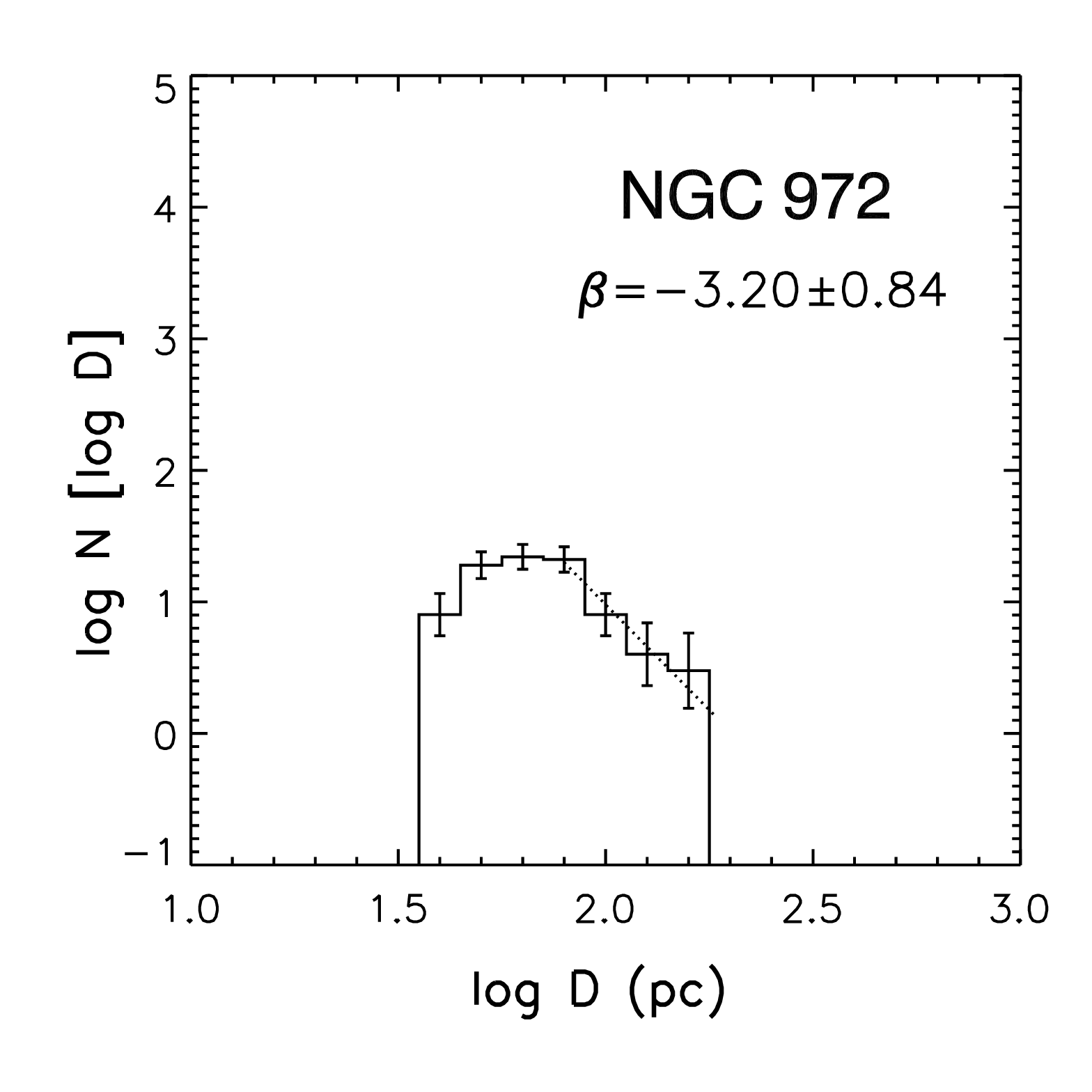}%
    \includegraphics[origin=c,scale=0.26,trim=0cm 10mm 1cm 5mm]{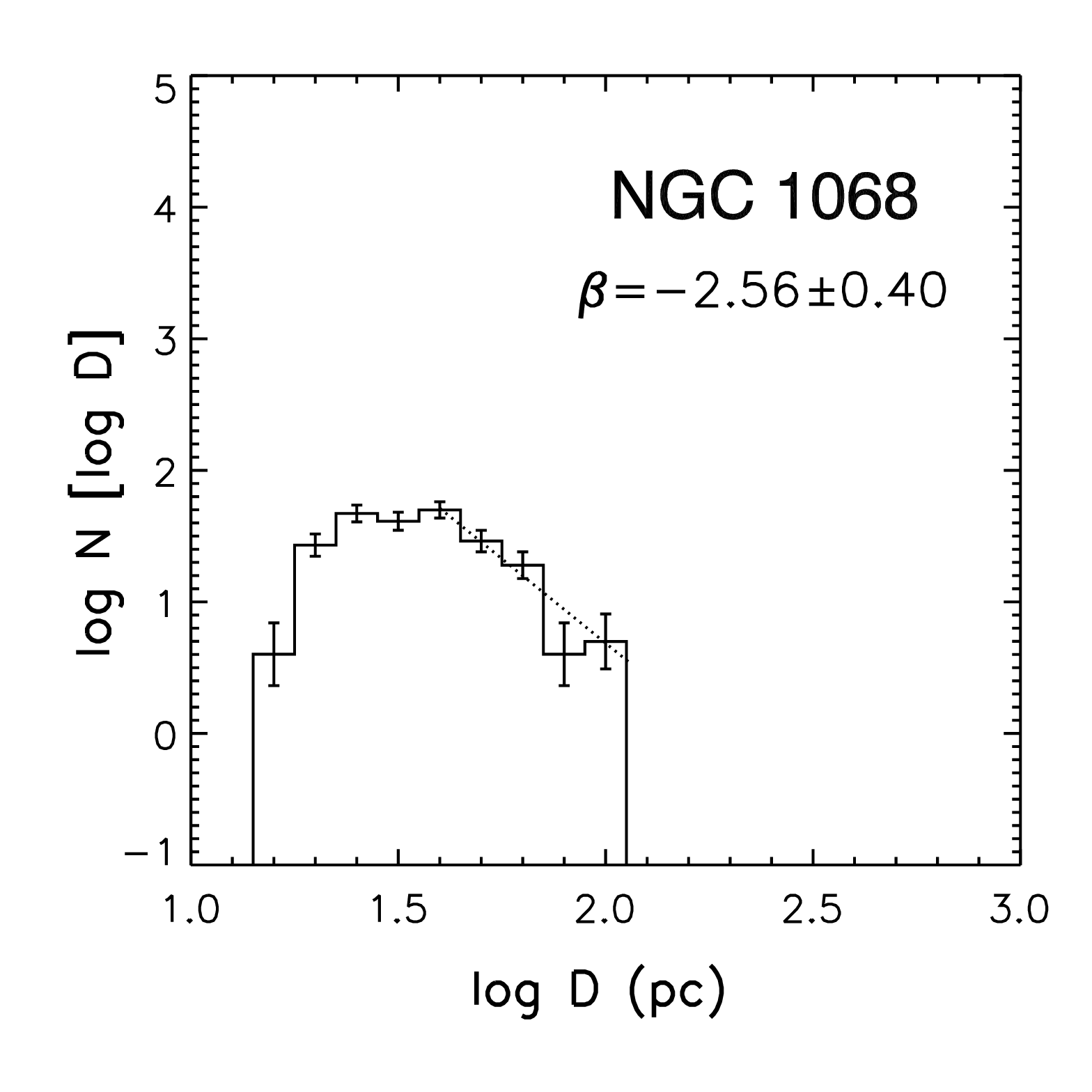}
    \includegraphics[origin=c,scale=0.26,trim=0cm 10mm 1cm 5mm]{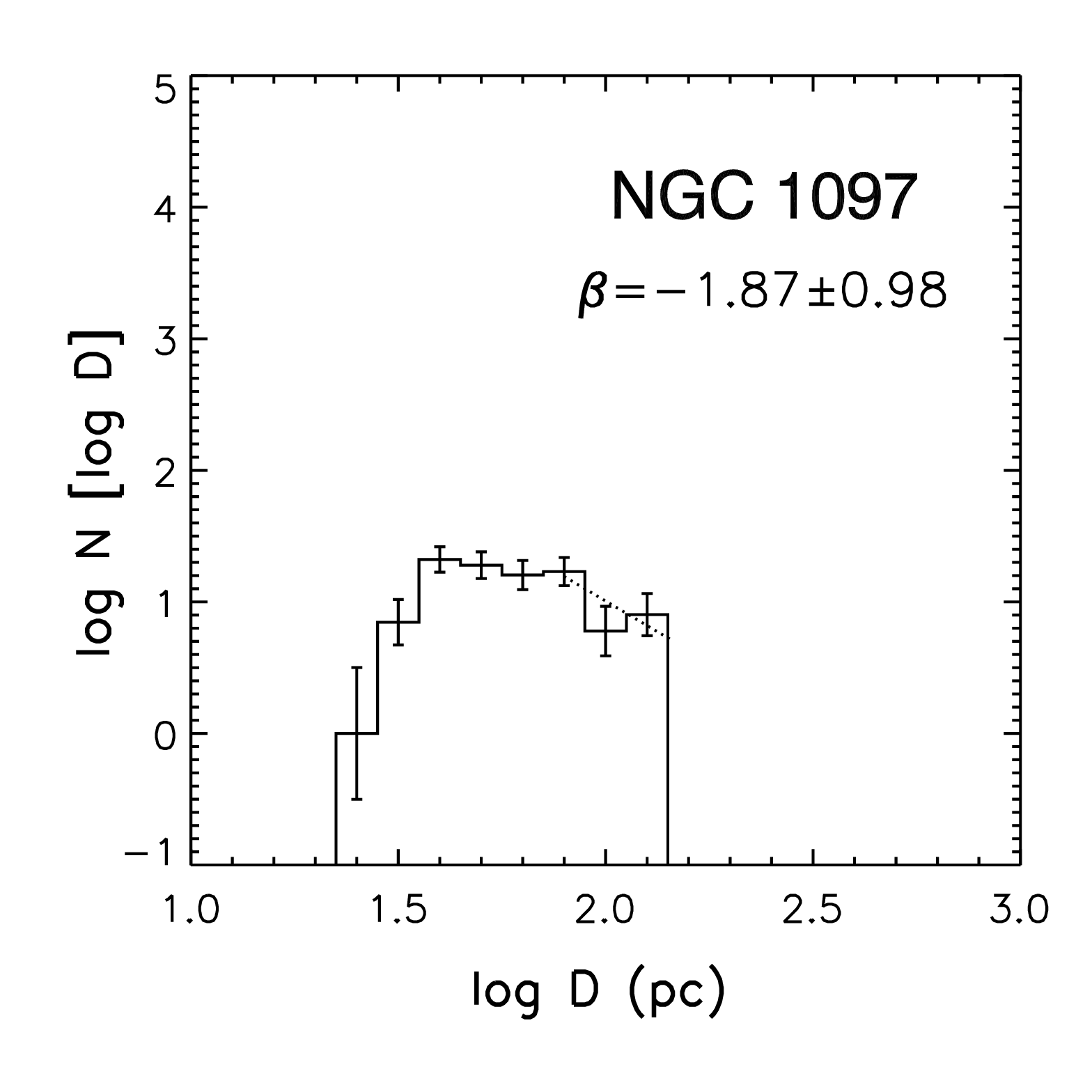}%
    \includegraphics[origin=c,scale=0.26,trim=0cm 10mm 1cm 5mm]{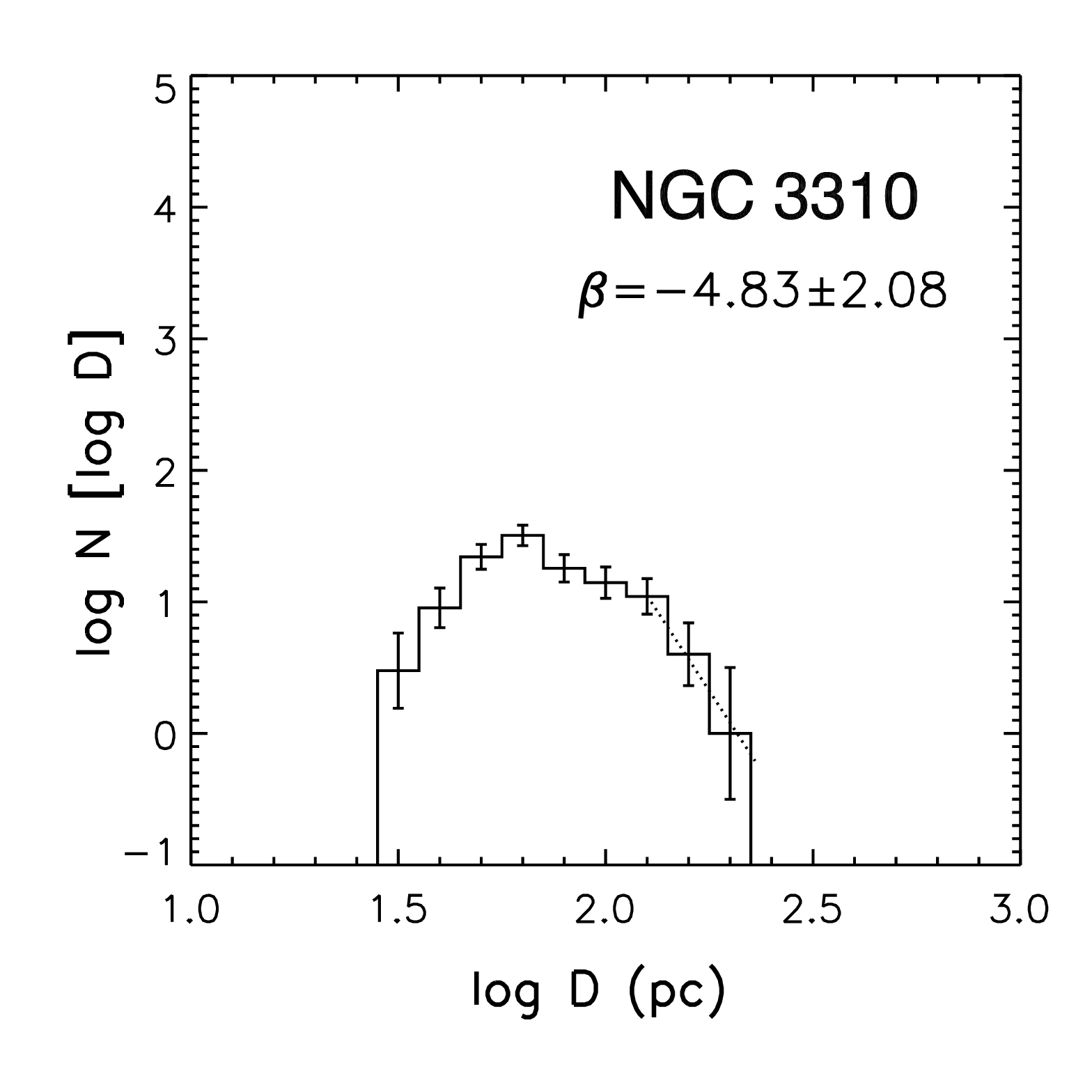}%
    \includegraphics[origin=c,scale=0.26,trim=0cm 10mm 1cm 5mm]{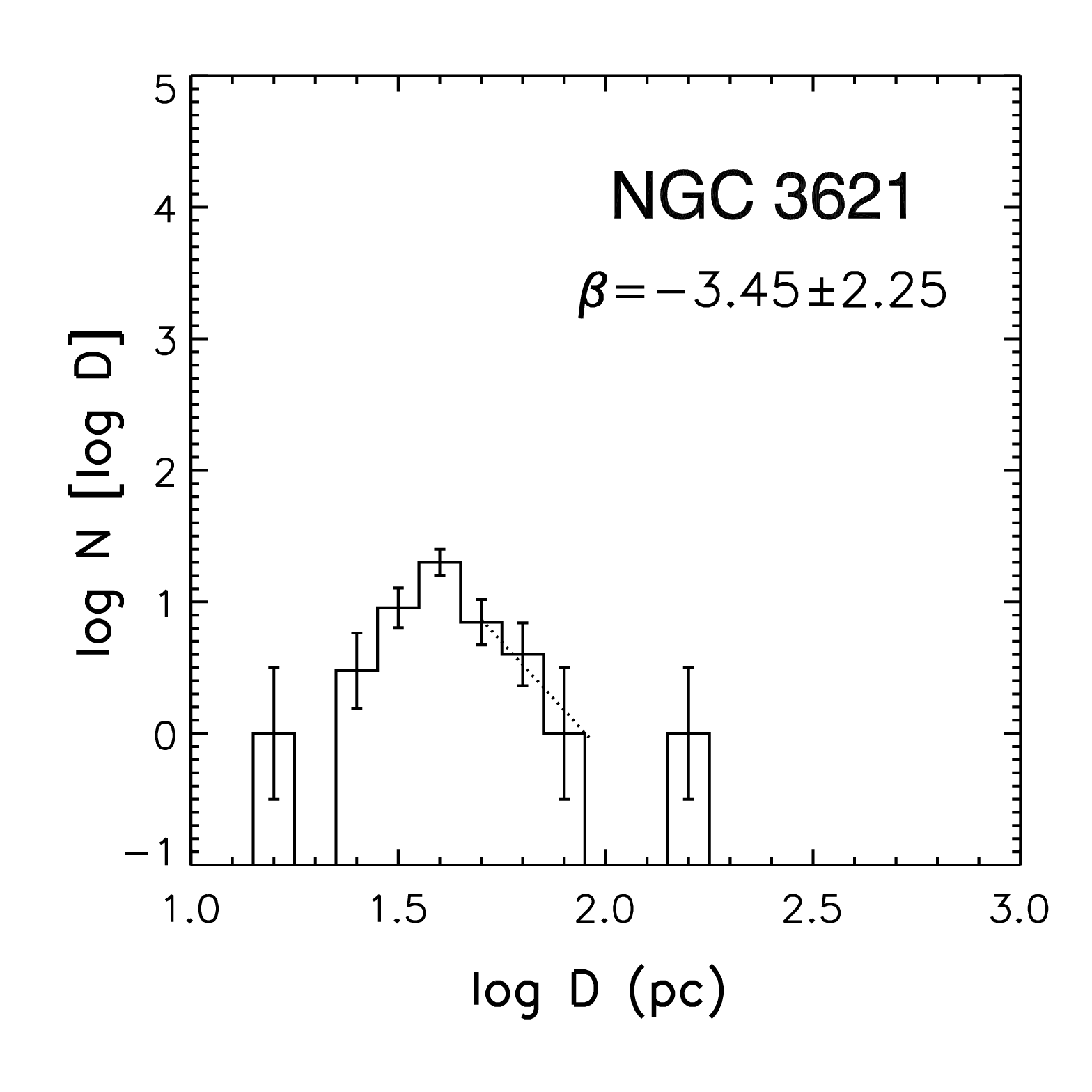}
    \includegraphics[origin=c,scale=0.26,trim=0cm 10mm 1cm 5mm]{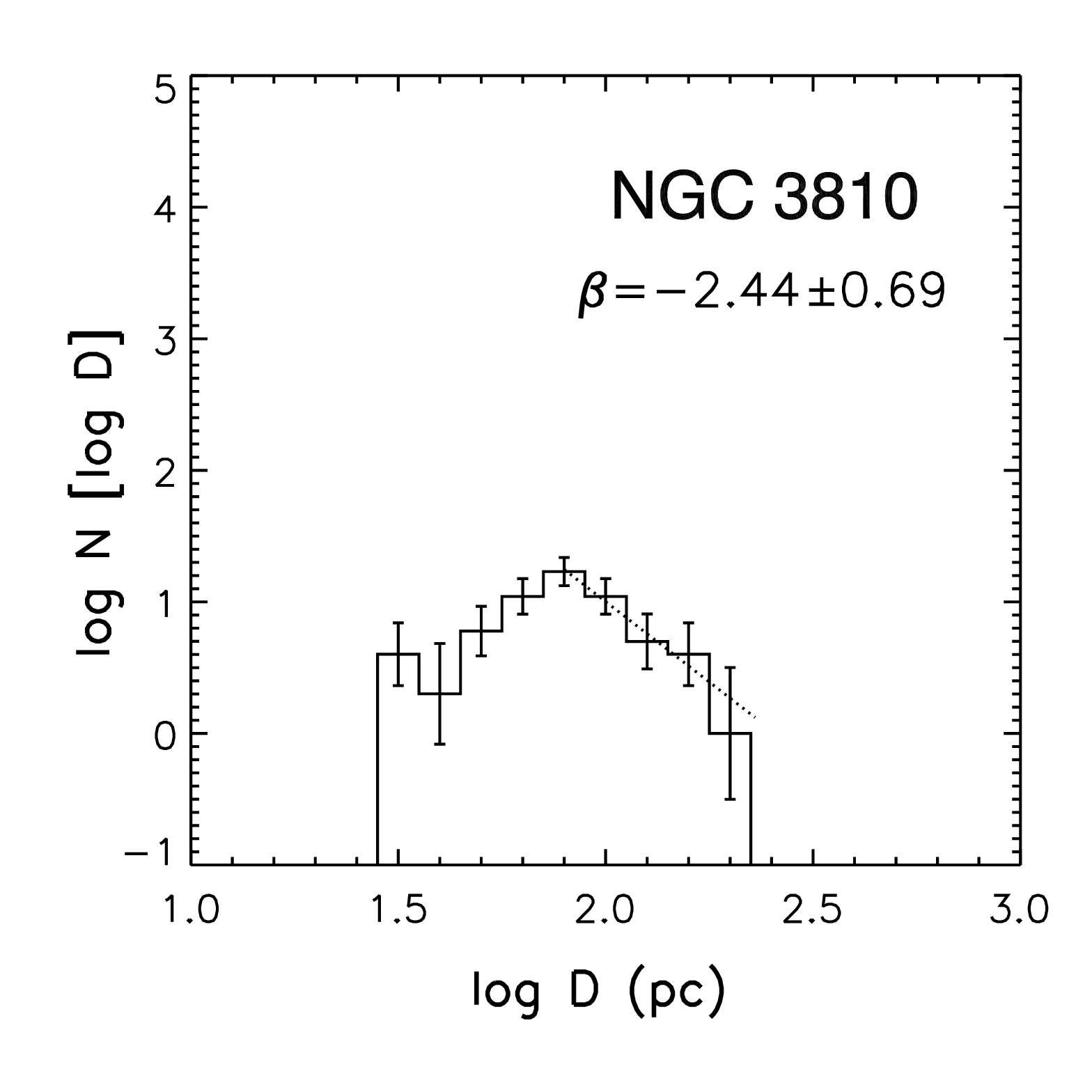}%
    \includegraphics[origin=c,scale=0.26,trim=0cm 10mm 1cm 5mm]{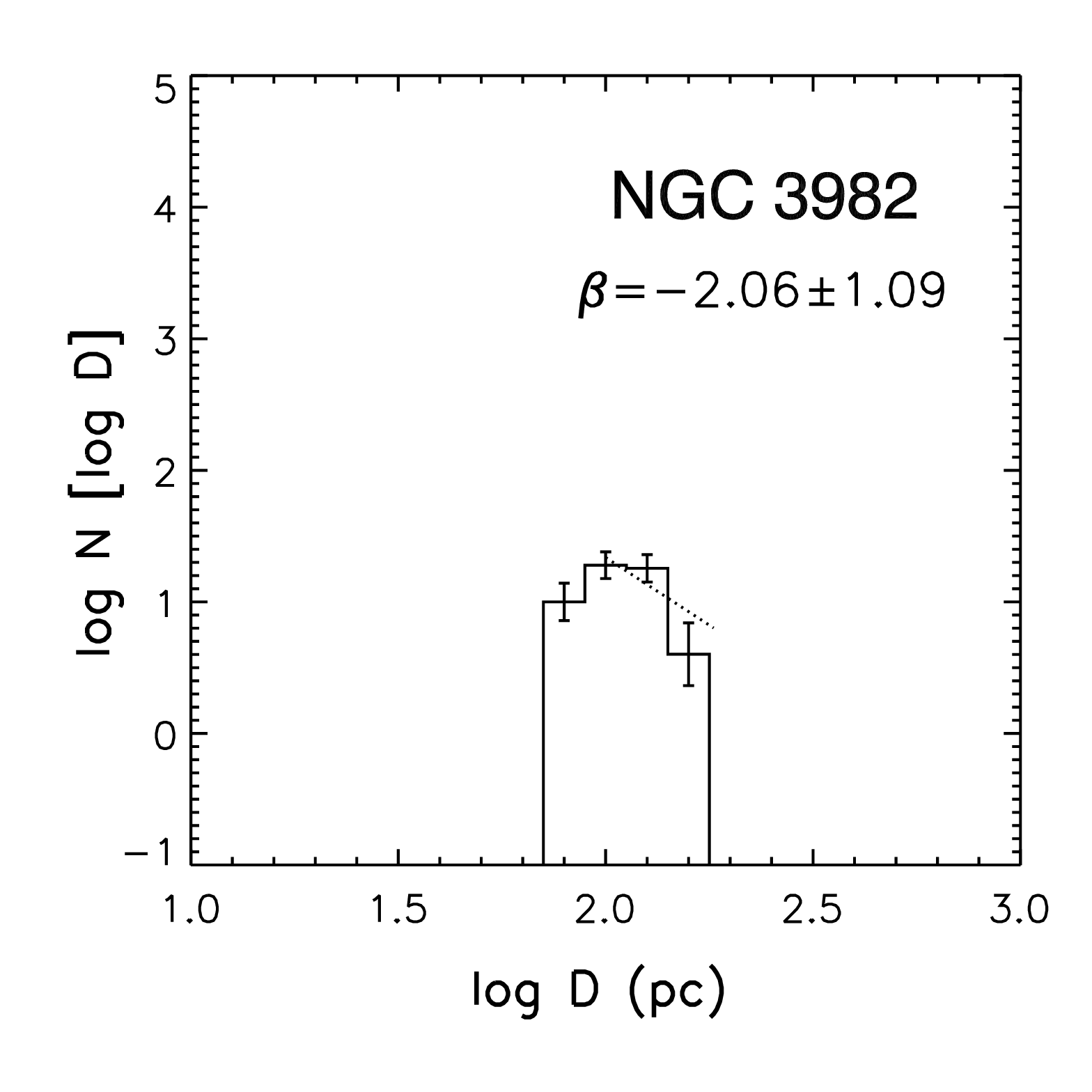}%
    \includegraphics[origin=c,scale=0.26,trim=0cm 10mm 1cm 5mm]{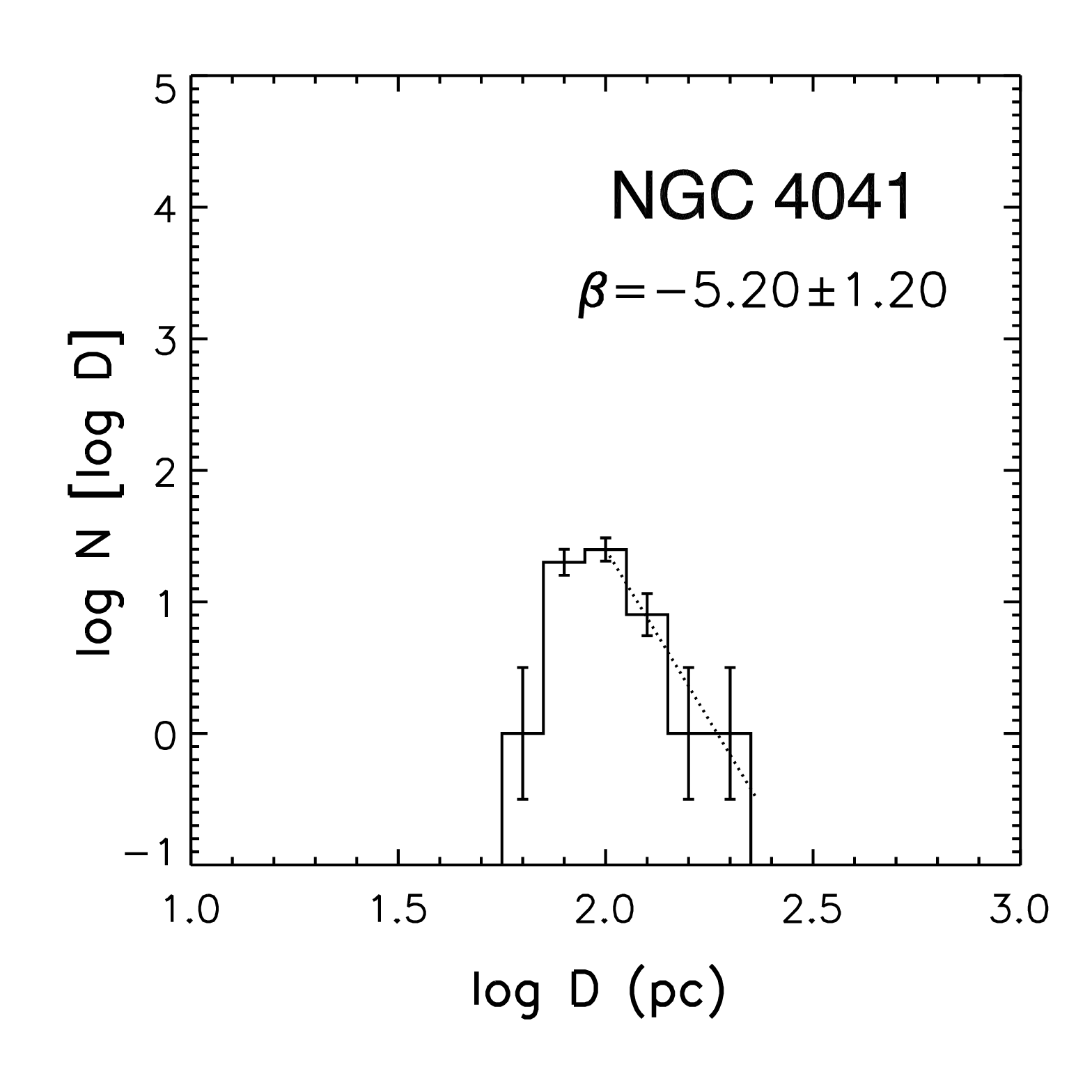}
    \includegraphics[origin=c,scale=0.26,trim=0cm 10mm 1cm 5mm]{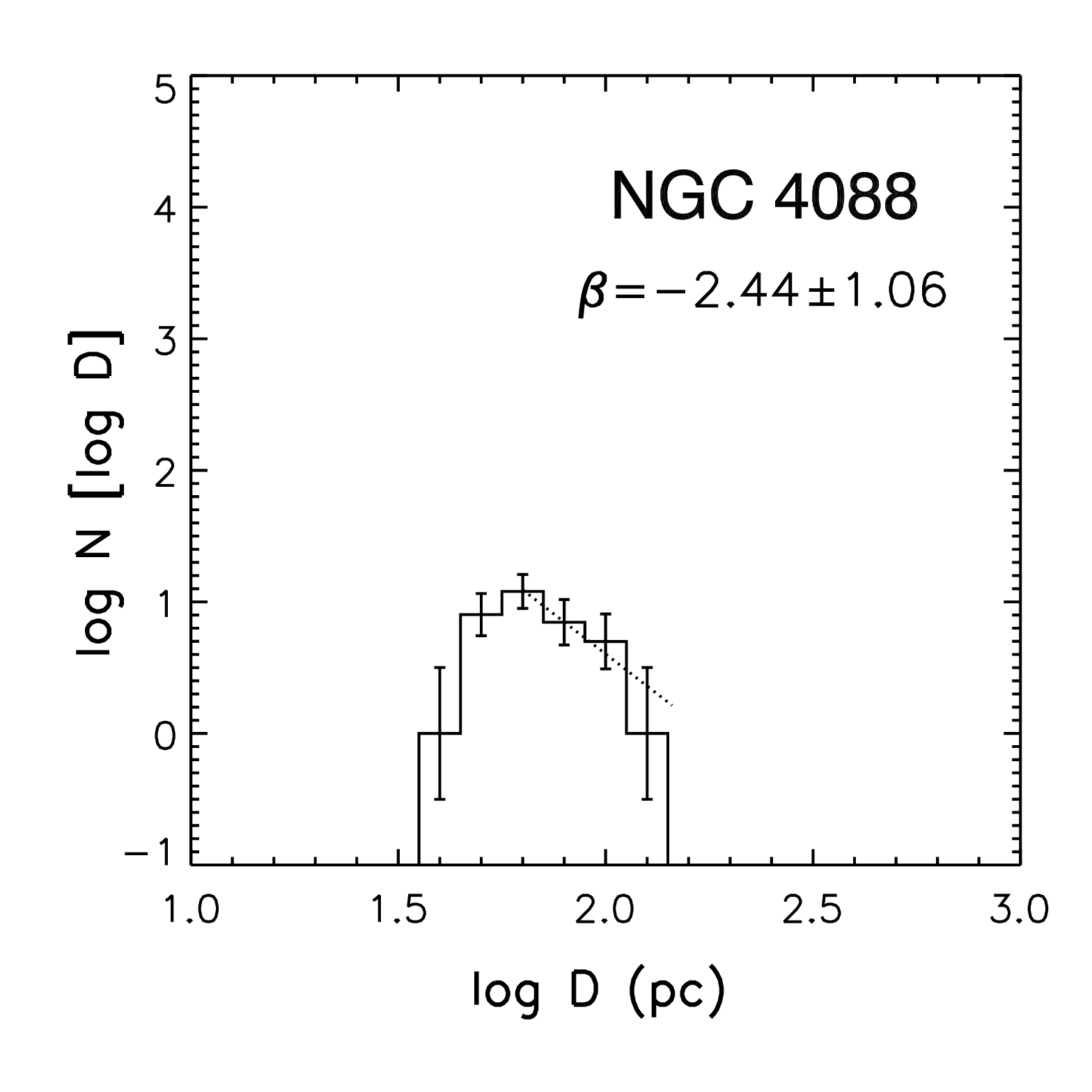}%
    \includegraphics[origin=c,scale=0.26,trim=0cm 10mm 1cm 5mm]{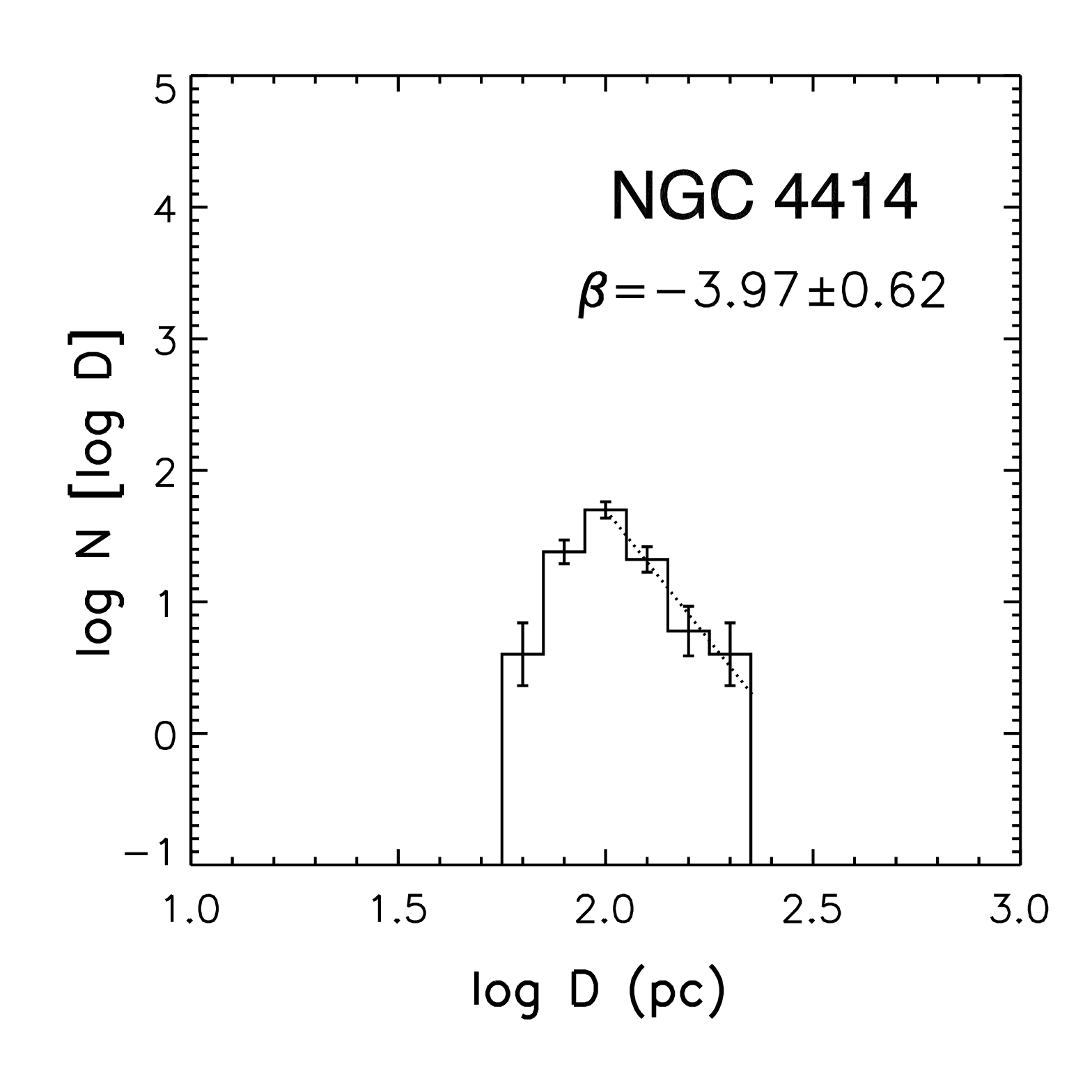}%
    \includegraphics[origin=c,scale=0.26,trim=0cm 10mm 1cm 5mm]{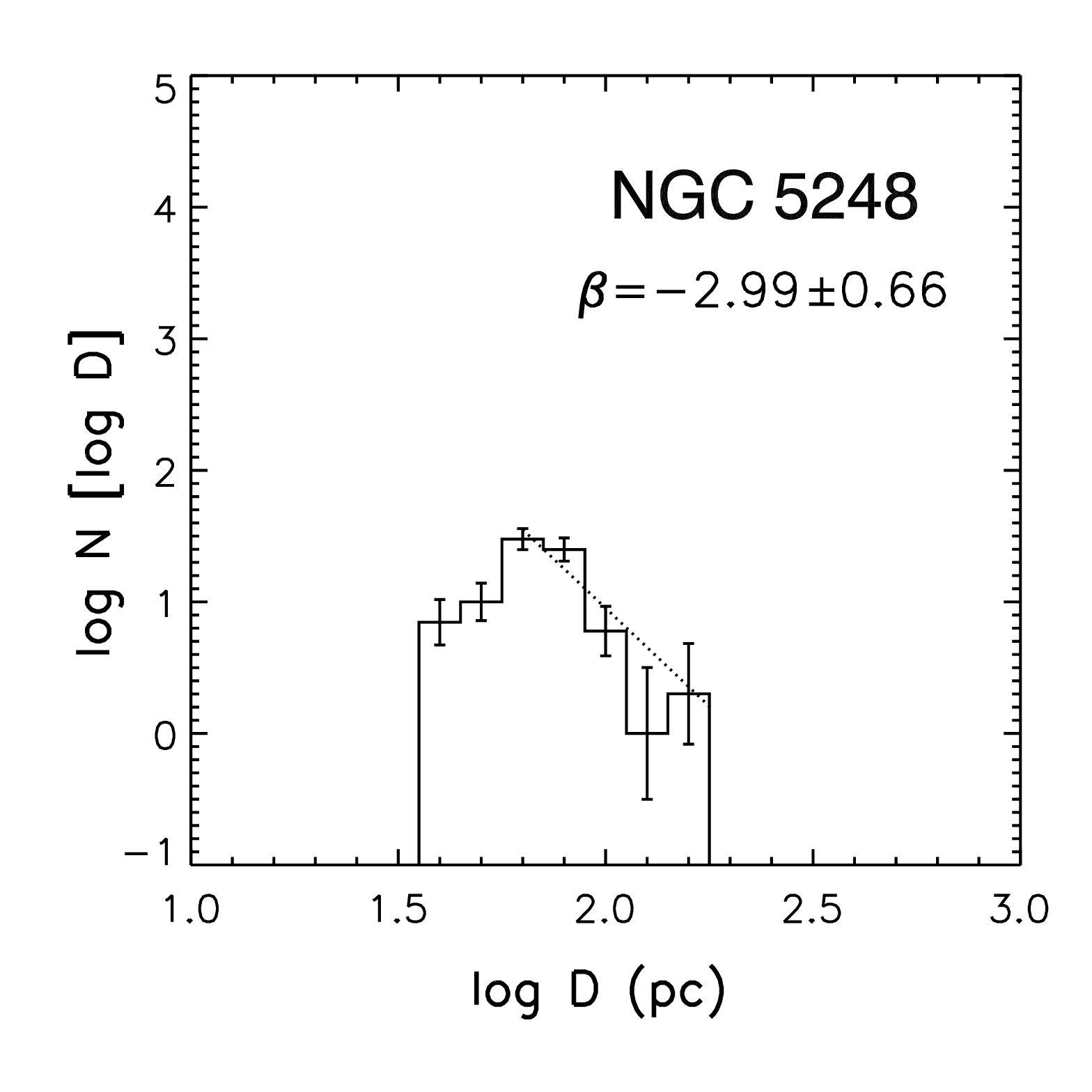}\\
\caption{The H {\sc ii} region size distribution of the 12 sample galaxies. The power-law
fits are performed in a manner similar to Figure~\ref{fig:lf_gold}. Only regions with 
$S/N\geqslant5$ are shown, of which the large-size end is fitted.}
\label{fig:sf_gold}
\end{figure*}

\subsection{Size and electron density distribution}
\label{sec:n_e}

The H {\sc ii} region size distribution for the sample galaxies are plotted in Figure~\ref{fig:sf_gold}
for regions with S/N$\ge$5, and the large-size end is fitted with a power law formulated similarly to 
that of the Pa$\alpha$ line luminosity:

\begin{equation}
dN(D)/d\ln D \propto D^{\beta}.
\end{equation}

For isolated sources, $L\propto D^3$ is expected, and the power index in 
the above equation will take the value $\beta=-3$ \citep[][note $\beta$ is linked to the parameter $b$ 
therein, which satisfies $N(D)~dD \propto D^{-b}~dD$, through $\beta=1-b$]{Oey03}. In case of  
$L\propto D^2$ as often seen in observations, $\beta=-2$ instead. The H {\sc ii} region size distribution in 
most of our galaxies are consistent with a slope $\beta\sim -3$ (Figure~\ref{fig:sf_gold}), but variations 
from galaxy to galaxy are significantly larger than for the LF power index $\alpha$. For instance, the 
size distributions of NGC 1097 and NGC 3982 do not appear to follow a power 
law behavior at all. Averaging over the sample without including these two objects, the mean value
of the exponent is $\langle\beta\rangle=-3.42$, and the median $\tilde{\beta}=-3.10$, with an uncertainty 
of $\pm1.04$, consistent with $\beta=-3.12\pm0.90$ found by \citet{Oey03}. 
The largest spatial scale beyond which the power-law scaling truncates (presumably because of small number
statistics), $D_{\rm up}$, varies from $\sim$90 pc to $\sim$200 pc.

The remarkably larger uncertainty on $\beta$ is partly caused by the fact that the H {\sc ii} region
sizes are much more weakly correlated with $S/N$ than the luminosities. For this reason, applying a $S/N$ threshold 
only leads to a sharp cut-off at the faint end of the LF, but changes the size distribution {\it globally}
or even artificially destruct its power-law scaling. This shortcoming can only be fixed with future deeper
data. On the other hand, one should be aware that the above simple calculations towards an expected behavior 
of the size distribution are based on an implicit assumption that the three relations (the H {\sc ii} LF, 
luminosity-size correlation and size distribution) are all dominantly driven by the same objects. However,
as can be seen in Figure~\ref{fig:lf_gold} and Figure~\ref{fig:ld_gold}, the values of $\alpha$ and $\eta$
are basically determined by the full sample of all detected H {\sc ii} regions (the power-law scaling roughly 
extends to completeness limits in most galaxies, see Figure~\ref{fig:lf_gold}), but $\beta$ is only determined 
by a fraction (often $<$50\%) of the largest regions (Figure~\ref{fig:sf_gold}). We will see in the co-added
analysis of the H {\sc ii} region size distribution (\S \ref{sec:coadd} and Figure~\ref{fig:sf_coadd}) that the largest
H {\sc ii} regions (some possibly blends) form a power law with $\beta=-3.89\pm0.48$. 

\begin{figure*}
\centering
    \includegraphics[origin=c,scale=0.26,trim=0cm 10mm 1cm 5mm]{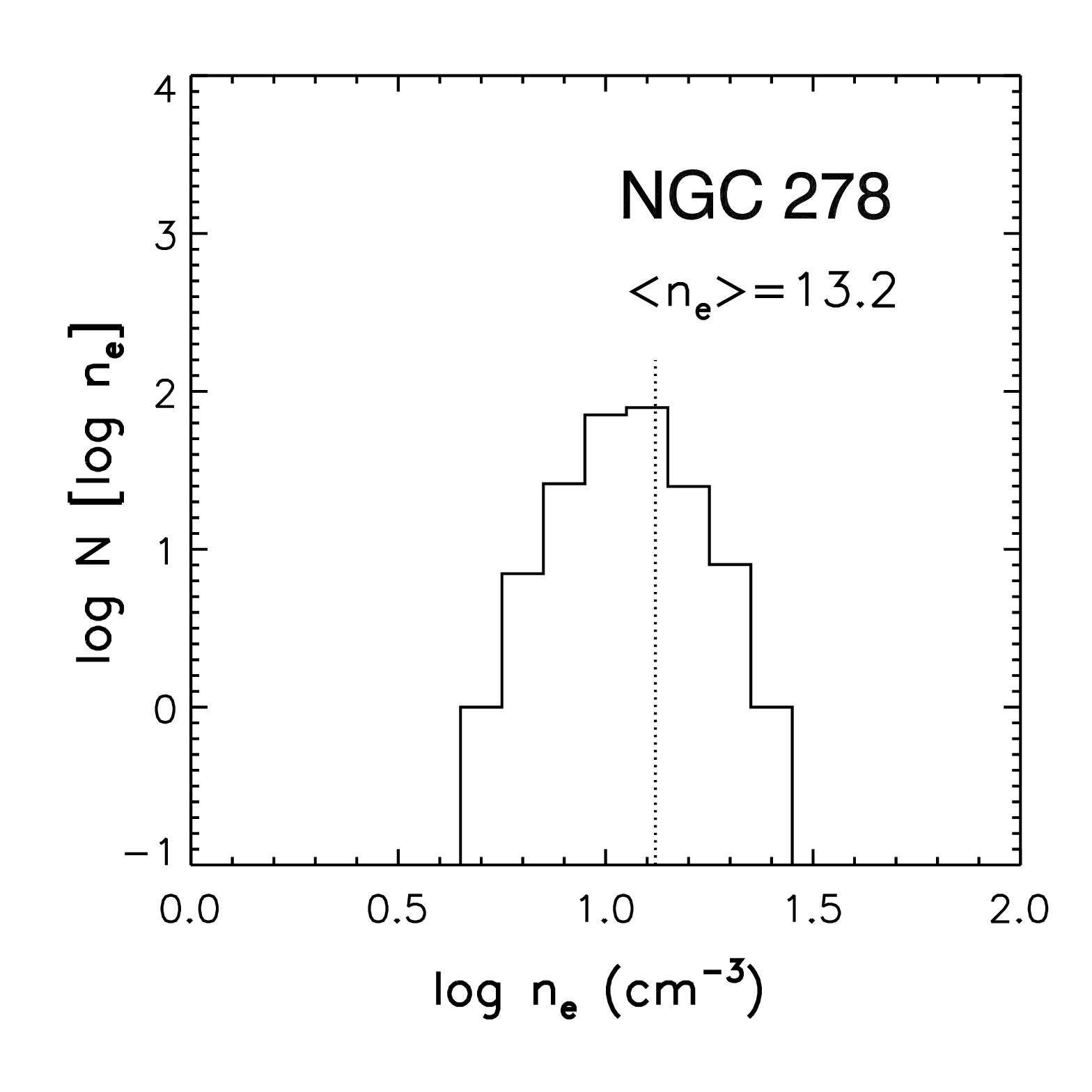}%
    \includegraphics[origin=c,scale=0.26,trim=0cm 10mm 1cm 5mm]{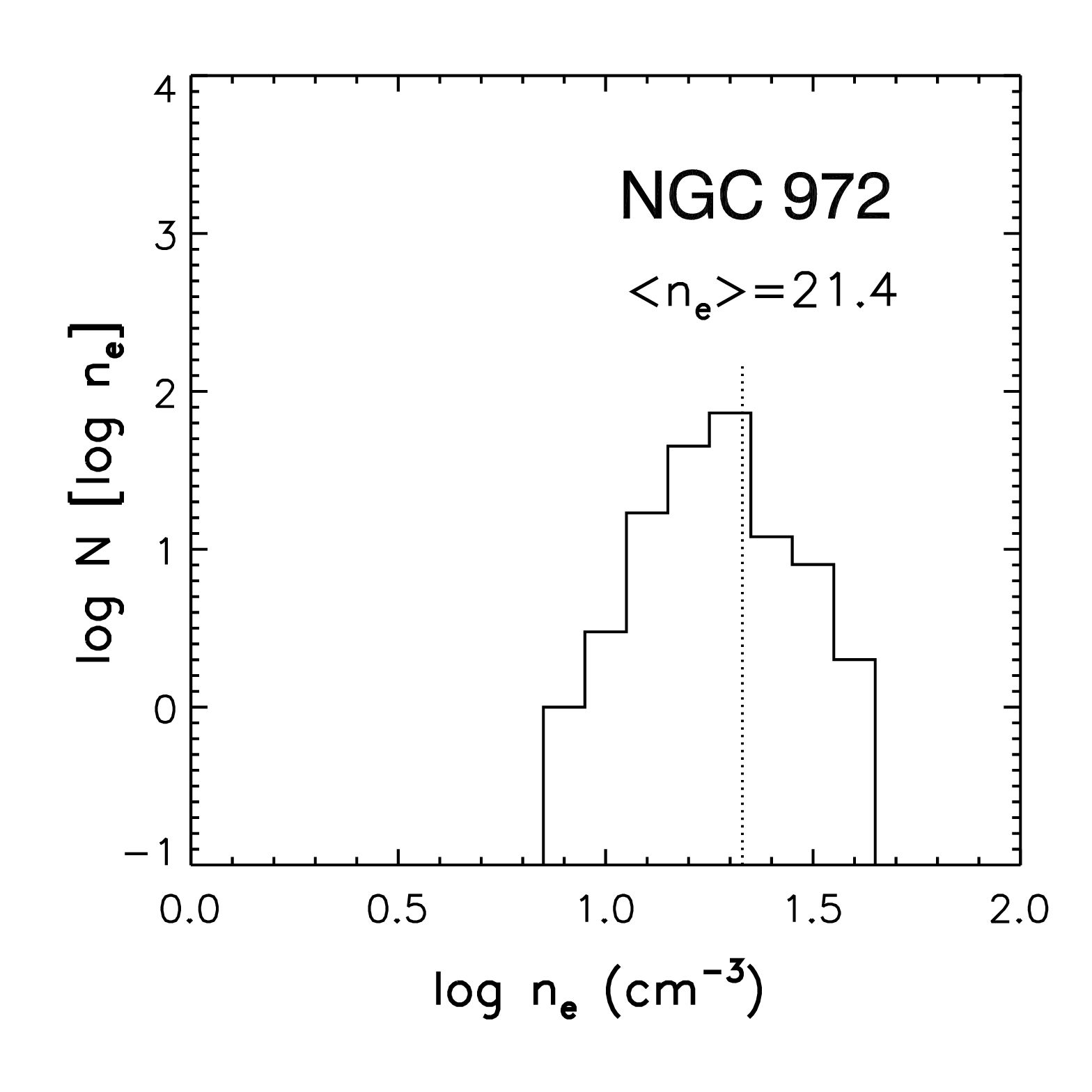}%
    \includegraphics[origin=c,scale=0.26,trim=0cm 10mm 1cm 5mm]{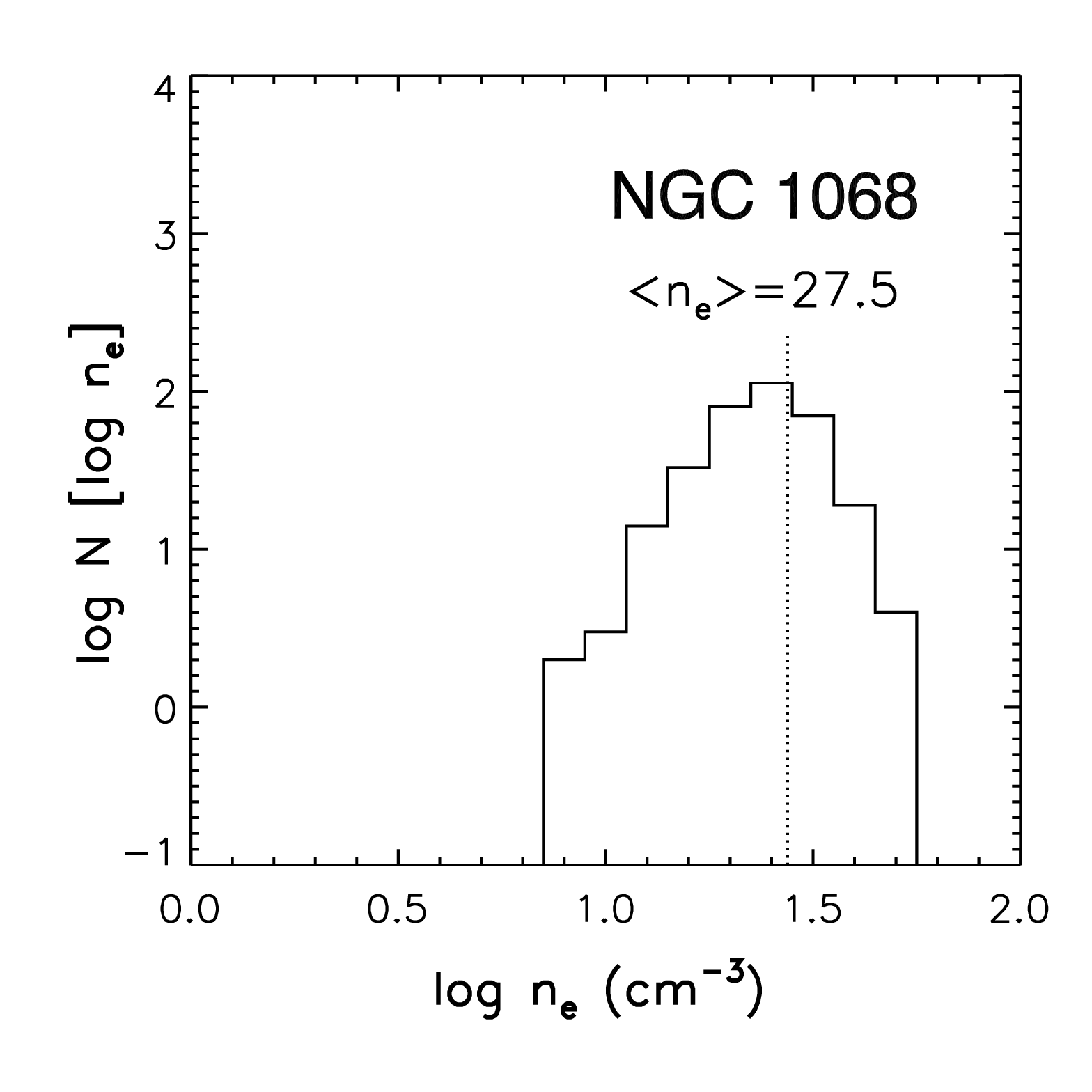}
    \includegraphics[origin=c,scale=0.26,trim=0cm 10mm 1cm 5mm]{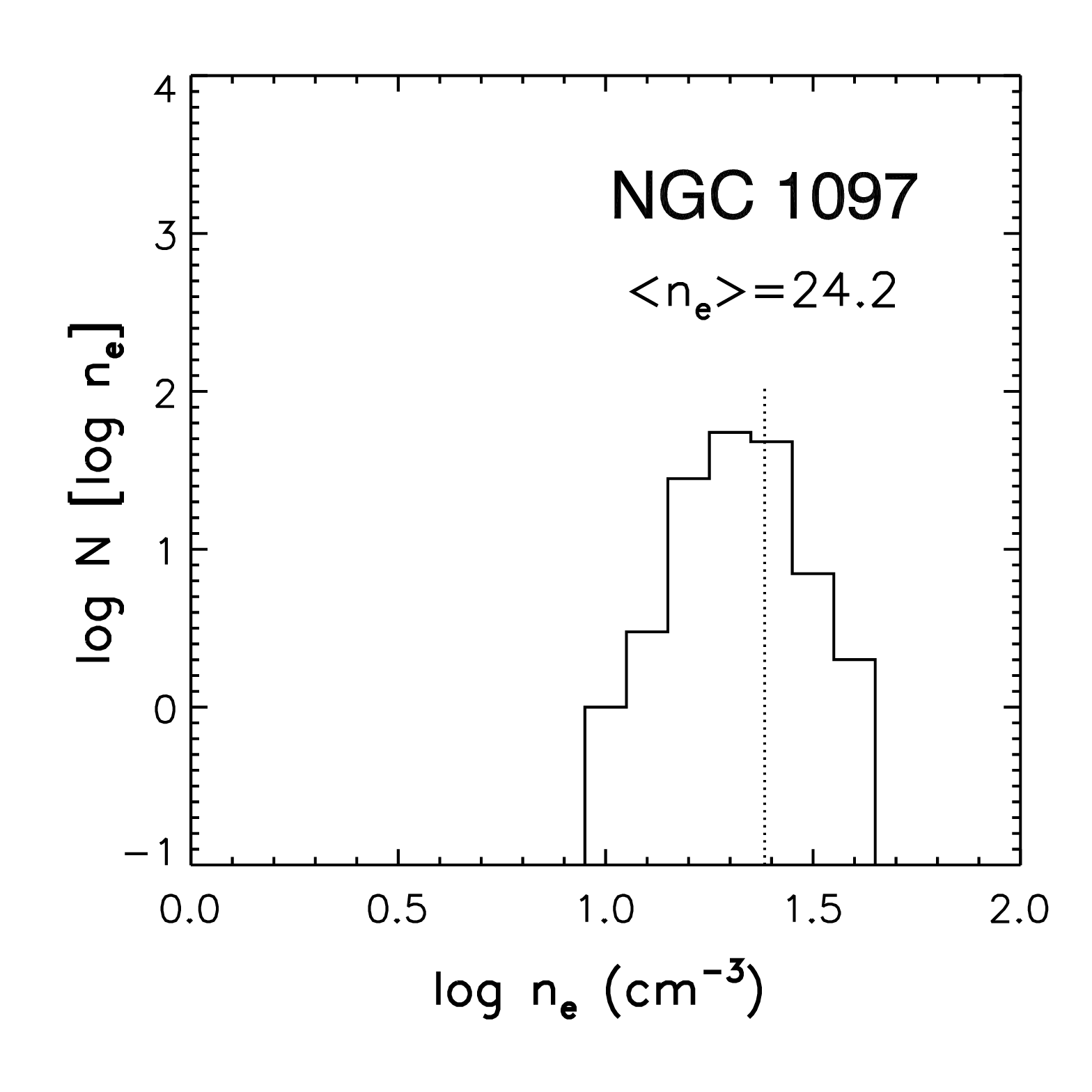}%
    \includegraphics[origin=c,scale=0.26,trim=0cm 10mm 1cm 5mm]{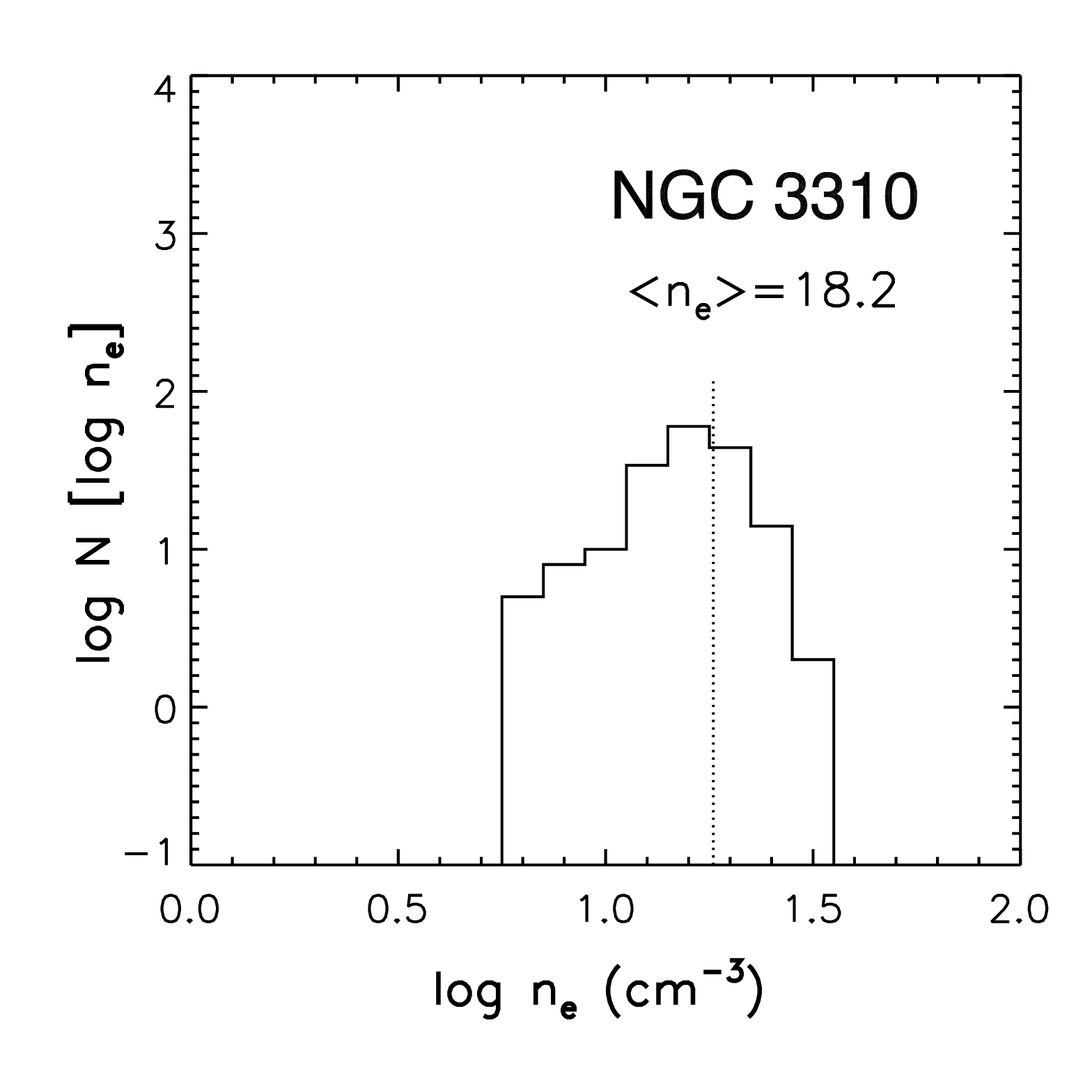}%
    \includegraphics[origin=c,scale=0.26,trim=0cm 10mm 1cm 5mm]{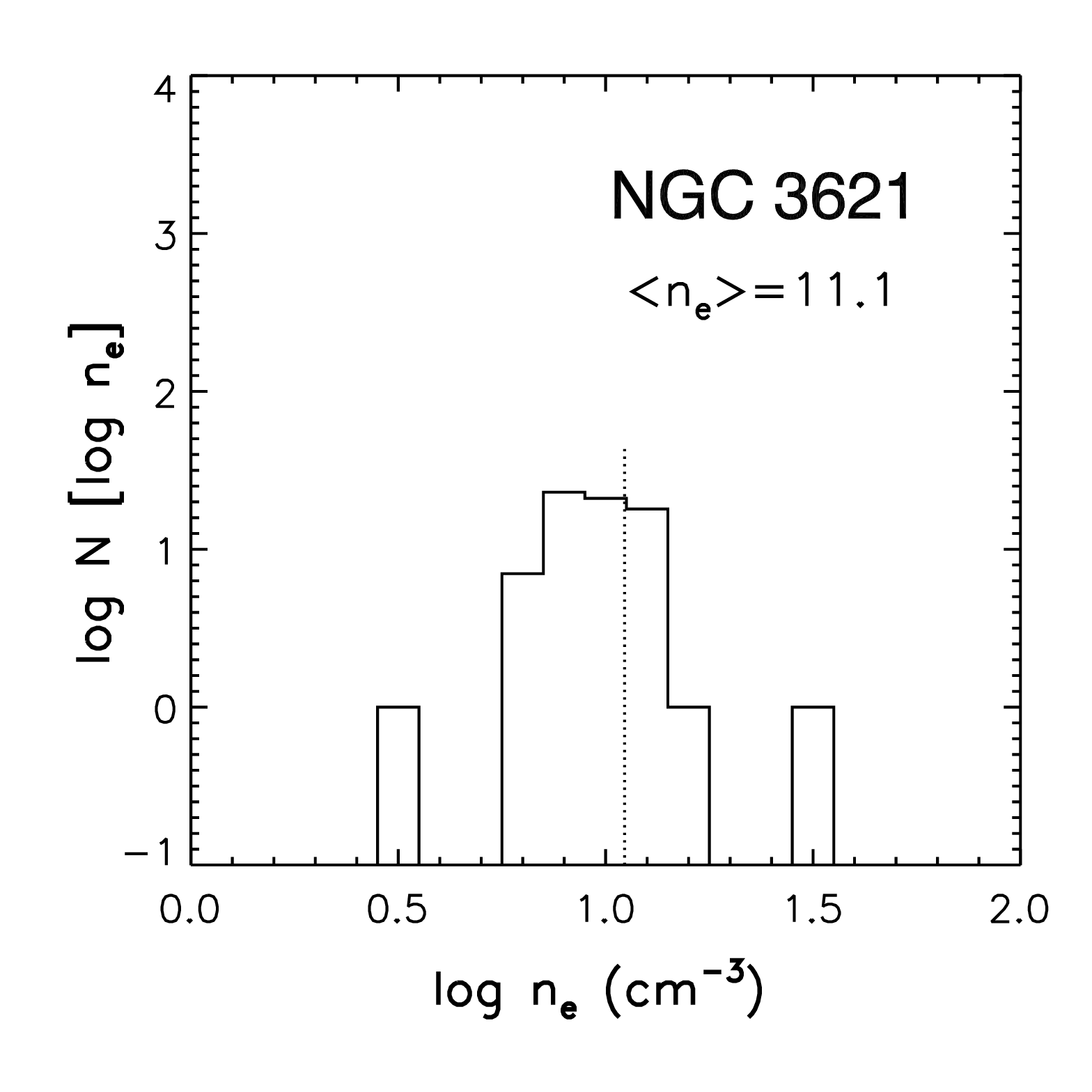}
    \includegraphics[origin=c,scale=0.26,trim=0cm 10mm 1cm 5mm]{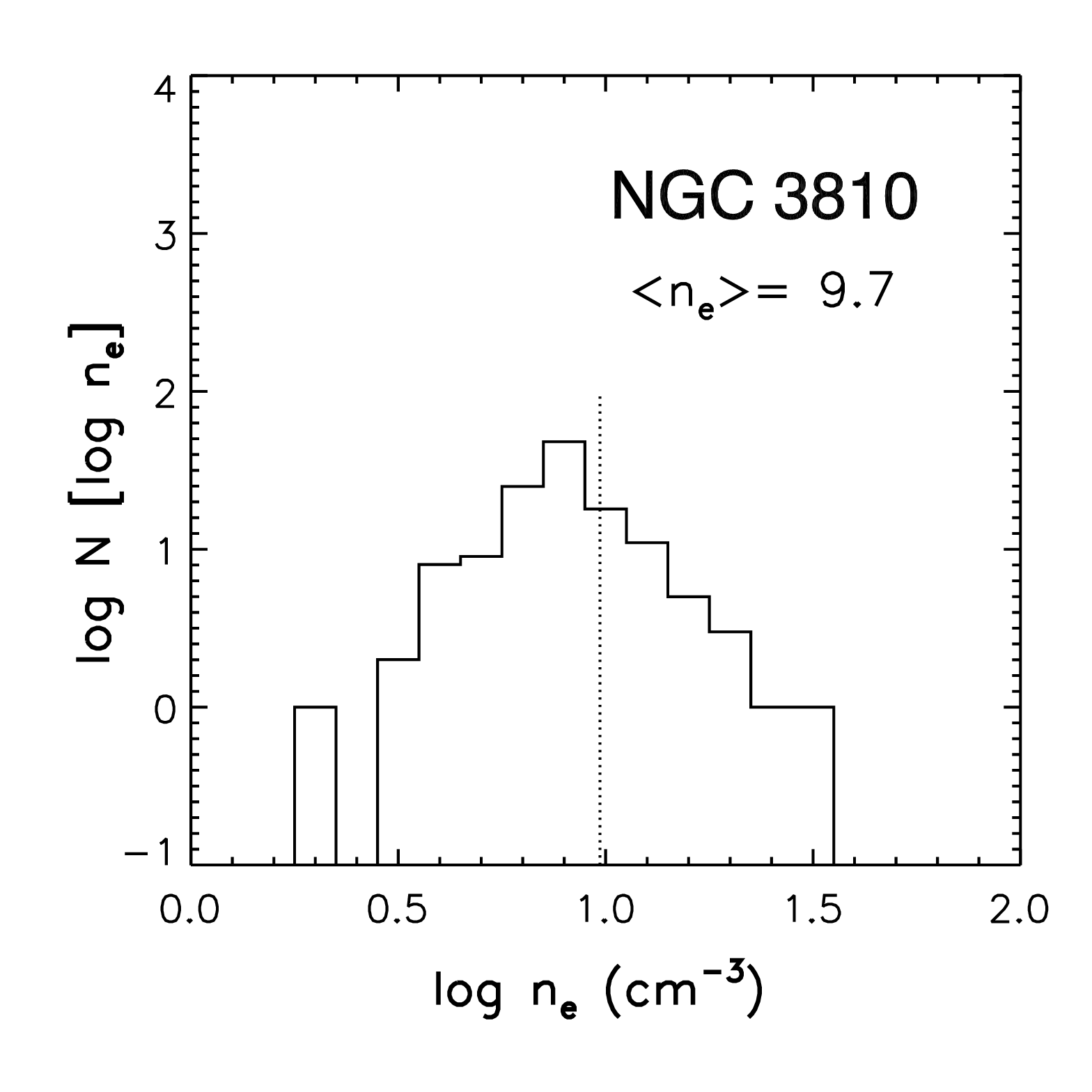}%
    \includegraphics[origin=c,scale=0.26,trim=0cm 10mm 1cm 5mm]{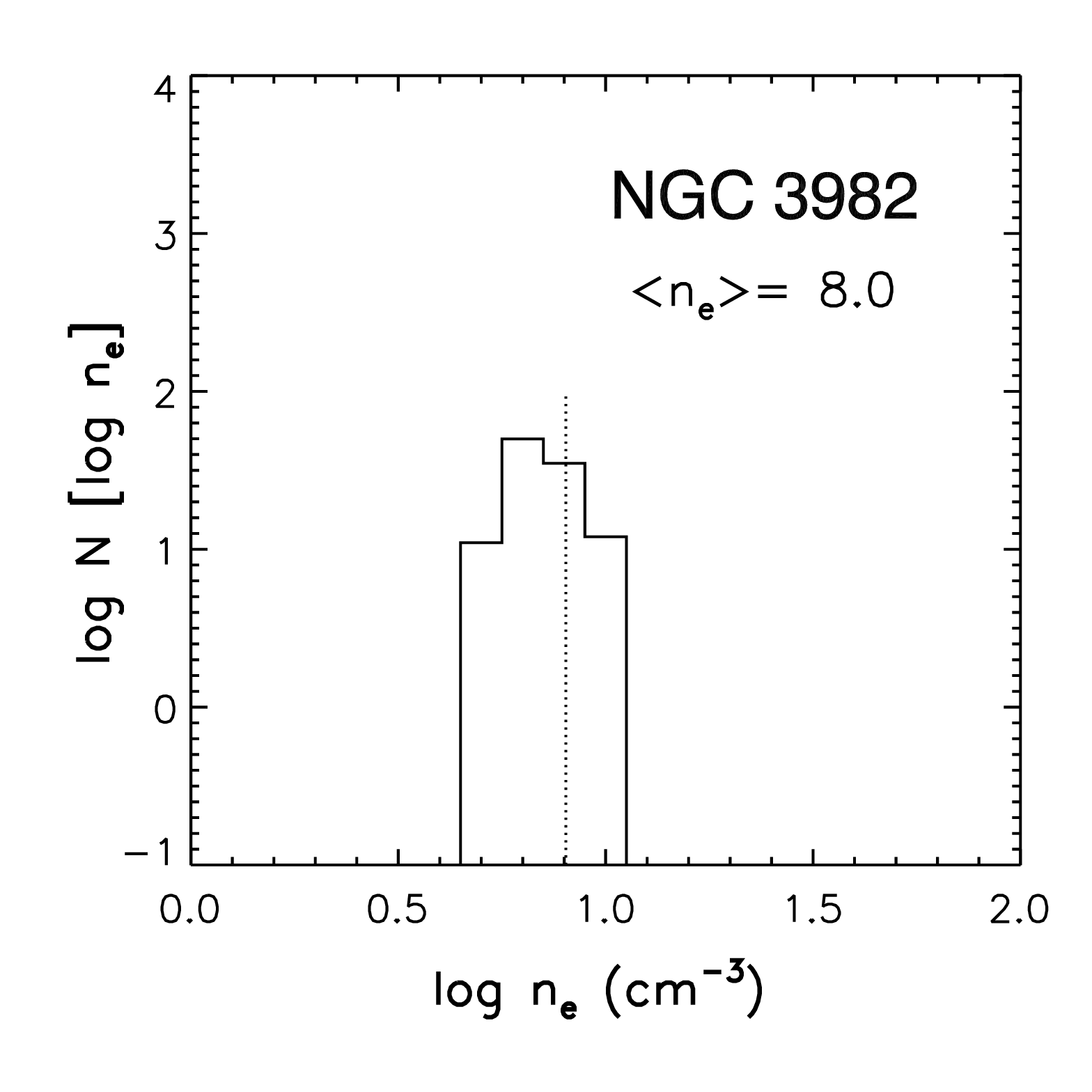}%
    \includegraphics[origin=c,scale=0.26,trim=0cm 10mm 1cm 5mm]{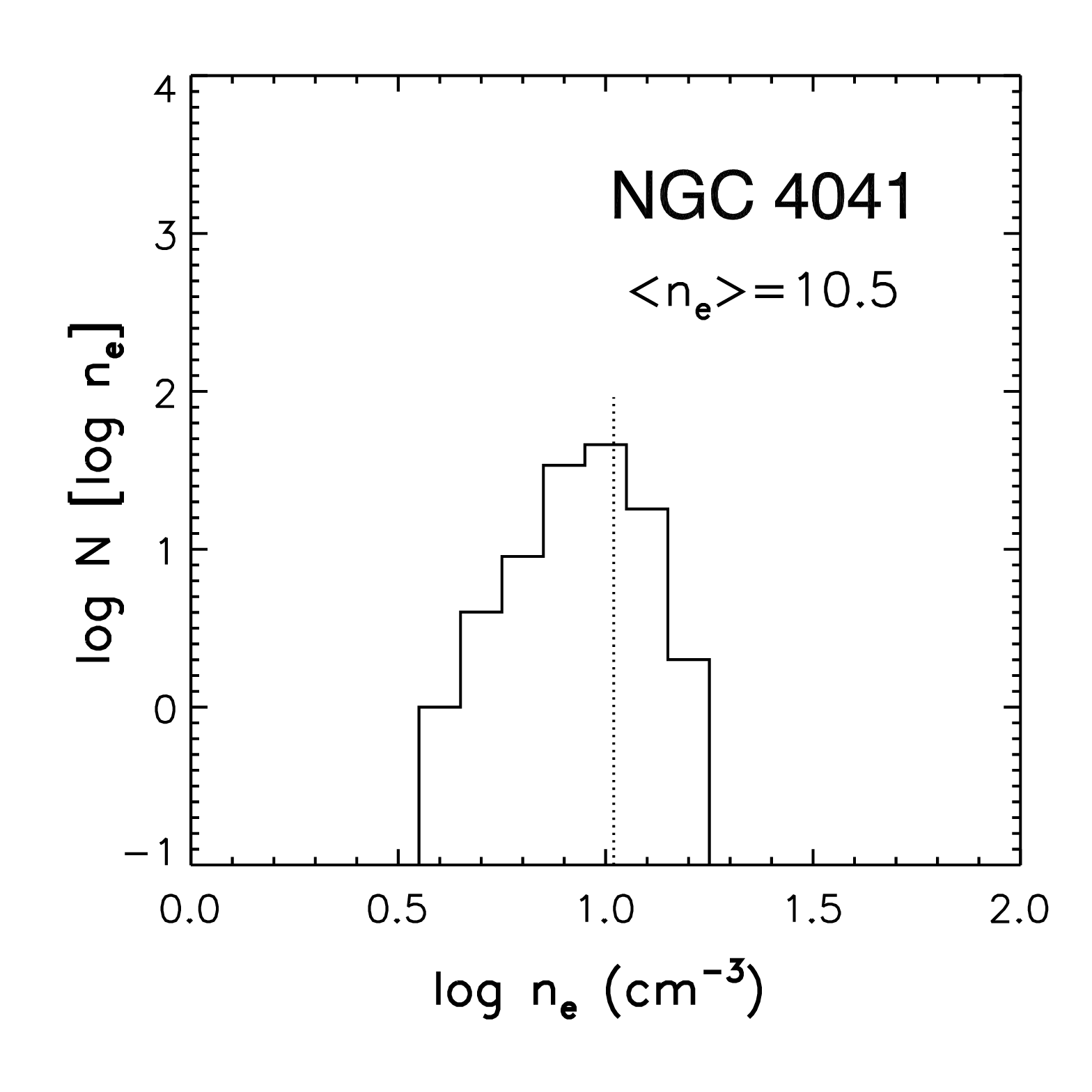}
    \includegraphics[origin=c,scale=0.26,trim=0cm 10mm 1cm 5mm]{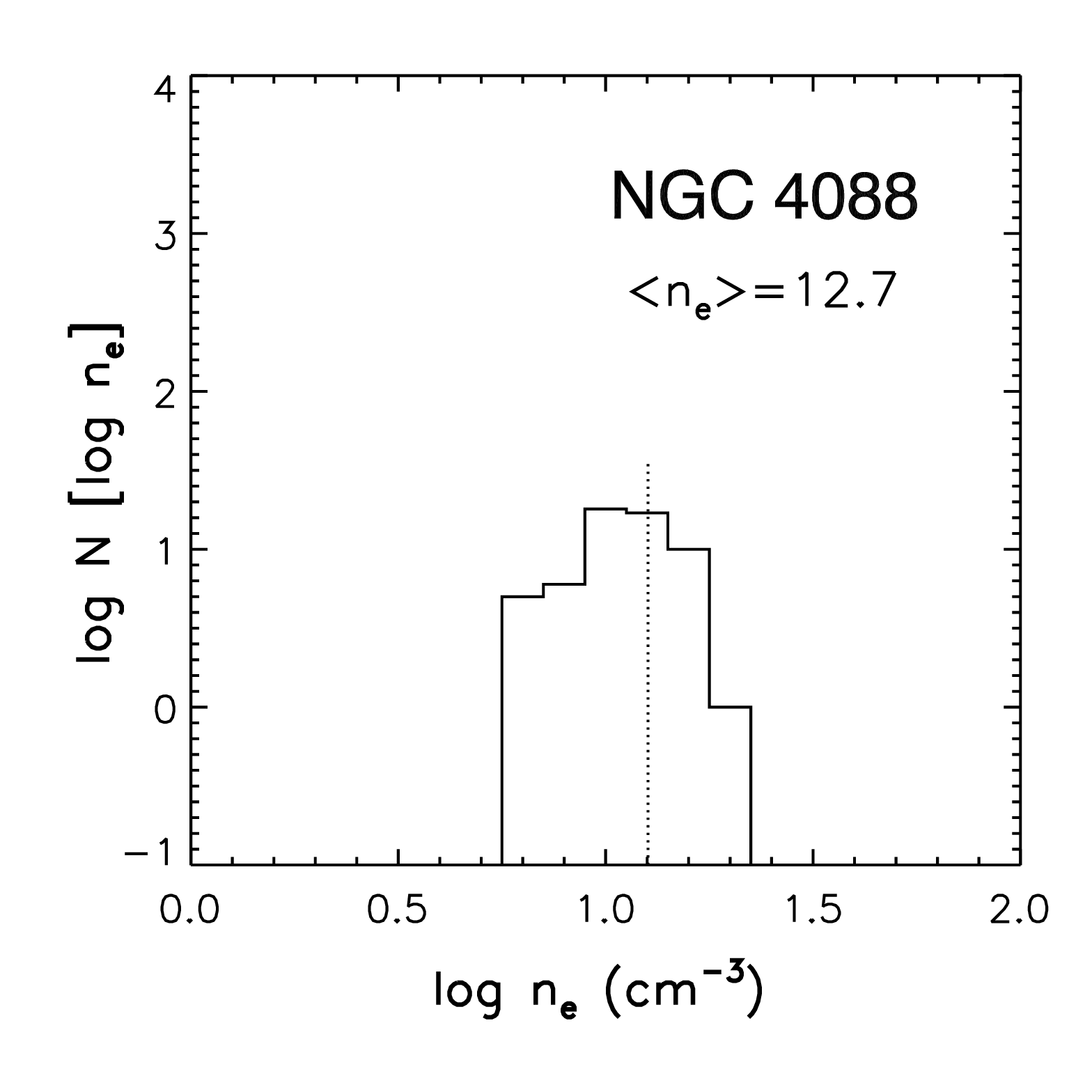}%
    \includegraphics[origin=c,scale=0.26,trim=0cm 10mm 1cm 5mm]{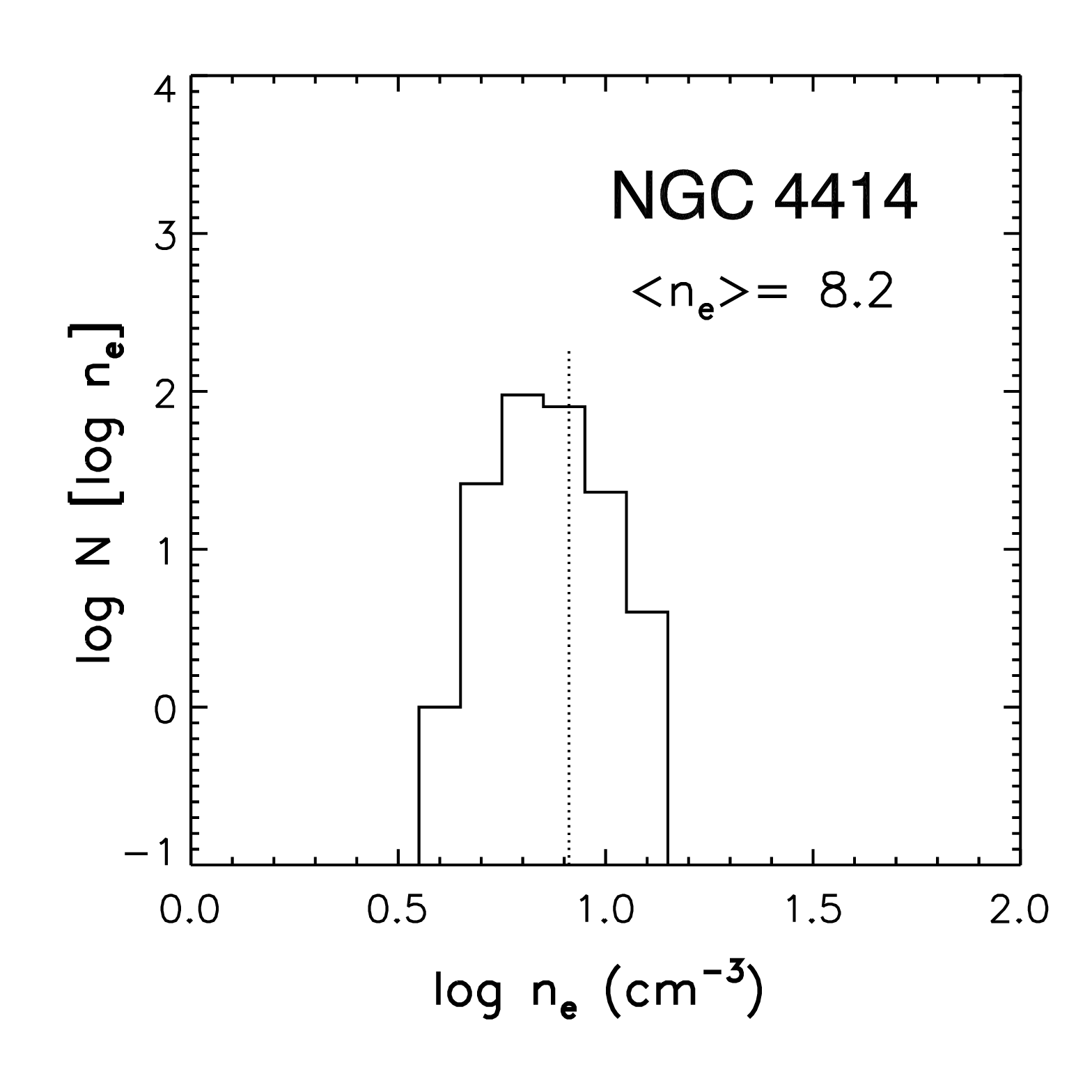}%
    \includegraphics[origin=c,scale=0.26,trim=0cm 10mm 1cm 5mm]{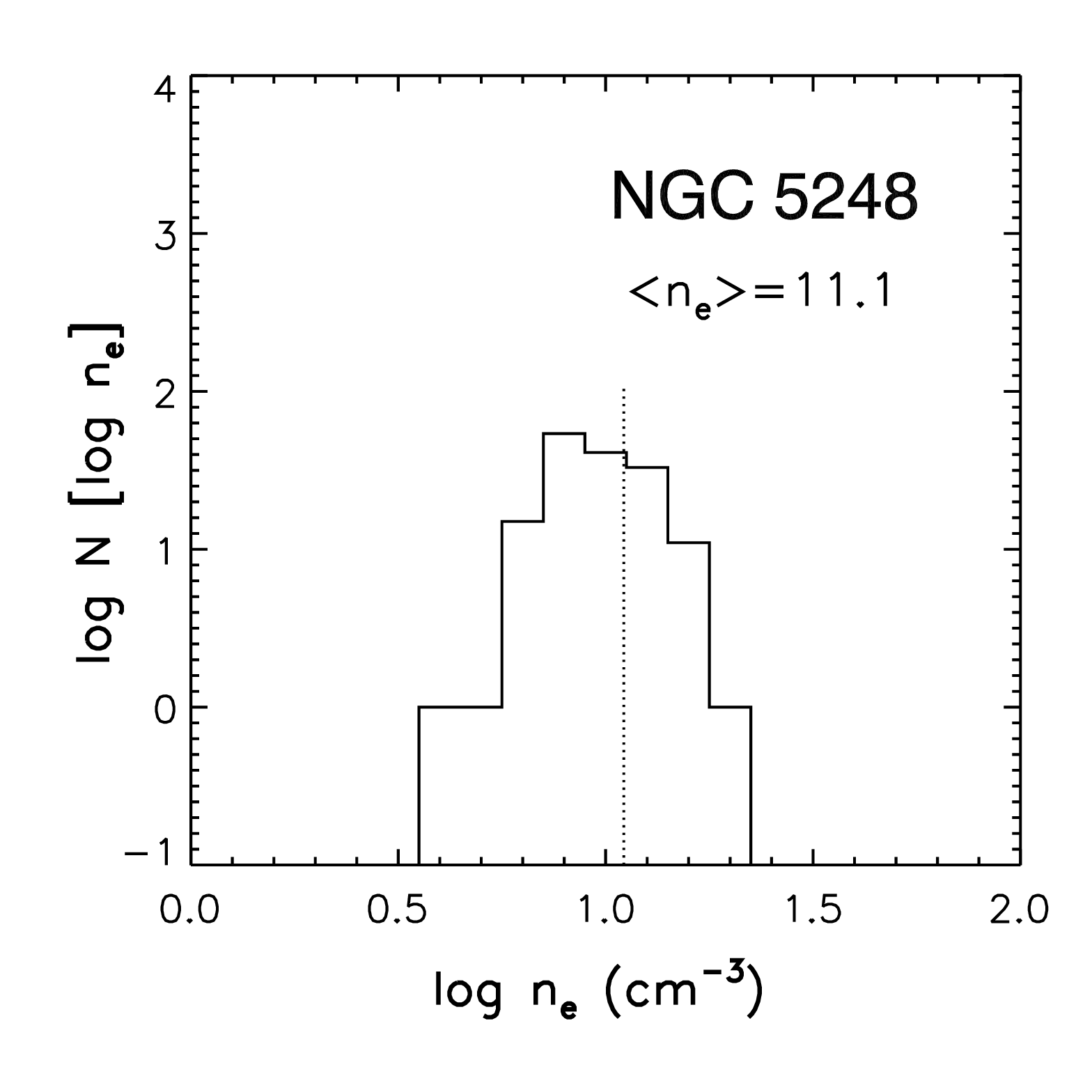}\\
\caption{The distribution of the mean electron density ($n_e$) of H {\sc ii} regions in 
individual galaxies. Note that the gas in H {\sc ii} regions is clumpy and filamentary
with local densities often much higher than the r.m.s. densities measured
here. The electron density averaged over each galaxy ($\langle n_e\rangle$)
is depicted by vertical dotted lines.}
\label{fig:nf_gold}
\end{figure*}

Assuming an intrinsic line ratio H$\alpha$/Pa$\alpha$=8 and an electron temperature $T_e=7500$ K (see \S
\ref{sec:h2_iden}), we express the average electron density of an H {\sc ii} region as \citep[][Equation 5]{Scoville01}
\begin{equation}
\frac{n_e}{\rm cm^{-3}}=34\left(\frac{L_{\rm Pa\alpha, intr}}{10^{36}~{\rm ergs~s^{-1}}}\right)^{1/2} \left(\frac{D}{\rm 10~pc}\right)^{-3/2},
\end{equation}
where the intrinsic Pa$\alpha$ luminosity is employed. As shown in Figure~\ref{fig:nf_gold}, the mean number 
density of electrons in each galaxy spans over the range $\langle n_e \rangle$=8--28 cm$^{-3}$ in our sample. 
With a typical dust extinction in H {\sc ii} regions of $A_V=2.2$ mag \citep{Calzetti07}, we expect the 
intrinsic $n_e$ to be larger by a factor of $\sim$1.2 assuming the Milky Way extinction curve \citep{Cardelli89}. 
The intrinsic mean electron density of the H {\sc ii} 
regions in our sample galaxies is therefore in the range 10--33 cm$^{-3}$, similar to the 
extinction corrected $n_e$=15--60 cm$^{-3}$ found in M51 \citep{Scoville01}.
However, we emphasize that the electron densities derived here are r.m.s. densities averaged over the entire
H {\sc ii} regions. The actual local densities are often significantly higher than these values because H {\sc ii}
regions tend to be highly clumpy and filamentary with filling factors of gas of order 1\% or less.

\section{Co-added Analysis}
\label{sec:coadd}

In H {\sc ii} region analyses of individual galaxies, one of the primary disadvantages of using Pa$\alpha$ 
compared to H$\alpha$ data is the smaller number statistics, driven by the small FoV of HST/NICMOS. 
Aiming for a better statistical significance, we co-add the H {\sc ii} region catalogues of our galaxies
and analyze the combined sample. However, we should caution that the co-added analyses implicitly employ
a strong assumption that a single universal luminosity function exists in all involved galaxies. 
For our galaxy sample with a median value for the LF power index $\tilde{\alpha}=-0.98\pm0.19$, a 
Kolmogorov-Smirnov test implies a probability of 85.1\% that the 12 measured $\alpha$ values are drawn 
from a Gaussian probability density function with $\mu=-0.98$ and $\sigma=0.19$.

\begin{figure*}
\centering
    \includegraphics[scale=0.35,clip,trim=0cm 0cm 0cm 0cm]{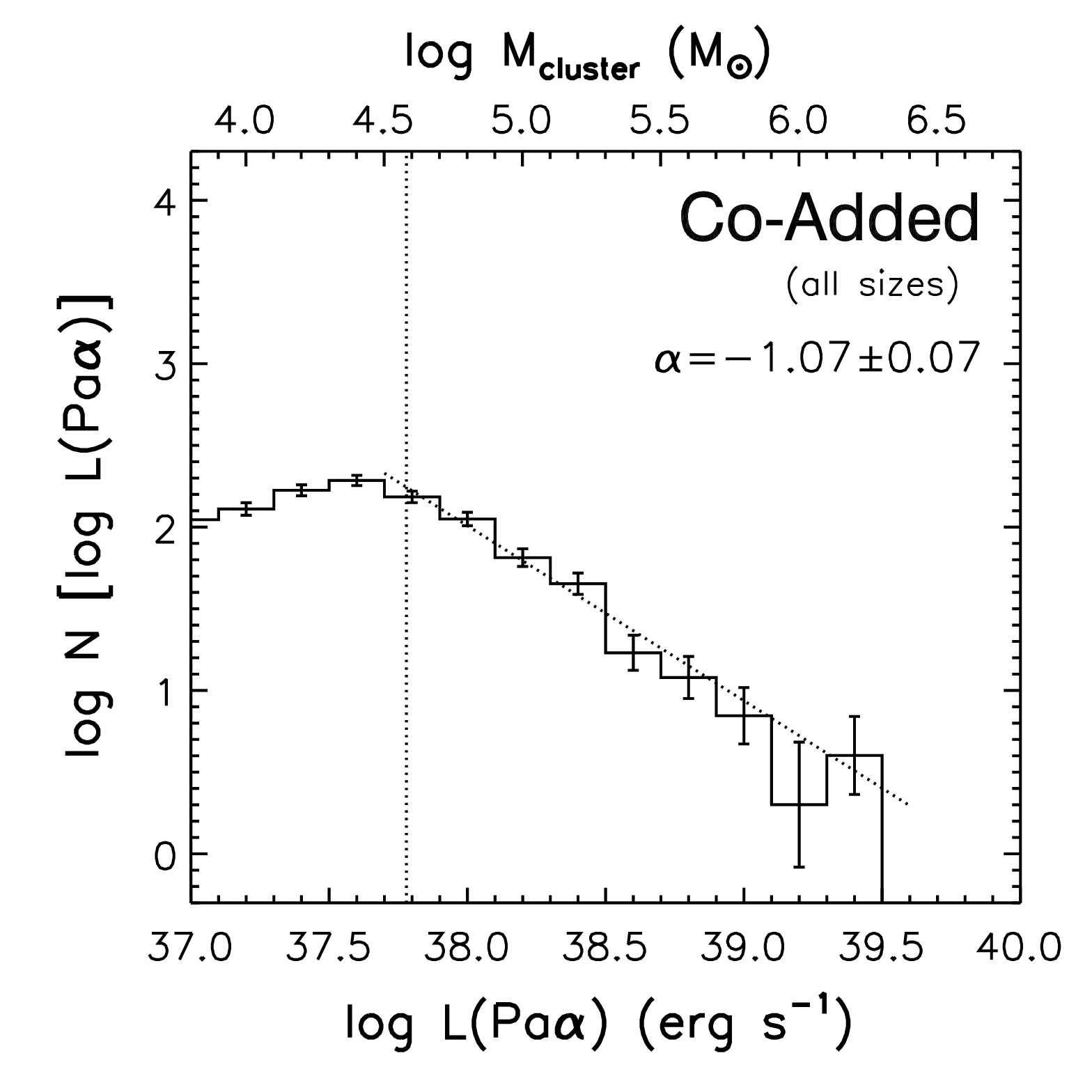}
    \includegraphics[scale=0.35,clip,trim=0cm 0cm 0cm 0cm]{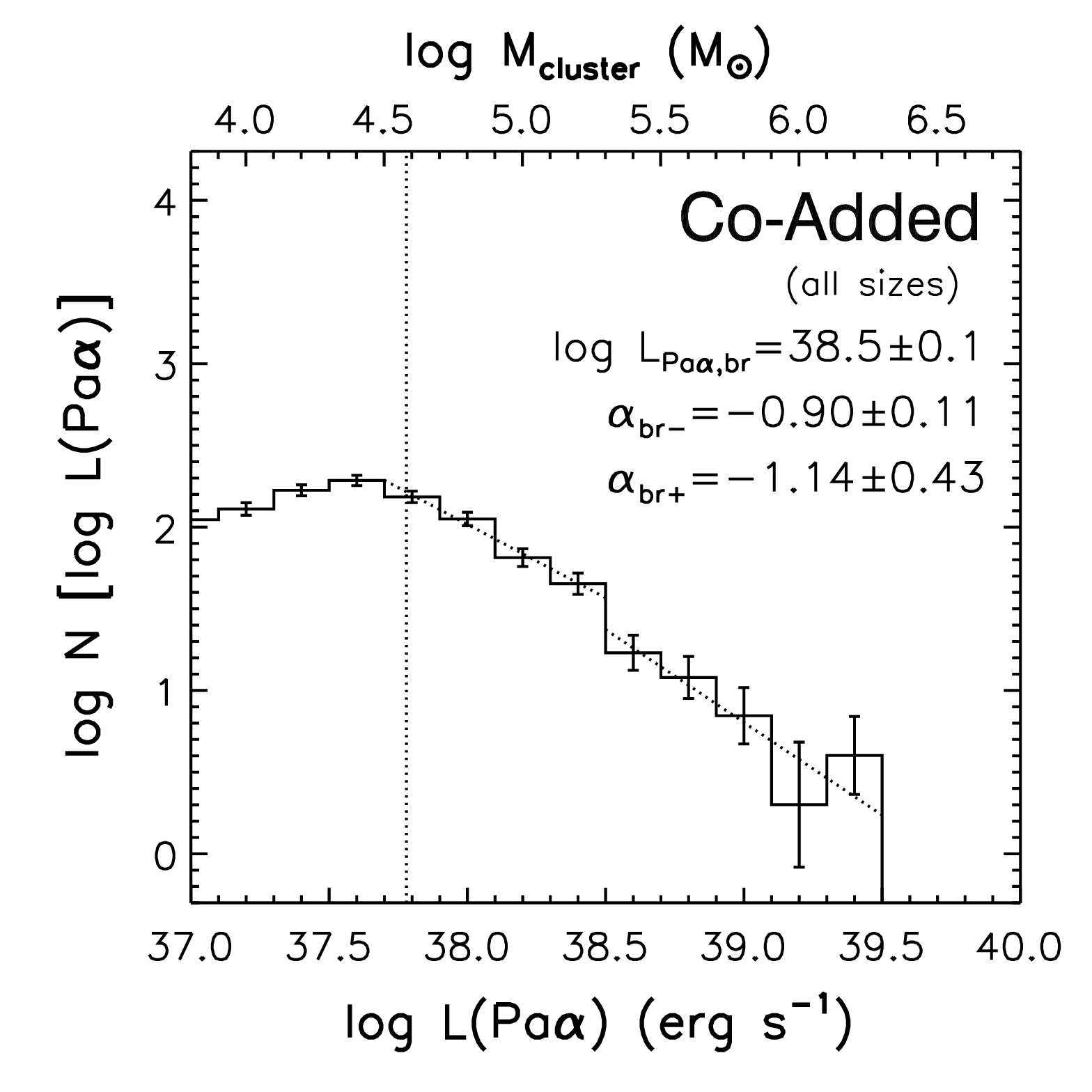}\\
\caption{The co-added Pa$\alpha$ luminosity function of H {\sc ii} regions for all objects 
above the completeness limit as denoted by the vertical dotted lines, fitted
by a single (left), or double (right) power law. The mass of the star
cluster as a function of $L_{\rm Pa\alpha}$ is calculated using the stellar population 
synthesis model {\sl Starburst99} \citep{Leitherer99} assuming solar metallicity, an age of 
4 Myr and the IMF given by \citet{Kroupa01} in the range 0.1--120 $M_{\odot}$.}
\label{fig:lf_coadd}
\end{figure*}

The co-added LF is shown in Figure~\ref{fig:lf_coadd}. The completeness limit of this combined H {\sc ii} 
catalogue, 10$^{37.78}$ erg s$^{-1}$, is set by NGC 4041 which has the shallowest image amongst the involved 
galaxies (in terms of H {\sc ii} region luminosity). Above this limit, the locus of the distribution can be
fitted by a single power law with an exponent of $\alpha=-1.07\pm0.07$ spanning over the range 
10$^{37.8\mhyphen39.4}$ erg s$^{-1}$ (Figure~\ref{fig:lf_coadd}-a). Although we do not observe convincing
evidence for the presence of a broken power law in our data, a potential break is seen at
$L_{\rm Pa\alpha,br} \sim 10^{38.5}$ erg s$^{-1}$, corresponding to an observed (i.e., not corrected 
for extinction) H$\alpha$ luminosity of $L_{\rm H\alpha,br} \sim 10^{38.6\mhyphen38.8}$ erg s$^{-1}$, 
consistent with the transition point $L_{\rm H\alpha,br}=10^{38.6\pm0.1}$ determined by \citet{Bradley06}. 
In Figure~\ref{fig:lf_coadd}-b we adopt this break luminosity and fit the regimes above and below this point, 
separately. We emphasize that this translation depends on our assumed constant extinction value. That is, if 
the extinction is lower than $A_V$=2.2 (see \S~\ref{sec:h2_iden}) then the Pa$\alpha$ luminosity break
will be considerably lower than the H$\alpha$ one for the same population.

We find that the best-fit power index is $\alpha_{\rm br+}=-1.14\pm0.43$ for the regions 
beyond $L_{\rm Pa\alpha,br}$, and $\alpha_{\rm br-}=-0.90\pm0.11$ for the fainter ones. 
These two indices are both approximately $-1$ when the uncertainties are 
taken into account, steeper (especially $\alpha_{\rm br-}$) than those found by \citet{Bradley06}, 
$\alpha_{\rm br+}=-0.86$ and $\alpha_{\rm br-}=-0.36$. The one order of magnitude lower angular resolution
of their data (seeing up to 3.7\arcsec~or $\sim300$ pc) is likely responsible for this difference. 
Another difference relative to previous results is that, despite that the breaks in our co-added LF and that of 
\citet{Bradley06} are both located at roughly the same luminosity, our two power laws are not only ``broken'',
but also ``discontinuous''. This discontinuity is sometimes also vaguely seen in the H$\alpha$ 
analyses of some individual galaxies \citep[e.g., NGC 5068, NGC 6300 etc, see][]{Helmboldt05}.

\begin{figure}
\centering
    \includegraphics[scale=0.35,clip,trim=0cm 0cm 0cm 0cm]{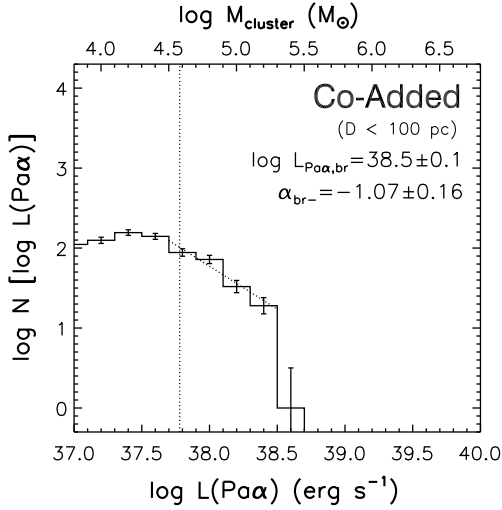}
    \vspace{-3mm}
\caption{The co-added Pa$\alpha$ luminosity function of H {\sc ii} regions for all objects 
as significant as in Figure~\ref{fig:lf_coadd}, but with a diameter smaller than 100 pc only.
The mass of star cluster is calculated the same way as Figure \ref{fig:lf_coadd}.}
\label{fig:lf_coadd_cut}
\end{figure}

It would be intriguing if the parameter(s) that may differentiate the H {\sc ii} regions that populate the nearly 
parallel double power laws could be tracked down.
As an exploratory experiment, we set a cut-off scale $D_{\rm cut}$, construct the luminosity function excluding 
regions with $D>D_{\rm cut}$, and observe its corresponding variation by successively changing $D_{\rm cut}$. 
When $D_{\rm cut}$ decreases from the largest observed value of $D$, we see that the regions of the LF locus 
that gets more drastically depopulated are those at the bright end, until nearly all regions disappear 
when $D_{\rm cut}$=100 pc. During this process, the power scaling on the faint side of $L_{\rm Pa\alpha,br}$ 
remains almost unchanged, both in shape and in index. When $D_{\rm cut}$ decreases further, the fainter locus starts 
to lose objects in a significant manner that starts changing the power law distribution. We show in 
Figure~\ref{fig:lf_coadd_cut} the co-added LF populated by H {\sc ii} regions smaller than 100 pc in diameter. 
We therefore consider $\sim$100 pc the characteristic scale that differentiates the two H {\sc ii} region populations 
which construct their respective power scalings on either side of the break luminosity, if the break is real. 
Excluding regions that have sizes $>100$ pc, the power index of the co-added luminosity function for sub-break regions 
becomes slightly steeper ($\alpha_{\rm Pa\alpha,br-}=-1.07\pm0.16$) than otherwise ($\alpha_{\rm Pa\alpha,br-}=-0.90\pm0.11$), 
and slightly more consistent with the mean slope for individual galaxies, $\langle\alpha\rangle=-1.05$. 
However, the discrepancy of the data from a single power law is at the level of $\sim$1$\sigma$ uncertainty of a 
typical data point and is therefore insignificant. Although a fraction of the regions larger than $\sim$100 pc 
are probably blends, the existence of large H {\sc ii} regions are unquestionable and we cannot arbitrarily attribute 
the break to source blending effects.

We have mentioned in Section \ref{sec:h2lf_ind} that H {\sc ii} regions more luminous than $\sim$10$^{38.5}$ 
erg s$^{-1}$ are not seen in four galaxies. Although excluding these objects from the co-added analysis will 
impose a further criterion that may bias our sample, including them for testing the existence of the break 
raises concerns, as well. In fact, we find that removing these galaxies from the sample will only flatten
the sub-break power index slightly ($\alpha_{\rm br-}=-0.82\pm0.11$), and the (in)significance of the break
and discontinuity remains unchanged, because the majority of the H {\sc ii} regions contributed by these four
galaxies are fainter than the completeness limit (10$^{37.78}$ erg s$^{-1}$).

\begin{figure}
\centering
\includegraphics[scale=0.35,clip,trim=0cm 0cm 0cm 0cm]{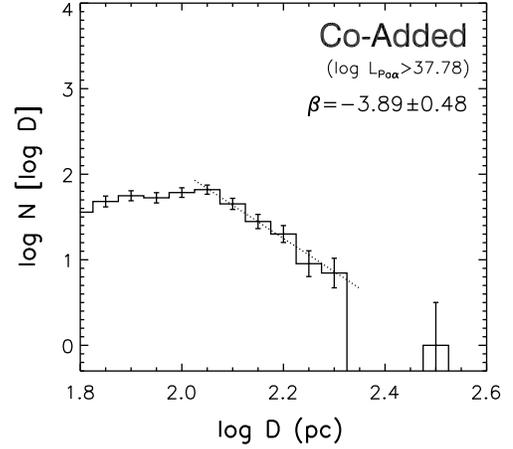}
\vspace{-3mm}
\caption{The co-added size distribution of H {\sc ii} regions for all objects above the 
completeness limit. A well-shaped power law exists over the range 110--220 pc although 
some of the largest regions are possibly blends.}
\label{fig:sf_coadd}
\end{figure}

We show in Figure~\ref{fig:sf_coadd} the co-added size distribution of the H {\sc ii} regions in our
sample galaxies, excluding all the regions fainter than the completeness limit, $10^{37.78}$ erg s$^{-1}$,
set by the shallowest image, i.e., that of NGC 4041. A well-shaped power law exists over the range 110--220 pc,
with a slope $\beta=-3.89\pm0.48$, consistent with the individual-galaxy analysis presented 
above. Although some regions at the large-size end are possibly blended regions or complexes, this 
locus may still reflect the intrinsic clustering properties of compact H {\sc ii} regions to some extent.

\section{DISCUSSION}

The LF of H {\sc ii} regions ionized by OB star associations are physically related to the mass 
function of compact star clusters and the mass spectrum of giant molecular clouds in galaxies, 
because young stars are formed predominantly in Giant Molecular Clouds (GMCs) and ionize the 
surrounding interstellar medium (ISM) to produce gaseous nebulae. 

For a constant and universal stellar Initial Mass Function (IMF), the number of ionizing photons emitted 
by the short-lived, massive stars is proportional to the total mass of such stars, which in turn is 
proportional to the total mass of the star cluster, if the latter is sufficiently young that no 
supernovae have exploded yet (age $<$3--4 Myr). In this case, the compact cluster mass function should 
follow the same power law as the H {\sc ii} region LF. Recent HST studies find that 
$dN/dM_{\star} \propto M_{\star}^{-2}$ \citep{Chandar10,Chandar11}, for clusters around $10^8$ Myr,
which is a power law similar to that found for the H {\sc ii} region LF. This result would argue 
for cluster destruction/dissolution mechanisms that are mass-independent \citep{Chandar10}, in order 
to preserve the constant power law shape between the younger H {\sc ii} regions LF and the older 
clusters mass function.

The relation between H {\sc ii} LFs and GMC mass spectra (expressed as 
$dN/dM_{\rm GMC} \propto M_{\rm GMC}^{\zeta}$) is more sophisticated, because the efficiency of 
star formation is likely a function of the local environment. Empirically, the star 
formation rate is linked to the density of cold gas through a single power-law scaling relation
(the Schmidt-Kennicutt law, abbreviated as the S-K law; see \citet{Kennicutt98} and references therein) 
which takes the form $\Sigma_{\rm SFR}\propto\Sigma_{gas}^{\gamma}$. Ideally, the extragalactic 
S-K law should be studied for each individual cluster and its parent GMC, and the {\it molecular} 
S-K law be expressed as ${\rm SFR} \propto L_{\rm Pa\alpha} \propto M_{\rm GMC}^{\gamma_{\rm H2}}$, if 
observed in Pa$\alpha$. For the power-law Pa$\alpha$ H {\sc ii} LF that we find in this work, the GMC 
mass spectra would have an exponent of $\zeta=-(1+\gamma_{\rm H2})$. Although $\gamma_{\rm H2}$ depends 
on the sampling scale as shown by \citet{Liu11} and \citet{Calzetti12}, for the case of $\gamma_{\rm H2}=1.4$ 
as found in M51 at 750 pc resolution by \citet{Kennicutt07}, this relation predicts $\zeta=-2.4$, in rough 
agreement with the observed exponent for the GMC mass function. Recent observations have found the value of 
$\zeta$ to be $-$(1.5--2.1) in the Galaxy, $-1.7$ in LMC, and $-2.9$ in M33, therefore varies 
significantly across the Local Group, but a truncation at the maximum mass of $10^{6.5}~M_{\odot}$ 
exists \citep{Rosolowsky05}. The above calculation, assuming all molecular clouds can be well detected, 
well separated, are all forming stars, and each can be clearly identified as the parent cloud of a 
star cluster, is far from being realistic, and is presented as heuristic discussion only.

For an H {\sc ii} region with a Pa$\alpha$ luminosity of 10$^{38.5}$ ergs s$^{-1}$, the corresponding
star cluster has a mass of (2--3)$\times10^5 M_{\odot}$ for ages 4--5 Myr, if calculated using the stellar 
population synthesis model {\sl Starburst99} \citep{Leitherer99}, assuming the IMF given by \citet{Kroupa01}
in the range 0.1--120 $M_{\odot}$ and solar metallicity. If such a star cluster is formed in a
GMC with the maximum mass, $10^{6.5}~M_{\odot}$, the star formation efficiency will be 10\% or less,
consistent with \citet{Leroy08} and \citet{Bigiel08} (note that the bright end of the S-K law is 
considered here, where the influence of preserving the diffuse stellar/dust emission mostly due to 
old stellar populations is insignificant, see \citet{Liu11} and \citet{Calzetti12} for detailed discussion 
on the influence of the local diffuse emission on deriving spatially-resolved star formation properties 
of nearby galaxies).

\section{SUMMARY}

We have analyzed the HST NICMOS Pa$\alpha$ images of twelve nearby galaxies and studied the 
luminosity function (LF) and size distribution of the H {\sc ii} regions 
both in individual galaxies and cumulatively. 
Our analysis is the first using a sample of galaxies, rather than individual objects, with HST data.
The use of high-angular resolution data offers an advantage over ground-based investigations (mainly 
in H$\alpha$) because it avoids severe blending (spatially or in projection) of H {\sc ii} regions.

The near-IR Pa$\alpha$ hydrogen line is virtually an extinction-free trace of newly formed massive stars 
and star associations. The resolution of our Pa$\alpha$ data (0.26\arcsec) translates to 9--29 pc at the 
distances of our sample galaxies (7--23 Mpc), satisfying the requirement for resolution to reliably measure 
the properties of H {\sc ii} regions \citep[$<$40 pc,][]{Pleuss00}. Employing the {\sl HIIphot} IDL software
for the identification and photometry of H {\sc ii} regions, we show that the luminosity-diameter relation 
of our catalogued H {\sc ii} regions, characterized by $L\propto D^{\eta}$ with $2.5<\eta<3$ for 10 galaxies.

In the individual-galaxy analysis, we do not confirm the double power-law LF found by typical ground-based 
works. Instead, all the twelve galaxies exhibit Pa$\alpha$ LFs that follow a single power 
law $dN(L_{\rm Pa\alpha})/d\ln L_{\rm Pa\alpha}\propto L_{\rm Pa\alpha}^{\alpha}$ with $\alpha \simeq -1$, 
consistent with the investigations of both Galactic radio H {\sc ii} regions and the HST studies of M51 
and M101. For the whole sample, we find a median value for the LF power index
$\tilde{\alpha}=-0.98\pm0.19$. The size distribution of H {\sc ii} regions, fitted to
$dN(D)/d\ln D \propto D^{\beta}$, shows significantly larger variation of the exponent than the LF, with 
two galaxies showing no apparent power scaling at all. Excluding these two objects, we find the median value
$\tilde{\beta}=-3.10\pm1.04$, consistent with $\beta=-3.12\pm0.90$ found by \citet{Oey03}. 

The co-added LF can be fitted by a single power law with an exponent of $\alpha=-1.07\pm0.07$ over 
a luminosity range of 1.6 orders of magnitude, but a possible break is seen at
$L_{\rm Pa\alpha,br} \sim 10^{38.5}$ erg s$^{-1}$, roughly corresponding to the transition luminosity
determined by ground-based H$\alpha$ observations. When fitted with a disconnected double power law, 
the co-added LF has a best-fit power index $\alpha_{\rm br+}=-1.14\pm0.43$ for the brighter regions 
and $\alpha_{\rm br-}=-0.90\pm0.11$ for the fainter ones. The power scaling of the brighter regions 
is mainly populated by regions with diameters larger than 100 pc, which may be contaminated with blends. 
Excluding these objects results in a single power law with $\alpha=-1.07\pm0.16$. 
We conclude that, the LFs, both individual and co-added, are consistent with a single power law with 
$\alpha=-1$, as expected from the mass function of star clusters in nearby galaxies. The co-added size 
distribution shapes a power law with a slope $\beta=-3.89\pm0.48$ over the range 110--220 pc. 
This scaling, in spite of its consistency with the individual-galaxy analysis, is probably contaminated 
by blended regions or complexes.

\acknowledgments

We thank the anonymous referee for a careful reading of the manuscript.
Partial support for program GO-11080 was provided by NASA through a grant from the Space Telescope 
Science Institute, which is operated by the Association of Universities for Research in Astronomy, Inc., 
under NASA contract NAS 5-26555.
This research has made use of the NASA/IPAC Extragalactic Database (NED) which is operated by the 
Jet Propulsion Laboratory, California Institute of Technology, under contract with the National 
Aeronautics and Space Administration. We also acknowledge the usage of the HyperLeda database 
(http://leda.univ-lyon1.fr).



{\it Facilities:} \facility{HST (NICMOS)}.

\end{document}